\documentclass[11pt]{article}

\usepackage{graphicx}
\usepackage{amsmath}
\usepackage{dcolumn}
\usepackage{bm}
\usepackage{slashed}
\usepackage[bookmarks,bookmarksnumbered,colorlinks=true,anchorcolor=blue,
linkcolor=blue,urlcolor=blue,citecolor=blue,breaklinks=true]{hyperref}
\usepackage[utf8]{inputenc}
\usepackage{authblk}
\usepackage{cite}
\usepackage{titlesec}
\usepackage{caption}
\usepackage{subcaption}
\usepackage{float}
\usepackage{color}
\usepackage[normalem]{ulem}
\usepackage{amssymb}
\usepackage{booktabs}
\usepackage{array}
\usepackage{tabularx}
\usepackage[table]{xcolor}
\usepackage{stmaryrd}

\newcommand{\be}{\begin{equation}}
\newcommand{\ee}{\end{equation}}
\newcommand{\ba}{\begin{eqnarray}}
\newcommand{\ea}{\end{eqnarray}}
\def\bea{\begin{eqnarray}}
\def\eea{\end{eqnarray}}

\newcommand{\gsim}{\mathrel{\hbox{\rlap{\lower.55ex \hbox {$\sim$}}
                   \kern-.3em \raise.4ex \hbox{$>$}}}}
\newcommand{\lsim}{\mathrel{\hbox{\rlap{\lower.55ex \hbox {$\sim$}}
                   \kern-.3em \raise.4ex \hbox{$<$}}}}

\def\roughly#1{\mathrel{\raise.3ex\hbox{$#1$\kern-.75em%
\lower1ex\hbox{$\sim$}}}}
\def\lsim{\roughly<}
\def\gsim{\roughly>}

\def\({\left(}
\def\){\right)}
\def\[{\left[}
\def\]{\right]}
\def\<{\langle}
\def\>{\rangle}

\usepackage{xcolor}

\begin{document}
\title{\bf Quantum Information Flow under String-Diagram Rewriting}

\author[]{Yi-Yu Lin$^{1,2}$ \thanks{yiyu@simis.cn}}
\author[]{Chen-Ye Li$^{3}$ \thanks{endgamelichenye@gmail.com}}

 \affil{${}^1$Fudan Center for Mathematics and Interdisciplinary Study, Fudan University, Shanghai, 200433, China}
\affil[]{${}^2$Shanghai Institute for Mathematics and Interdisciplinary Sciences (SIMIS), Shanghai, 200433, China}
 \affil{${}^3$Department of Physics , Fudan University, Shanghai, 200433, China}


\maketitle

\begin{abstract}
We revisit the notion of ``quantum information flow'' introduced in Bob Coecke's early work~ \cite{Coecke:2004sxv,Coecke:2005bin} and seek to give it an explicit string-diagrammatic formalization. 
Given a semantics-preserving string-diagram rewriting sequence, we first distinguish apparent through-paths, which depend on the current graphical presentation, from genuine through-paths, which can be compatibly inherited through successive rewrites to a terminal decoupled bare wire factor. We then formally define a ``Coecke flow line’’ in terms of this compatible-inheritance relation.
 In particular, we use the ZX calculus, i.e., the ZX string-diagram rewriting system, to illustrate the resulting formalism. 
In the physical setting of quantum protocols, a Coecke flow can be interpreted as a constrained, quasi-local, line-like presentation of a target bare wire morphism factor within the protocol string diagram.
\end{abstract}

\tableofcontents

\newpage

\section{Background and Introduction}

The notion of ``quantum information flow'' that lies at the centre of this paper should be traced back to Bob Coecke's reinterpretation of celebrated protocols such as quantum teleportation in \cite{Coecke:2004sxv,Coecke:2005bin}. This reinterpretation aims to reveal the internal information-flow structure of such protocols in a process-theoretic language, which can ultimately be made rigorous by category theory. Traditional quantum information theory has already understood quantum teleportation as a rigorously constructed protocol, and has proved that its overall input--output action is equivalent to an identity channel, $\mathcal{E} = \mathrm{id}$\cite{Bennett:1992tv, Nielsen:2012yss }. What categorical quantum mechanics\cite{Abramsky:2004doh, Abramsky:2008qkz, Coecke:2005clw }, developed by Bob Coecke, Samson Abramsky, and others in the early 2000s, adds to this kind of traditional proof of channel equivalence is, one may say, an intrinsic process perspective. More specifically, it transfers the correctness of a quantum protocol from matrix calculations in Hilbert space into string-diagrammatic reasoning under categorical semantics. In this graphical syntax, a ``list of protocol steps'', originally consisting of local operations, classical communication, measurements, and so on, is organized into a string diagram with a rigorous categorical semantics, thereby further revealing the internal process structure behind the overall channel equivalence certified by traditional quantum information theory.

The present paper is devoted specifically to formalizing the notion of ``quantum information flow''. This is because, although Coecke explicitly used the intuitive notion of quantum information flow in his early work to describe the information-flow structure supporting those celebrated protocols~\cite{Coecke:2004sxv,Coecke:2005bin}, later developments placed this notion in a rather special state. It soon inspired more systematic and more powerful research directions, namely categorical quantum mechanics~\cite{Abramsky:2004doh, Abramsky:2008qkz, Coecke:2005clw } and, later, the ZX calculus~\cite{Coecke:2008lcg, Duncan:2009ocf, vandeWetering:2020giq, Coecke:2017dti}. In other words, the original intuition of ``quantum information flow'' successfully gave birth to more mature formal tools, but precisely for this reason, the name itself was not continuously polished as an independent object of definition. It largely receded, rather, into an heuristic or explanatory expression: when a string-diagrammatic rewriting is finally revealed as a bare wire, one says that the information flow inside it has been graphically characterized. 
In a certain sense, the present paper attempts to bring this early notion back to the foreground and to endow it with an explicit and testable form. What concerns us is not merely the global semantic fact that a complicated protocol string diagram can eventually be reduced, through a sequence of string-diagrammatic rewrites, so as to reveal a distinguished ``bare wire''. Rather, we ask whether the bare-wire factor ultimately revealed by the rewriting can be assigned a constrained line-like representative in the initial complicated presentation.
For this purpose, we first distinguish the apparent through-goingness temporarily possessed by a line in a particular string-diagrammatic presentation from the genuine through-goingness, which should be acquired through semantics-preserving string-diagrammatic rewriting. Consider a sequence of semantics-preserving rewrites. Suppose that an apparent through-path in the initial diagram can be tracked step by step in the following sense: at every rewriting step, the through-path representative in the preceding intermediate diagram admits a compatible successor representative in the next intermediate diagram; and suppose that this tracking can be continued until the final representative is carried simply by a bare-wire factor tensor-decoupled from the remainder of the diagram. We then say that the initial apparent through-path realizes genuine through-goingness.
Furthermore, in the physical setting of a quantum-protocol process diagram with classical measurement branches, if the same initial through-path, chosen independently of the classical branch, can be certified in this manner in every fixed branch, we say that it defines a ``quantum information flow''. To emphasize the relation between this reformalization and the intellectual lineage of Coecke's early thought, we shall also call such a flow a Coecke flow.

It is worth noting that, in Coecke's original context\cite{Coecke:2004sxv}, quantum information flow is mainly used to explain the mechanism of information transmission inside certain concrete protocols, such as teleportation and entanglement swapping, and thereby to reveal the intrinsic mechanism by which these protocols work. 
After the formalization developed in the present paper, however, quantum information flow may be particularly well suited to studying a quantum state itself that possesses a nontrivial intrinsic entanglement structure, especially when that structure admits a nontrivial and manifestly geometrical representative organization.
Particularly interestingly, in another paper~\cite{companion}, we find that, for a holographic tensor-network state~\cite{Swingle:2009bg, Swingle:2012wq, Pastawski:2015qua, Hayden:2016cfa, Bao:2018pvs}, Coecke flows have the same characteristic features as the bit threads~\cite{Freedman:2016zud, Cui:2018dyq, Headrick:2017ucz, Headrick:2022nbe} that explain the Ryu--Takayanagi (RT) formula for holographic entanglement entropy~\cite{Ryu:2006bv, Ryu:2006ef, Hubeny:2007xt}. Therefore, if this observation is taken seriously, bit threads thereby acquire a genuinely concrete candidate physical interpretation, rather than being regarded merely as purely mathematical objects arising from convex-programming duality.

The remainder of this paper is organized as follows. Section~\ref{sec2} reviews the historical background of quantum information flow. Starting from Coecke's early entanglement-specification networks, we explain how this picture acquired a more rigorous process-theoretic formulation in categorical quantum mechanics and in the subsequently developed ZX string diagrams. 
Section~\ref{sec3} presents the main formal development of this work. We introduce apparent through-paths, compatible inheritance, and genuine through-goingness, and on this basis give a formal definition of a Coecke flow line. 
Section~\ref{sec4} further localizes compatible inheritance to boundary fibres associated with elementary string-diagram rewrite rules, establishes the local-to-global gluing principle and the backward-generation algorithm, and illustrates the construction using concrete rewrite rules of the ZX calculus. 
Section~\ref{sec5} turns to the physical interpretation of Coecke flow, regarding it as a constrained, quasi-local, line-like presentation of a target bare wire morphism factor within the complete protocol diagram, and discusses this interpretation through examples including teleportation, entanglement swapping, and GHZ-assisted teleportation.
Finally, Section~\ref{sec6} discusses entanglement distillation, holographic bit threads, and possible extensions beyond the dagger-compact and ZX frameworks, before concluding the paper. 
Necessary background on category theory, CQM, quantum teleportation, and the ZX calculus, together with the ZX rewrite rules used in the text, is collected in the appendices.


\section{A Rudimentary Description of Quantum Information Flow}\label{sec2}

\subsection{The Entanglement Specification Network Perspective}\label{sec21}

In this subsection, we begin with a rudimentary description of the idea of quantum information flow. In particular, we use quantum teleportation as a pedagogical example. In this example, the resource state is a simple bipartite Bell state, and the projective operations involved are also at most bipartite.

\begin{figure}[htbp]
    \centering
    \includegraphics[width=0.9\textwidth]{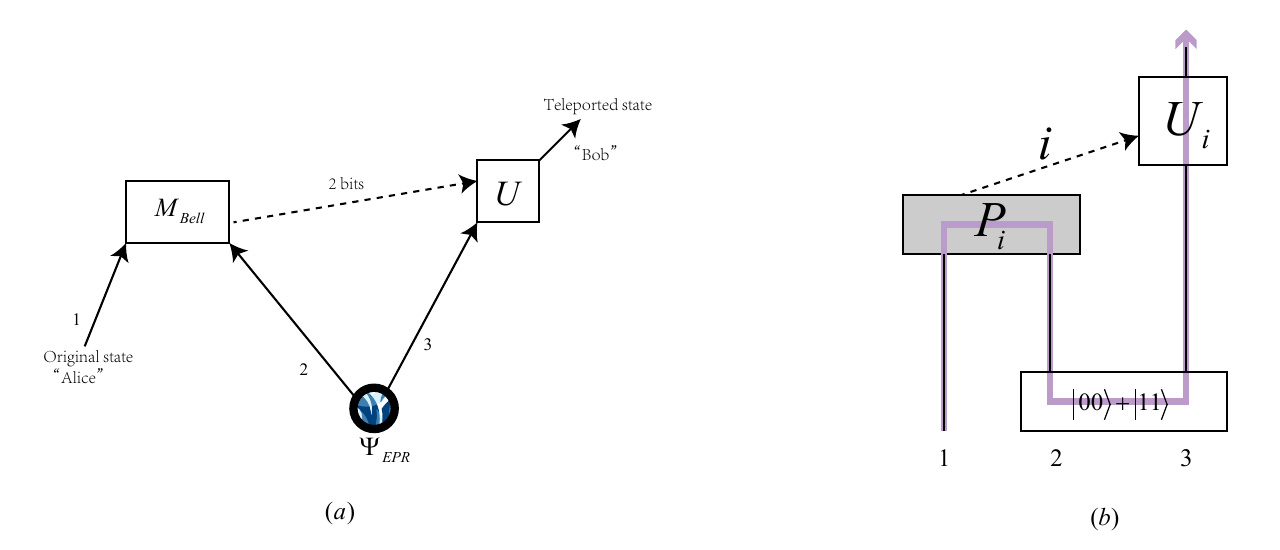}
    \caption{ (a) Quantum teleportation. 
(b) The corresponding entanglement specification network and the quantum information flow on it, shown as the purple line.}
    \label{fig-tele}
\end{figure}

A detailed review of quantum teleportation is given in Appendix~\ref{appd3}. In Figure~\ref{fig-tele}, the standard experimental arrangement of quantum teleportation on the left is rewritten by Coecke into the so-called $entanglement$ $specification$ $network$ on the right. In this entanglement specification network, the dashed lines represent the ``classical information flow'' associated with the propagation of two classical bits. By contrast, the purple line marks what Coecke defines as the ``quantum information flow''. The quantum information flow is assigned a direction, as indicated by the arrows in the figure. However, even in this extremely simple example, the careful reader will immediately notice that this flow direction sometimes runs against the direction of time. This is one of the main reasons why we need to spend more effort reviewing the notion of ``quantum information flow''.

\begin{figure}[htbp]
    \centering
    \includegraphics[width=0.8\textwidth]{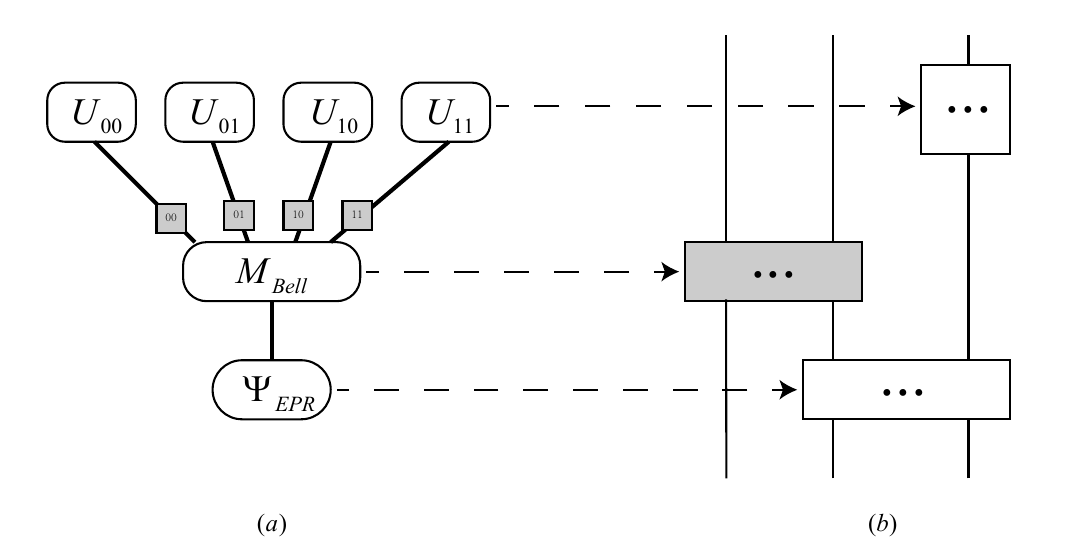}
    \caption{ (a) The tree decomposition of quantum teleportation. 
(b) The configuration picture to be filled into an entanglement specification network.}
    \label{fig-tree}
\end{figure}

Figure~\ref{fig-tree} introduces the concrete construction of an ``entanglement specification network''. The trick is to first use a $tree$ to extract away the ``classical information flow'', so that the quantum information flow can be studied more cleanly. The left panel is the tree. Its nodes represent operations that occur successively, including both ordinary unitary operations and measurement operations. However, each measurement gives rise to several branches, each branch representing one actual measurement outcome, which we denote by a token $i$. In the present example, $i$ is precisely the value $xz=00,01,10,11$ arising from the Bell measurement. In this way, classical communication is encoded in the tree as the dependence of certain operations on the branch tokens below them. For example, the dependence of Bob's unitary operation $U_{xz}$ on Alice’s Bell-measurement token $xz$ represents the two-bit classical channel required for teleportation. On the other hand, the right panel of Figure~\ref{fig-tree} represents the $configuration$ $picture$, in which the involved operations are arranged according to the time at which they are applied and the subsystem on which they act. Now, for each path in the tree from the ``root” to a ``leaf”, if we correspondingly fill the operations associated with the nodes along that path into the corresponding boxes in the configuration graph, we define an ``entanglement specification network'' containing only unitary operations and projectors. For example, for each of the four possible values of $xz$, what we obtain is precisely the entanglement specification network shown in Figure~\ref{fig-tele}. It should be especially noted that the projectors associated with measurement in an entanglement specification network should all be understood conditionally. We therefore deliberately draw them as light-grey boxes, in order to distinguish them from ordinary unitary-operation boxes, which are drawn transparently.

\begin{figure}[htbp]
    \centering
    \includegraphics[width=0.8\textwidth]{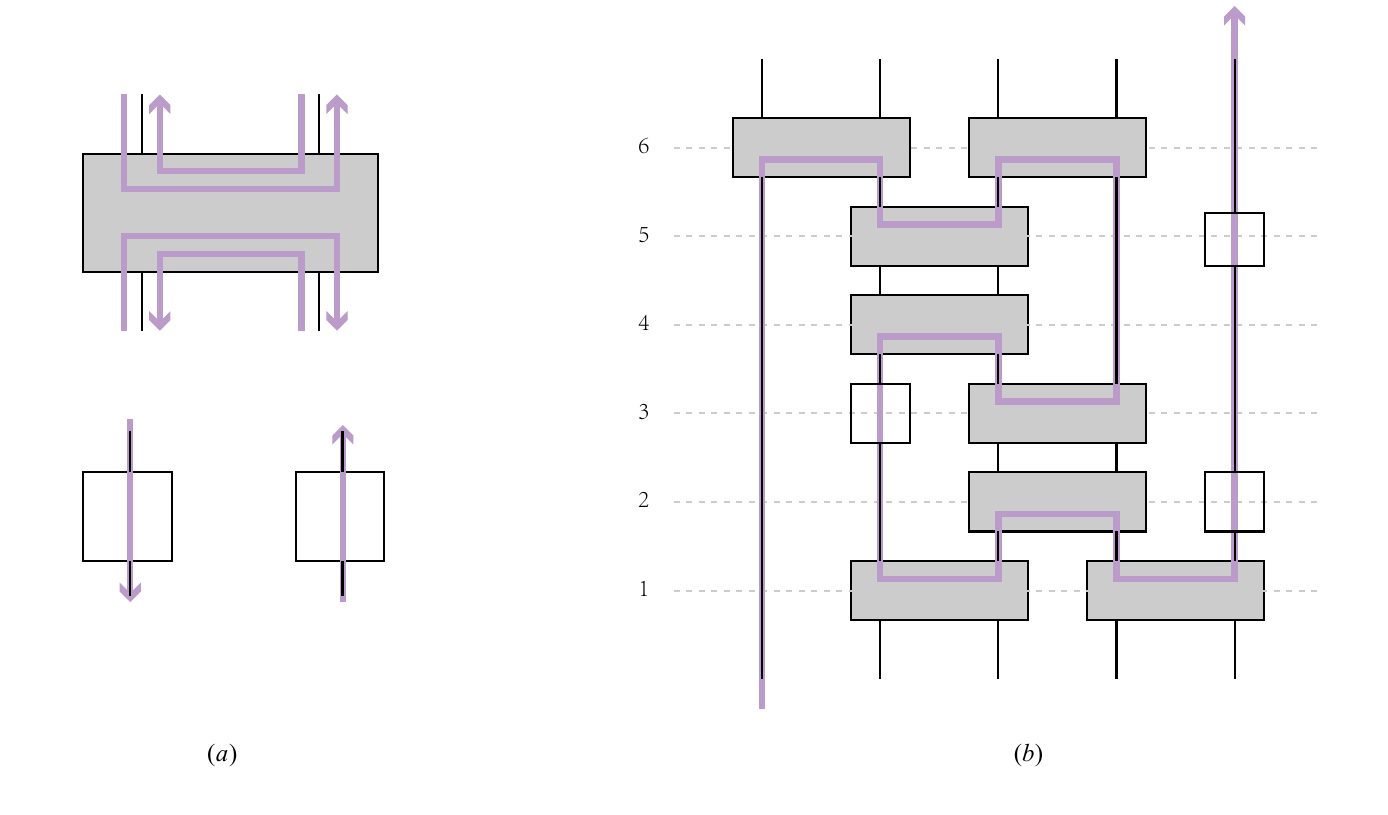}
    \caption{ (a) The traversal rules for quantum information flow. 
(b) An example of quantum information flow.}
    \label{fig-flow}
\end{figure}
It is precisely in such entanglement specification networks, after the classical information flow has been ``extracted away'', that quantum information flow is stated in \cite{Coecke:2004sxv} as follows. A quantum information-flow path is a line proceeding along such a trajectory, which, with respect to the actual physical time, may move either forward or backward, and which satisfies the following conditions:
\begin{enumerate}
    \item When it passes through a bipartite projector, represented by a grey box, it obeys the four possible ways of ``entering'' and ``leaving'' shown in Figure~\ref{fig-flow} (a).
    \item When it passes through a local unitary operation, it does not change direction.
    \item It cannot end at a time before any other time which it covers.
\end{enumerate}
The purple line in Figure~\ref{fig-flow} (b) is an example of such a quantum information-flow path. The simplest example is the one already presented at the beginning in the case of quantum teleportation, shown in Figure~\ref{fig-tele}.

\subsection{The CQM Perspective and the ZX Perspective}\label{sec22}

Up to this point, we obtain at least the following naive and intuitive impression: an entanglement specification network is like a map of scenic spots, while a quantum information flow is like a traceable tour route in it. Formally, we may write such a route as
\begin{equation}
R=
\bigl(
e_{\mathrm{in}};
\,(v_1:i_1\to o_1),\,
(v_2:i_2\to o_2),\,
\ldots,\,
(v_m:i_m\to o_m);
\,e_{\mathrm{out}}
\bigr).
\end{equation}
It starts from an input boundary leg $e_{\mathrm{in}}$, passes successively through a sequence of local scenic spots $v_1,\ldots,v_m$, and finally arrives at an output boundary leg $e_{\mathrm{out}}$. 
More precisely, each scenic spot $v_k$ must also be equipped with a through-choice, 
specifying through which leg $i_k$ the route enters that spot and through which leg $o_k$ it leaves.
However, this  picture still leaves a genuinely crucial question: when the route reaches a projector box, why is it allowed to ``pass through'' the box in some ways rather than others? In other words, it gives us heuristic threading rules and the resulting intuition of a ``flow'', but it has not yet explained the structural origin of this through-choice. 
Moreover, the careful reader should notice that, in our discussion of quantum information flow, so far, we have only temporarily dealt with bipartite entangled resource states, bipartite measurements, and single-leg local unitary operations, as presented in Figures~\ref{fig-tele} and~\ref{fig-flow}. In other words, we have not yet truly presented the notion of quantum information flow in the presence of more general resource states and multipartite operations.

\begin{figure}[htbp]
    \centering
    \includegraphics[width=1\textwidth]{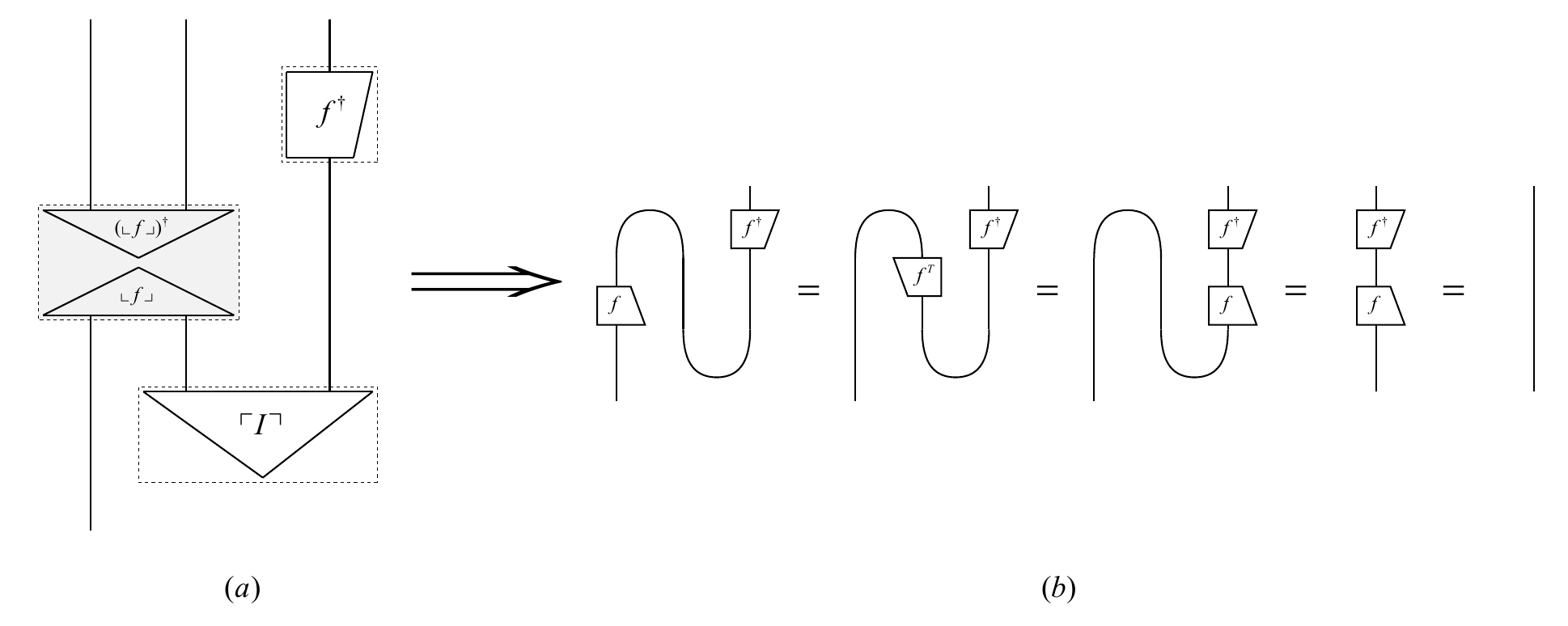}
    \caption{(a) The string diagram corresponding to the entanglement specification network of teleportation in a fixed branch. (b) String-diagrammatic calculus of teleportation.  }
    \label{fig-cqm}
\end{figure}

For a more essential and general purpose, it is necessary to examine the somewhat phenomenological description above from the perspective of categorical quantum mechanics (CQM)~ \cite{Abramsky:2004doh, Abramsky:2008qkz, Coecke:2005clw }.
Categorical quantum mechanics is a research programme that uses the structural language of category theory~\cite{MacLane:1998} to characterize quantum processes. It shifts quantum mechanics from a ``state-centred'' perspective to a ``process-centred'' perspective. In fact, an entanglement specification network in a fixed branch is precisely an early, pre-formalized version of the language of string diagrams in category theory, placed in the context of quantum information protocols. A string diagram is a graphical representation of morphisms in a monoidal category: wires represent objects, boxes represent general morphisms, vertical stacking represents sequential composition, and horizontal juxtaposition represents tensor product. 
For a review of categorical quantum mechanics and string diagrams, see Appendix~\ref{appa1}. More explicitly, Appendix~\ref{appb3} explains how a fixed-branch entanglement specification network can be formalized as a string diagram in a dagger compact closed category. For the moment, let us accept this point and use teleportation as an example to experience the string-diagrammatic representation of a quantum protocol and the corresponding string-diagrammatic calculus, as shown in Figure~\ref{fig-cqm}.

Figure~\ref{fig-cqm} (a) is the string diagram corresponding to the entanglement specification network of teleportation in a fixed branch. Throughout this paper, we read diagrams from bottom to top. In this diagram, the projector describing the measurement operation has already been decomposed into a ket, also called a state, followed by a bra, also called an effect. The former is represented string-diagrammatically by a downward-pointing triangle, while the latter is represented by an upward-pointing triangle. 
The key point is that the Bell effect can be understood, through the C-J duality, as the coname of a morphism $f$, usually denoted by $\llcorner f \lrcorner$. In this concrete example, $f$ is precisely one of the four familiar Pauli maps, $f\in\{I,Z,X,XZ\}$. In other words,
\begin{equation}
    \llcorner I \lrcorner=\langle\Phi^+|,\qquad
    \llcorner Z \lrcorner=\langle\Phi^-|,\qquad
    \llcorner X \lrcorner=\langle\Psi^+|,\qquad
    \llcorner XZ \lrcorner=\langle\Psi^-| .
\end{equation}
In this way, the outcome of each measurement branch is turned into a concrete morphism label $f$. Teleportation thereby becomes a story about the propagation of this morphism label $f$, as shown by the string-diagrammatic calculus in Figure~\ref{fig-cqm} (b). Interestingly, the whole string-diagrammatic calculation is highly intuitive: a morphism ``ID card'' $f$ starts from Alice's side, first slides through the cap bend, then slides through the cup bend, and finally arrives on Bob's side, where it meets the corresponding correction operation. In the end, the snake tail is straightened, while the morphism originally on Alice's side cancels with Bob's correction morphism. We are therefore left with a clean straight line running from bottom to top, representing an identity morphism. A more detailed and complete algebraic derivation is given in Appendix~\ref{appc2}, where one can see that each intermediate string diagram corresponds, in a completely non-mysterious way, to one step of algebraic transformation. It is worth noting that, in this special setting, quantum information flow appears to admit an intuitive interpretation as the transmission of morphism-valued data through a protocol process diagram. This picture, however, depends on the exceptionally clean dagger-compact structure of the protocol. A deeper discussion of the general physical meaning of quantum information flow will be given in Section~\ref{sec5}.

\begin{figure}[htbp]
    \centering
    \includegraphics[width=1\textwidth]{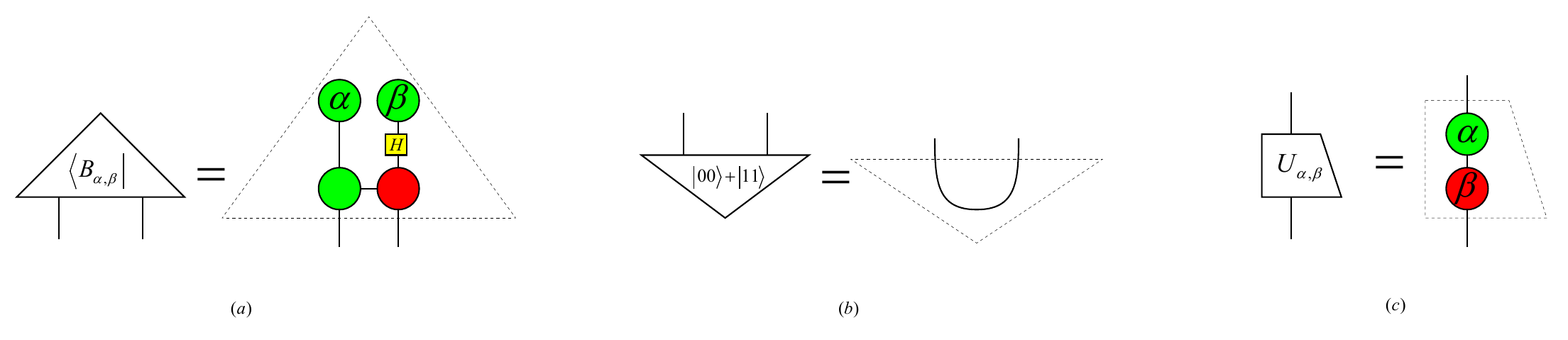}
    \caption{(a) ZX-spider representation of the Bell effect (bra). 
(b) ZX representation of the Bell state (ket). 
(c) ZX representation of Bob's correction operation. }
    \label{fig-fine}
\end{figure}
\begin{figure}[htbp]
    \centering
    \includegraphics[width=0.55\textwidth]{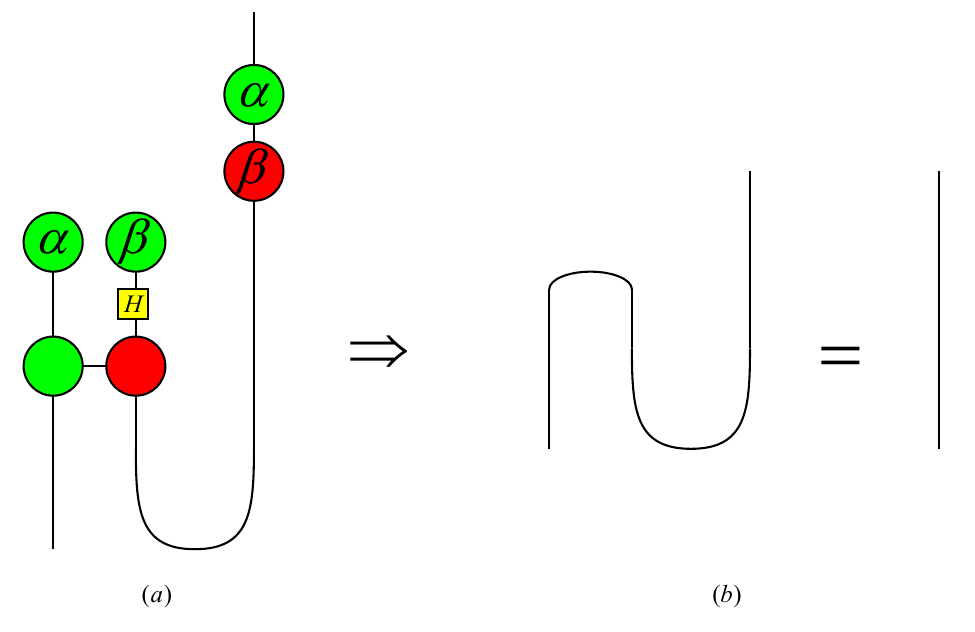}
    \caption{ (a) Complete ZX string-diagrammatic representation of the teleportation protocol.
(b) Figure (a) can be simplified, by means of ZX rewriting rules, into a snake-shaped bare wire, which is then rewritten into a straight wire by the snake equation.}
    \label{fig-zxte}
\end{figure}

In this paper, we shall also be particularly concerned with a concrete class of dagger compact closed string diagrams, called ZX string diagrams, or red-green spider string diagrams~\cite{Coecke:2008lcg, Duncan:2009ocf, vandeWetering:2020giq, Coecke:2017dti}. The corresponding string-diagrammatic rewriting is usually called the ZX calculus. We introduce it in Appendix~\ref{appd}. ZX diagrams are not another graphical language detached from the CQM string-diagrammatic context discussed above. Rather, they are still built on the same dagger compact closed string-diagrammatic skeleton, but provide a finer-grained graphical calculus.
In a certain sense, the ZX calculus can be viewed as further decomposing certain relatively coarse ``black boxes’’ in entanglement specification networks or CQM string diagrams into finer-grained string diagrams built from three elementary components: red spiders, also called $X$-spiders; green spiders, also called $Z$-spiders; and $H$-boxes, representing Hadamard gates. For the example of quantum teleportation, we display this ``opening up of black boxes'' for the relevant components in Figure~\ref{fig-fine}. We thereby obtain the ZX string-diagrammatic representation of the quantum teleportation protocol shown in Figure~\ref{fig-zxte} (a).

More importantly, using the rigorous rewriting rules of the ZX calculus, reviewed in Appendix~\ref{appd2}, this initial diagram can be simplified step by step. The detailed simplification is given in Appendix~\ref{appd3}. Eventually, the diagram reveals the same semantic content as the clean snake-shaped bare wire, and further as a straight bare wire running from bottom to top, as shown in Figure~\ref{fig-zxte} (b).\footnote{Here our simplification order is slightly different from that in Figure~\ref{fig-cqm}. There, we first cut off the snake-shaped tail and then cancelled the correction. Nevertheless, this kind of different simplification orders do not lead to any essential difference.}


\section{A Formalization of Quantum Information Flow}\label{sec3}

\subsection{Motivation}\label{sec31}

Before entering into the formal work of this paper, let us first clarify the historical background once more. In Ref.~\cite{Coecke:2004sxv}, 
Coecke explicitly studied the identification of a ``virtual flow of information'' in what he called entanglement specification networks, and argued that this flow embodies the transfer effected in well-known protocols such as quantum teleportation.
As reviewed in Section~\ref{sec21}, in somewhat concrete entanglement specification networks, he represented quantum information flow as a path through the graphical geometry of the protocol, traversing bipartite projection boxes according to certain local rules.~\footnote{In fact, Ref.~\cite{Coecke:2004sxv} also proves the nontrivial result that, if such a path successively traverses projection boxes labelled by $f_1,\ldots,f_k$, then the action between the input and output of the path is
\begin{equation}
    \phi_{\mathrm{out}}
    =
    (f_k\circ\cdots\circ f_1)(\phi_{\mathrm{in}}).
\end{equation}
Here the order of composition is not the physical-time order of the boxes, but the order in which the path traverses them. We shall discuss this aspect in greater detail in Section~\ref{sec5}, where we examine the physical meaning of quantum information flow.}

On the other hand, in the categorical string-diagrammatic calculi that developed subsequently, Coecke and his collaborators continued to unravel the graphical structures of quantum protocols in the spirit of ``quantum information flow''. Refs. ~\cite{Abramsky:2004doh, Coecke:2005bin}  point out (see also~\cite{Abramsky:2009jjw}), that after entanglement specification networks are recast at the abstract level of compact closed categories, several basic lemmas governing names, conames, and their rules of composition ``capture'' the quantum information flow in protocols such as quantum teleportation, logic-gate teleportation, and entanglement swapping.
In the corresponding categorical pictures, quantum information is described as seeming to flow ``following the line'', while being acted on successively by the morphisms labelling the boxes encountered along the way.
In the wording of Ref.~\cite{Abramsky:2008qkz}, this abstract categorical framework ``recovers'', at a new level, the results of the earlier work~\cite{Coecke:2004sxv} concerning information-flow paths in networks of projections.
Furthermore, as string-diagrammatic calculi became increasingly mature and technically specialized, particular attention came to be paid to the following feature of their graphical structure: a seemingly complicated string diagram for a quantum protocol can often, after semantics-preserving graphical calculation, eventually be reduced to a strikingly simple bare-wire pattern. 
In the case of teleportation, for example, Ref.~\cite{Coecke:2005clw} directly regards as quantum information flow such a continuous line --a ``black line’’ that runs through components of the string diagram, such as states and effects, and may assume different graphical presentations as the graphical calculation proceeds--and is ultimately revealed, in the semantics-preserving minimal diagram, as a simple bare wire.
 This manner of speaking was retained in the more technical ZX calculus developed subsequently (see e.g.~\cite{Dundar-Coecke:2023xyf, Dundar-Coecke:2025shi, Duncan:2009ocf, Coecke:2008nxx}): The complete teleportation diagram is reduced step by step, by means of the ZX rules, to a bare identity wire, and this proof is explicitly said to ``elucidate the flow of quantum information''. In particular, the step of ``yanking’’ the wires straight is said to highlight an important property of information flow. Section~\ref{sec22} above provided the necessary review of this part of the historical background.

In summary, within this historical development, ``quantum information flow'' has in fact left us with two related graphical descriptions. The first is a path identified in the initial entanglement network according to certain local traversal rules. The second is the clean line-like pattern revealed after the protocol diagram has undergone semantics-preserving string-diagrammatic calculation. Indeed, in the historical formulations of the relevant literature, the highly suggestive path in a projection network and the information flow captured by the more rigorous categorical compact-closed graphical calculus were explicitly regarded as two successive stages in the same development. In other words, the latter was usually understood as an abstraction and axiomatization of the former.
This continuity of intellectual lineage, however, does not by itself amount to an explicit formal correspondence between the two linear descriptions. In the simplest prototypical diagrams, the relation between the initial path and the simple line obtained after reduction may perhaps be read off directly from the diagram itself. But when the initial entanglement specification network is sufficiently complicated, as it may be in more general protocols of quantum information theory, and when its reduction requires a sequence of local string-diagrammatic rewrites with a large number of steps, the correspondence between the two is no longer directly visible.
Indeed, for a given initial path, it is not self-evident either how a compatible successor representative should be assigned to it at each subsequent string-diagrammatic rewriting step, or even whether such a natural chain of inheritance always exists in principle.

This also raises a more general problem concerning string-diagrammatic calculation. A semantics-preserving string-diagrammatic rewrite guarantees that the morphism represented by the entire diagram remains unchanged, but it does not necessarily at the same time endow the local line-like structures within the diagram with any predetermined continuity of identity. Whether such a continuation can be characterized rigorously and testably, without relying on purely visual intuition, is a direct motivation for the formalization developed in this paper.

This raises a more general question concerning string-diagrammatic calculation. Although a semantics-preserving string-diagrammatic rewrite guarantees that the morphism denoted by the diagram as a whole remains unchanged, it does not by itself provide a predetermined rule by which local line-like structures are inherited from one diagram to the next. Whether such inheritance can be characterized in a rigorous and testable manner, without relying on purely visual intuition, is a direct motivation for the formalization developed in this paper.

\subsection{Formalization}\label{sec32}

In this section, as promised in the introductory discussion, we shall provide a dedicated formalization of the notion of ``quantum information flow'' (in somewhat particular sense required for our purposes). To emphasize the relation between this reformalization and the intellectual lineage of Coecke's early work, we shall also refer to such a flow as a \emph{Coecke flow}. 
Our formalization begins with the naive picture provided by entanglement specification networks; we shall, however, make this picture rigorous in the language of string diagrams. The correspondence between entanglement specification networks and string diagrams is discussed in Appendix~\ref{appb3}. We shall begin with the notion of a continuous apparent through-path in a rewritable string diagram. On the basis of this rigorously formulated notion, we shall ultimately give an explicit formalization of quantum information flow.

\subsubsection{Apparent Through-Paths and Compatible Inheritance}\label{sec321}

Throughout this paper, we shall refer pictorially to a process of string-diagrammatic rewriting as a \emph{string-diagram rewriting movie}.

\paragraph{Definition (Apparent through-path).}

Given a string-diagram rewriting movie
\begin{equation}
    W:
    D_0
    \Rightarrow
    D_1
    \Rightarrow
    \cdots
    \Rightarrow
    D_n,
\end{equation}
we call $D_k$ the $k$th frame diagram. For any frame diagram $D_k$, an \emph{apparent through-path} in $D_k$ is a continuous path that, in the current graphical presentation, crosses no visible tensor-product gap; that is, it never passes directly between subdiagrams that occur as distinct tensor factors rather than being diagrammatically connected. We denote the set of all apparent through-paths in $D_k$ by $\mathcal{L}(D_k)$.

\begin{figure}[htbp]
    \centering
    \includegraphics[width=1\textwidth]{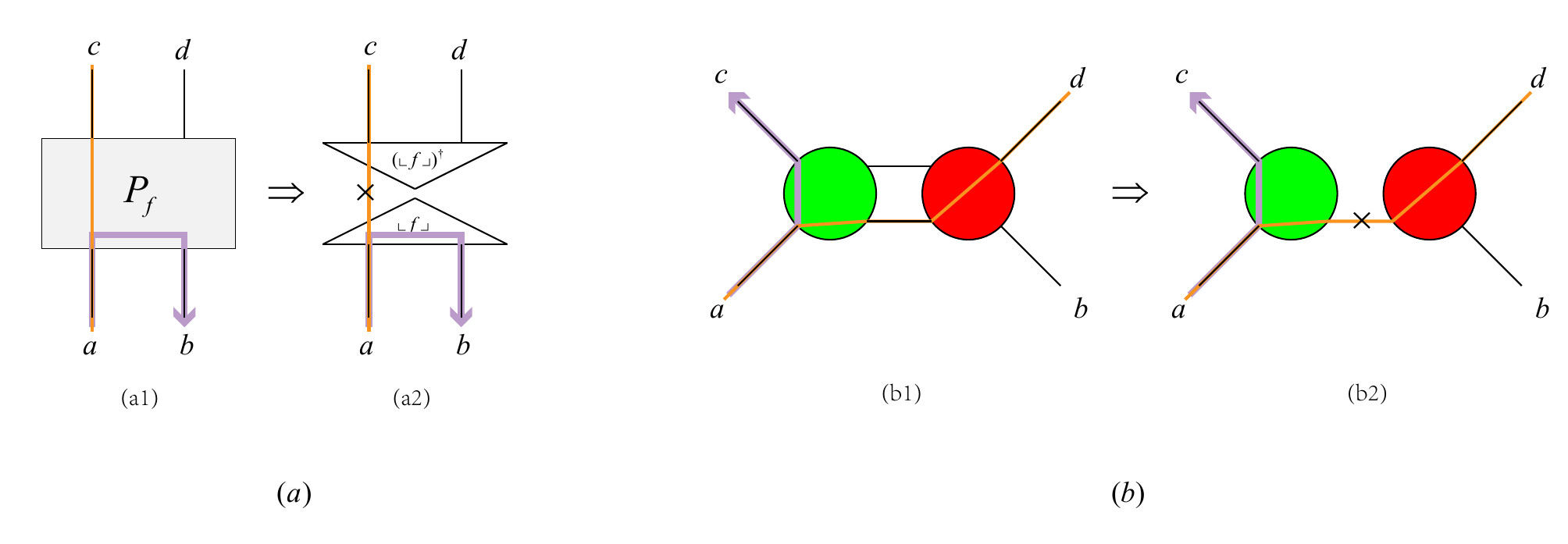}
   \caption{Two minimal examples illustrating the presentation dependence of apparent through-paths.
(a) A projection morphism $P_f$, initially represented as a coarse-grained black box, is refined into a state and an effect. In the black-box presentation, both the purple and orange paths appear to be through-paths. After the refinement, the purple path remains an apparent through-path, whereas the orange path is revealed to cross the tensor-product gap between the state and the effect, as marked by $\times$.
(b) An analogous phenomenon under a nontrivial string-diagram rewrite in the ZX-calculus. Both highlighted paths are apparent through-paths in the initial diagram. After the rewrite, the purple path admits compatible inheritance, whereas the orange path does not, since it is revealed to cross a tensor-product gap. }
    \label{fig-fake}
\end{figure}

Placed back into our physical setting, an apparent through-path can be understood intuitively as follows. We first formalize the entanglement specification network representing a fixed branch of a protocol process diagram as a string diagram, in which wires represent systems and boxes represent operational processes. An apparent through-path always enters a local process box along some wire, exits the box along another wire attached to it, and then follows this wire into a further box. To understand why we qualify such a path as merely ``apparent'', consider the two minimal examples shown in Figure~\ref{fig-fake}.

In Figure~\ref{fig-fake}(a), we begin with a morphism representing a projection operator, drawn at a coarse-grained level as a ``black-box'' operation. In this temporary graphical presentation, both $ab$ and $ac$ satisfy the requirements for apparent through-paths. We know, however, that a projector morphism can be further decomposed into a state and an effect, represented by the two oppositely oriented triangles, as shown in Figure~\ref{fig-fake}(a2). In this refined presentation, $ab$ still satisfies the definition of an apparent through-path, whereas $ac$ no longer does, because it has now been revealed to cross the tensor-product gap between the state and the effect. We may, of course, say that $ac$ had already been crossing this gap ``secretly'' in Figure~\ref{fig-fake}(a1), although the gap had not yet been exposed by the graphical presentation.
Consider next the slightly subtler example in Figure~\ref{fig-fake}(b). It gives a simple ZX string-diagram rewriting movie arising, in fact, from a nontrivial complementarity rule of the ZX calculus. In the first frame, Figure~\ref{fig-fake}(b1), both $ac$ and $ad$ satisfy the requirements for apparent through-paths in $D_0$. After one string-diagrammatic rewrite, however, we obtain the second frame $D_1$, Figure~\ref{fig-fake}(b2). There, $ac$ remains an apparent through-path, whereas $ad$ is now revealed to cross a tensor-product gap and therefore no longer satisfies the definition.

These examples show that the notion of an apparent through-path is relative: it depends on the current graphical presentation of the string diagram. Whether by refining coarse-grained diagrammatic components or by performing more nontrivial string-diagrammatic rewrites, one may expose the possibility that a path had been secretly crossing a latent tensor-product gap. Notice, however, that neither refinement nor rewriting changes the morphism semantics of the string diagram. This gives rise to a natural mathematical question: can one define a notion of a \emph{genuine through-path} that preserves its apparent through-goingness across all equivalent string-diagrammatic presentations of a given morphism?
It must be admitted that, for morphism diagrams in the most general sense, this demand need not automatically have a well-behaved answer. A string diagram may continually change its presentation under rewriting, and there is in principle no guarantee that further tensor-product structure will not be exposed in a later presentation.

Nevertheless, Coecke's early graphical analyses of quantum information flow repeatedly display the following phenomenon: a seemingly complicated string diagram for a quantum protocol is often shown, after graphical reduction, to have morphism semantics entirely equivalent to those of a trivial diagram consisting of a clean bare wire representing an identity morphism; moreover, this bare wire is interpreted as the result of quantum information flow. See, for example, Ref.~\cite{Coecke:2005clw}. This suggests that, if we restrict attention to morphisms whose semantics can ultimately be exhibited in the final frame of a string-diagrammatic simplification as
\begin{equation}
    D_n
    \simeq
    B_{a\to b}\otimes R,
\end{equation}
where $B_{a\to b}$ denotes a bare identity wire and $R$ is a remaining subdiagram tensor-decoupled from it, then the desideratum of a genuine through-path admits an elegant realization.
The intuitive reason is that $B_{a\to b}$ is an exceptionally simple diagrammatic component with no further internal structure and is already tensor-decoupled from the residual subdiagram $R$. It therefore supplies an evident endpoint for genuine through-goingness: when an apparent through-path can be compatibly inherited all the way through the rewriting movie until its final representative is carried simply by this decoupled bare wire, it may finally be said to have passed all the tests posed by the successive graphical presentations.

We call the intuitive phenomenon in Figure~\ref{fig-fake} that an apparent through-path survives a single rewriting step, the \emph{compatible inheritance} of an apparent through-path. We now introduce the basic intuition behind this notion.
First of all, for a more general and possibly complicated string diagram, we have
\begin{equation}
    D_k=C[P],
    \qquad
    D_{k+1}=C[P'].
\end{equation}
This means that the $k$th rewriting step in a string-diagram rewriting movie merely replaces a local subdiagram $P$ by another local subdiagram $P'$, while leaving the ambient context $C[-]$ unchanged. Since $P$ and $P'$ represent the same typed morphism, the wire ports through which they connect to the ambient context remain fixed.
Therefore, suppose that an apparent through-path $\Gamma_k$ in $D_k$ passes through the local region $P$ that is going to be rewritten, then the part of $\Gamma_k$ lying outside this region remains unchanged; only the local apparent through-segment lying inside $P$ needs to be updated. This suggests the following reasonable expectation of locality: provided that the rewritten local subdiagram is sufficiently small, the inheritance of the local apparent through-segment may be determined primarily by the external wire ports to which it is attached, even if we impose only the minimal requirement that it must not cross a tensor-product gap.
This expectation is natural because the smaller the local subdiagram is, the more constrained its internal freedom becomes. Once the two wire ports $a$ and $b$ have been fixed, there will usually be only a limited number of admissible apparent traversals between them. We therefore have reason to expect that compatible inheritance can be decided locally:
\begin{equation}
    \ell_{a\to b}\subset P
    \;\rightsquigarrow\;
    \ell'_{a\to b}\subset P'.
\end{equation}
This expectation of locality is not, however, self-evident. Merely knowing that $P$ and $P'$ represent the same morphism and possess the same external wire ports does not automatically guarantee that every local apparent through-segment admits compatible inheritance. Which pairs of ports permit inheritance, which original apparent through-segments fail because an internal tensor-product gap becomes visible in the next frame, and even whether the inheritance relation is unique all depend on the particular specialized string-diagrammatic calculus under consideration. This is precisely what we shall discuss later in the specialized setting of ZX string diagrams.

In any case, consider a string-diagram rewriting movie
\begin{equation}
    W:
    D_0
    \Rightarrow
    D_1
    \Rightarrow
    \cdots
    \Rightarrow
    D_n,
\end{equation}
together with an apparent through-path $\Gamma_k$ in each intermediate frame diagram $D_k$. For the $k$th rewriting step
\begin{equation}
    \rho_k:
    D_k
    \Rightarrow
    D_{k+1},
\end{equation}
we say that the apparent through-path $\Gamma_k$ in the preceding frame is \emph{compatibly inherited} by $\Gamma_{k+1}$ in the succeeding frame, and write
\begin{equation}
    \Gamma_k
    \;T_{\rho_k}\;
    \Gamma_{k+1},
\end{equation}
where $    T_{\rho_k}
    \subseteq
    \mathcal{L}(D_k)
    \times
    \mathcal{L}(D_{k+1}) $ denotes the through-path update relation induced by the rewriting step $\rho_k$.\footnote{Here $\times$ denotes the ordinary Cartesian product of sets. Equivalently, in terms of ordered pairs, $(\Gamma_k,\Gamma_{k+1})\in T_{\rho_k}$.}

We are now in a position to define a genuine through-path.

\subsubsection{Genuine Through-Paths and Quantum Information Flow}\label{sec322}

\paragraph{Definition (Genuine through-path).}

Let $D_0$ be a string diagram representing a morphism. Suppose that, through a sequence of string-diagrammatic rewrites
\begin{equation}
    W:
    D_0
    \Rightarrow
    D_1
    \Rightarrow
    \cdots
    \Rightarrow
    D_n,
\end{equation}
it is ultimately shown to be equivalent to a simple diagram of the form
\begin{equation}
    D_n
    \simeq
    B_{x\to y}\otimes R,
\end{equation}
namely, the tensor product of a bare wire representing a simple morphism and a residual subdiagram.

Let $ \Gamma_0\in L(D_0)$ be an apparent through-path. If there exists a sequence of representatives $    \Gamma_0,\Gamma_1,\ldots,\Gamma_n$ such that compatible inheritance holds at every rewriting step,
\begin{equation}
    \Gamma_k
    \;T_{\rho_k}\;
    \Gamma_{k+1},
    \qquad
    k=0,\ldots,n-1,
\end{equation}
and the terminal representative $\Gamma_n$ is carried simply by the bare wire diagram $B_{x\to y}$ in the final frame, then we say that $\Gamma_0$ realizes \emph{genuine through-goingness} in the string-diagram rewriting movie $W$, or equivalently that it is a \emph{genuine through-path} in $W$.

In the quantum-protocol process diagrams that inspired Coecke's work, the initial diagram generally contains measurement outcomes, classical controls, and conditional operations. Consequently, one usually obtains a fixed-branch initial string diagram $D_0^\omega$ only after fixing a classical branch $\omega$. Indeed, $D_0^\omega$ is precisely the formal string-diagrammatic realization of what Coecke called an entanglement specification network. The detailed correspondence between the two will be presented in Table~\ref{tab:string-zx-network}.

We can now define the notion of a Coecke flow line.

\paragraph{Definition (Coecke flow line).}

Let $D^{\mathrm{init}}$ be a string diagram, and let $\Omega$ be its set of fixed branches. For every fixed branch $\omega\in\Omega$, denote the corresponding string-diagram rewriting movie by
\begin{equation}
    W^\omega:
    D_0^\omega
    \Rightarrow
    D_1^\omega
    \Rightarrow
    \cdots
    \Rightarrow
    D_{n_\omega}^\omega,
\end{equation}
where $D_0^\omega$ is the instantiation of $D^{\mathrm{init}}$ in the branch $\omega$.
Let $\Gamma$ be an apparent through-path from two boundary wire ports $x$ to $y$ in the initial string diagram $D^{\mathrm{init}}$. Suppose that, for every fixed branch $\omega\in\Omega$, there exists a sequence of apparent through-path representatives $    \Gamma_0^\omega,
    \Gamma_1^\omega,
    \ldots,
    \Gamma_{n_\omega}^\omega$ satisfying the following conditions:
\begin{enumerate}
    \item $\Gamma_0^\omega$ is the instantiation of $\Gamma$ in the branch $\omega$.

    \item For every rewriting step $
        \rho_k^\omega:
        D_k^\omega
        \Rightarrow
        D_{k+1}^\omega $
      and the compatible-inheritance relation induced by it, $        T_{\rho_k^\omega}
        \subseteq
        L(D_k^\omega)\times L(D_{k+1}^\omega),$ 
    we require
    \begin{equation}
        \Gamma_k^\omega
        \;T_{\rho_k^\omega}\;
        \Gamma_{k+1}^\omega.
    \end{equation}
    That is, the apparent through-path $\Gamma_k^\omega$ in the $k$th frame is compatibly inherited by the apparent through-path $\Gamma_{k+1}^\omega$ in the $(k+1)$th frame under that rewriting step.

    \item In the final frame, there exists a bare-wire diagram tensor-decoupled from the residual subdiagram, namely
    \begin{equation}
        D_{n_\omega}^\omega
        \simeq
        B_{x\to y}^\omega\otimes R^\omega,
    \end{equation}
    and $\Gamma_{n_\omega}^\omega$ is carried simply by the bare-wire diagram $B_{x\to y}^\omega$.

    \item The preceding conditions hold for every fixed branch $\omega\in\Omega$.
\end{enumerate}

We then say that $\Gamma$ defines a \emph{Coecke flow line} from $x$ to $y$. Correspondingly, the family of rewriting movies covering all fixed branches, together with their respective sequences of through-path representatives, is called a \emph{string-diagrammatic witness} for this Coecke flow line.

Put simply, we define a Coecke flow as a branch-independent initial through-path that realizes genuine through-goingness in every fixed branch of the protocol string diagram.

The definition above remains, in itself, at the level of string-diagrammatic calculus. When such string diagrams are endowed with the physical interpretation of a quantum protocol, however, the appearance in every fixed branch of a tensor-decoupled bare wire represents, in many ordinary cases, an identity quantum channel from the input system $x$ to the output system $y$. Intuitively, it means that a quantum message can be successfully conveyed without loss through the protocol. In this physical setting, the corresponding Coecke flow may be given a more physically suggestive name: a \emph{quantum information flow}.
Quantum teleportation is the simplest and most typical example: the protocol string diagram in every fixed measurement branch can ultimately be rewritten into a bare identity wire. The gate-teleportation and entanglement-swapping protocols discussed in Coecke's early work possess the same graphical feature.
The present definition is a systematic generalization of this idea. First, the final diagram need not be reduced in its entirety to a bare wire. A Coecke flow can already be defined whenever the final diagram reveals a bare wire factor tensor-decoupled from the residual subdiagram. Second, several mutually decoupled bare-wire factors may occur
simultaneously in a single rewriting movie, as in protocols
implementing parallel quantum teleportation. To represent these
factors simultaneously by several flow lines, we impose a joint
compatibility condition, formulated in
Section~\ref{sec:joint-configurations}.
Finally, although we shall later make the compatible-inheritance relation concrete in the setting of ZX string diagrams, the definition above does not itself depend on the ZX calculus. In principle, the notion of a Coecke flow can be applied whenever a string-diagrammatic calculus possesses controllable local rewriting rules and terminal diagrams containing bare-wire factors.

\subsubsection{Joint Coecke-Flow Configurations}
\label{sec:joint-configurations}

The preceding definition concerns a single flow line.
When several distinct target bare-wire factors are represented
simultaneously, we impose a further compatibility condition:
each elementary system wire may be used by at most one line
within the same configuration. We formulate this condition at
a specified wire resolution and for a specified set of permitted
local rewrite steps. In qubit ZX diagrams, each elementary wire
represents one qubit. If a composite system wire is resolved
into several tensor factors, that resolution is included in
the chosen graphical presentation.

For an apparent through-path $\Gamma\in\mathcal{L}(D)$,
let $E_D(\Gamma)$ denote the set of wires traversed by $\Gamma$.
Define the space of candidate configurations by
\begin{equation}
    \operatorname{Conf}_m(D)
    :=
    \left\{
        (\Gamma_1,\ldots,\Gamma_m)\in\mathcal{L}(D)^m
        \,\middle|\,
        E_D(\Gamma_r)\cap E_D(\Gamma_s)=\varnothing
        \quad\text{for }r\neq s
    \right\}.
    \label{eq:joint-conf}
\end{equation}
The paths are labelled by their respective target factors.
They may pass through a common process node, provided that
they do not use a common wire.

For a rewriting step $\rho:D\Rightarrow D'$, we define
the joint compatible-inheritance relation
\begin{equation}
    \widehat{T}_{\rho}^{(m)}
    \subseteq
    \operatorname{Conf}_m(D)
    \times
    \operatorname{Conf}_m(D')
\end{equation}
by
\begin{equation}
    \boldsymbol{\Gamma}
    \;\widehat{T}_{\rho}^{(m)}\;
    \boldsymbol{\Gamma}'
    \quad\Longleftrightarrow\quad
    \Gamma_r\;T_\rho\;\Gamma_r'
    \quad\text{for every }r=1,\ldots,m,
    \label{eq:joint-inheritance}
\end{equation}
where both tuples are required to belong to their respective
configuration spaces. Thus the single-line inheritance
conditions are imposed together with the requirement that
the representatives coexist without sharing elementary wires.

\paragraph{Definition (Joint Coecke-flow configuration).}
Let $D^{\mathrm{init}}$ be a protocol string diagram with
classical branches $\Omega$, and choose a branch-independent
candidate
\begin{equation}
    \boldsymbol{\Gamma}
    =
    (\Gamma_1,\ldots,\Gamma_m)
    \in
    \operatorname{Conf}_m(D^{\mathrm{init}}).
\end{equation}
The boundary pair associated with each label $r$ is fixed
independently of the branch.
We call $\boldsymbol{\Gamma}$ a \emph{joint Coecke-flow
configuration} if, for every $\omega\in\Omega$, there exists
a rewriting movie
\begin{equation}
    W^\omega:
    D_0^\omega
    \Rightarrow\cdots\Rightarrow
    D_{n_\omega}^\omega
\end{equation}
and a sequence of configurations
$\boldsymbol{\Gamma}_i^\omega
=(\Gamma_{1,i}^\omega,\ldots,\Gamma_{m,i}^\omega)
\in\operatorname{Conf}_m(D_i^\omega)$ satisfying:
\begin{enumerate}
    \item $\boldsymbol{\Gamma}_0^\omega$ is the instantiation
    of $\boldsymbol{\Gamma}$ in branch $\omega$.

    \item All lines are inherited jointly in the same movie:
    \begin{equation}
        \boldsymbol{\Gamma}_i^\omega
        \;\widehat{T}_{\rho_i^\omega}^{(m)}\;
        \boldsymbol{\Gamma}_{i+1}^\omega,
        \qquad i=0,\ldots,n_\omega-1.
    \end{equation}

    \item The final frame has the form
    \begin{equation}
        D_{n_\omega}^\omega
        \simeq
        \left(
            \bigotimes_{r=1}^{m}B_r^\omega
        \right)\otimes R^\omega,
    \end{equation}
    where the $B_r^\omega$ are mutually tensor-decoupled
    bare-wire factors with the prescribed boundary pairs,
    and $\Gamma_{r,n_\omega}^\omega$ is the trivial path
    carried by $B_r^\omega$.
\end{enumerate}

The movies and the configuration sequences together constitute
a \emph{joint string-diagrammatic witness}.
For $m=1$, this definition reduces to the preceding single-line
definition. For $m>1$, the representatives must satisfy the
non-sharing condition throughout each witness movie.
Different candidate configurations may use the same wire;
the restriction concerns lines asserted to coexist within
one configuration. The existence of several terminal factors
does not, by itself, establish the existence of such a joint
witness in a given graphical presentation.

\section{ Compatible-Inheritance Relation}\label{sec4}

\subsection{Local Inheritance Relations and Boundary Fibres}\label{sec41}

In the preceding section, on the basis of the idea of compatible inheritance of apparent through-paths under string-diagrammatic rewriting, we defined in succession genuine through-paths and Coecke flow lines. The central idea was to examine whether an apparent through-path can be compatibly inherited throughout an entire string-diagram rewriting movie, until it finally coincides with a clean tensor-decoupled bare wire. 
These definitions were formulated at the global level of the entire string diagram. However, provided that no intermediate rewriting steps are suppressed--every one-step rewrite of the full diagram consists, in essence, merely in applying an elementary rewrite rule supported on a small local subdiagram, while leaving the remainder of the diagram unchanged.
Here we make a mild assumption, one that is ordinarily satisfied in the cases of interest: a specialized string-diagrammatic rewriting system, such as the ZX calculus considered below, possesses a reasonably small axiomatic set of elementary rewrite rules, each supported on a sufficiently small local diagram. It is then enough, in principle, to specify how local apparent through-segments are compatibly inherited under each of these elementary rules.

\subsubsection{Local Boundary Fibres}\label{sec411}

For a concrete string-diagrammatic calculus system, denote each elementary rewrite rule $\rho$ in its axiomatic rule set as
\begin{equation}\label{rhok}
    \rho:
    P_\rho
    \equiv_\rho
    Q_\rho
    :
    A_\rho\to B_\rho,
    \qquad
    \llbracket P_\rho\rrbracket
    =
    \llbracket Q_\rho\rrbracket .
\end{equation}

At the level of graphical syntax, $P_\rho$ and $Q_\rho$ are local diagrammatic expressions with the same boundary type $A_\rho\to B_\rho$. They therefore fit into the same typed hole: for any context $C[-]$ whose hole has this boundary type, both $C[P_\rho]$ and $C[Q_\rho]$ are well-formed string diagrams. Furthermore, $\llbracket-\rrbracket$ denotes the semantic interpretation functor, and the rewrite rule $\rho$ asserts that the two expressions are semantically equal. In other words, they are two distinct graphical presentations of the same semantic morphism.

We now define the notion of a boundary fibre. Consider the map
\begin{equation}
    \partial_P:
    \mathcal{L}(P)
    \longrightarrow
    \binom{\partial P}{2},
\end{equation}
which sends a local apparent through-path $\theta\in\mathcal{L}(P)$ to the pair of boundary legs (i.e., wire ports) that it meets:
\begin{equation}
    \partial_P\theta=e.
\end{equation}
Thus, for a boundary-leg pair $    e\in\binom{\partial P}{2},$
we define the corresponding \emph{boundary fibre} by 
\begin{equation}
    \mathcal{L}_P(e)
    :=
    (\partial_P)^{-1}(e)
    =
    \left\{
        \theta\in\mathcal{L}(P)
        \,\middle|\,
        \partial_P\theta=e
    \right\}.
\end{equation}
That is, $\mathcal{L}_P(e)$ is the set of all admissible local apparent through-paths in the local diagram $P$ that meet the same pair of boundary legs $e$.
Since a local diagram is finite, each boundary fibre is also a finite set.
We therefore consider its cardinality, denoted as $    \left|\mathcal{L}_P(e)\right|,$
which counts the admissible local apparent through-paths in $P$ connecting the same boundary-leg pair $e$. If
\begin{equation}
    \left|\mathcal{L}_P(e)\right|=0,
\end{equation}
then the pair of boundary legs admits no legal local through-segment in $P$.
If
\begin{equation}
    \left|\mathcal{L}_P(e)\right|=1,
\end{equation}
then the boundary-leg pair uniquely determines a local through-segment. If
\begin{equation}
    \left|\mathcal{L}_P(e)\right|>1,
\end{equation}
then the same boundary data admit more than one local traversal.

\subsubsection{The Local Inheritance Relation}\label{sec412}

We now return to the elementary rewrite rule~(\ref{rhok}), s ince $P_\rho$ and $Q_\rho$ have the same boundary $\partial\rho$, we may
compare their boundary fibres for each boundary-leg pair $    e\in\binom{\partial\rho}{2}.$
We define the \emph{local inheritance relation} induced by $\rho$ to be
\begin{equation}
    \tau_\rho
    :=
    \coprod_{e\in\binom{\partial\rho}{2}}
    \mathcal{L}_{P_\rho}(e)
    \times
    \mathcal{L}_{Q_\rho}(e)
    \subseteq
    \mathcal{L}(P_\rho)
    \times
    \mathcal{L}(Q_\rho).
\end{equation}
In other words,
\begin{equation}
    \theta
    \;\tau_\rho\;
    \theta'
\end{equation}
if and only if there exists a common boundary-leg pair
$e\in\binom{\partial\rho}{2}$ such that
\begin{equation}
    \theta\in\mathcal{L}_{P_\rho}(e),
    \qquad
    \theta'\in\mathcal{L}_{Q_\rho}(e).
\end{equation}
The meaning of this definition is quite straight: a local through-segment in the preceding frame and one in the succeeding frame are related by inheritance if and only if each is a legal apparent through-path in its own local diagram and both are attached to the same pair of external boundary legs.
If $    \theta\in\mathcal{L}_{P_\rho}(e)$ is given in the preceding frame, then all of its possible counterparts in the succeeding frame are precisely the elements of the boundary fibre $\mathcal{L}_{Q_\rho}(e)$. The success, failure, and multiplicity of local
inheritance are therefore completely determined by the cardinality of this fibre:
\begin{align}
    \left|\mathcal{L}_{Q_\rho}(e)\right|=0
    &\quad\Longrightarrow\quad
    \text{inheritance fails},
    \\
    \left|\mathcal{L}_{Q_\rho}(e)\right|=1
    &\quad\Longrightarrow\quad
    \text{inheritance succeeds uniquely},
    \\
    \left|\mathcal{L}_{Q_\rho}(e)\right|>1
    &\quad\Longrightarrow\quad
    \text{inheritance succeeds non-uniquely}.
\end{align}

If the same rewrite rule is used in the opposite direction at some step of a rewriting movie, namely from $Q_\rho$ to $P_\rho$, then the corresponding local inheritance relation is the converse relation of $\tau_\rho$. Thus, a rewrite rule itself carries no absolute direction of ``time’’: which side is the preceding frame and which is the succeeding frame is determined by the direction in which the particular rewriting movie is played.

\subsubsection{The Local-to-Global Gluing Principle}\label{sec413}

The definitions above concern only the local diagrams. We now explain how they induce the inheritance of an entire through-path.

Suppose that a rewrite of a larger diagram takes the form
\begin{equation}
    C[P_\rho]
    \longrightarrow
    C[Q_\rho],
\end{equation}
where
\begin{equation}
    \rho:
    P_\rho
    \equiv_\rho
    Q_\rho
    :
    A_\rho\to B_\rho,
    \qquad
    \llbracket P_\rho\rrbracket
    =
    \llbracket Q_\rho\rrbracket .
\end{equation}
Here $C[-]$ is a context with a hole, and the boundary of the hole is the common boundary $\partial\rho$ of the two local diagrams. For each boundary-leg pair $    e\in\binom{\partial\rho}{2},$ we already have the local boundary fibres:
\begin{align}
    \mathcal{L}_{P_\rho}(e)
    &=
    \left\{
        \theta\in\mathcal{L}(P_\rho)
        \,\middle|\,
        \partial_{P_\rho}\theta=e
    \right\},
    \\
    \mathcal{L}_{Q_\rho}(e)
    &=
    \left\{
        \theta'\in\mathcal{L}(Q_\rho)
        \,\middle|\,
        \partial_{Q_\rho}\theta'=e
    \right\}.
\end{align}

We now define a \emph{context fibre} $\mathcal{L}_C(e)$. It consists of all external through-path remnants in the context $C[-]$ that attach to the boundary-leg pair $e$ of the hole. Denote by $\mathcal{L}^{\circ}(C[P_\rho])$ the set of global apparent through-paths that pass through the replacement region. Each such global path decomposes uniquely into the gluing of an external remnant and a local segment inside
the hole:
\begin{equation}
    \mathcal{L}^{\circ}(C[P_\rho])
    \cong
    \coprod_{e\in\binom{\partial\rho}{2}}
    \mathcal{L}_C(e)
    \times
    \mathcal{L}_{P_\rho}(e).
\end{equation}
After the rewrite, one likewise has
\begin{equation}
    \mathcal{L}^{\circ}(C[Q_\rho])
    \cong
    \coprod_{e\in\binom{\partial\rho}{2}}
    \mathcal{L}_C(e)
    \times
    \mathcal{L}_{Q_\rho}(e).
\end{equation}
Notice that the external context is unchanged, and hence the same context fibre $\mathcal{L}_C(e)$ occurs on both sides. The only change takes place within the boundary fibre inside the hole:
\begin{equation}
    \mathcal{L}_{P_\rho}(e)
    \longrightarrow
    \mathcal{L}_{Q_\rho}(e).
\end{equation}
We can accordingly define the global inheritance relation. Suppose that a global path in the preceding frame is written as
\begin{equation}
    \Gamma=(\xi,\theta),
    \qquad
    \xi\in\mathcal{L}_C(e),
    \qquad
    \theta\in\mathcal{L}_{P_\rho}(e).
\end{equation}
Its possible inheritors in the succeeding frame are precisely
\begin{equation}
    \Gamma'=(\xi,\theta'),
    \qquad
    \theta'\in\mathcal{L}_{Q_\rho}(e).
\end{equation}
In other words,
\begin{equation} 
    (\xi,\theta)
    \longmapsto
    \left\{
        (\xi,\theta')
        \,\middle|\,
        \theta'\in\mathcal{L}_{Q_\rho}(e)
    \right\}.
  \end{equation}
This is the \emph{local-to-global gluing principle}: 
The one-step inheritance of an entire through-path consists of leaving its external remnant unchanged and replacing the local through-segment inside the hole by a segment belonging to the same boundary fibre.
Consequently, the inheritance relation at the level of the entire diagram requires no independent additional definition. Once every global rewrite is decomposed into an application of some local elementary rewrite rule within a context, the inheritance of the entire through-path is automatically induced by the local inheritance relation $\tau_\rho$ through gluing with that context.
If a global apparent through-path does not pass through the local region being replaced, then the rewriting step is transparent to it: in the succeeding frame, it is uniquely inherited as the same external path. The genuinely nontrivial case is the one considered above, in which the path passes through the replacement region. The external remnant $\xi$ is then preserved unchanged, while all questions of success, failure, and multiplicity are concentrated in the local boundary fibre
$\mathcal{L}_{Q_\rho}(e)$. 
We therefore immediately obtain the following three cases:
\begin{align}
    \left|\mathcal{L}_{Q_\rho}(e)\right|=0
    &\quad\Longrightarrow\quad
    \text{the global path dies},
    \\
    \left|\mathcal{L}_{Q_\rho}(e)\right|=1
    &\quad\Longrightarrow\quad
    \text{the global path is inherited uniquely},
    \\
    \left|\mathcal{L}_{Q_\rho}(e)\right|>1
    &\quad\Longrightarrow\quad
    \text{the global path undergoes branching inheritance}.
\end{align}
In the first case, the global path dies because there is no longer any legal local through-segment inside the hole with the same boundary data $e$. In other words, the fate of global inheritance is completely governed by the cardinality of the local boundary fibre.

\subsection{Further Clarification of the Definition of Quantum Information Flow}\label{sec42}

With the notion of compatible inheritance now fully clarified, we can obtain
a deeper understanding of our definition of quantum information flow. The
essential point is that compatible inheritance $    T_{\rho_k}
    \subseteq
    \mathcal{L}(D_k)
    \times
    \mathcal{L}(D_{k+1}),$ is a relation rather than an ordinary function. Consequently, after a single rewriting step, an apparent through-path may have no inheritor, a unique
inheritor, or several possible inheritors. More strictly speaking, $\Gamma_0$ is a Coecke
flow line certified by the rewriting movie $\mathcal{W}$ if and only if, for every fixed branch $    \omega\in\Omega,$ there exists an inheritance lineage $    \left(
        \Gamma_0^\omega,
        \Gamma_1^\omega,
        \ldots,
        \Gamma_{n_\omega}^\omega
    \right)$ 
such that
\begin{equation}
    \Gamma_k^\omega
    \;T_{\rho_k^\omega}\;
    \Gamma_{k+1}^\omega,
    \qquad
    k=0,\ldots,n_\omega-1,
\end{equation}
and
\begin{equation}
    \Gamma_{n_\omega}^\omega
    =
    \ell_{x\to y}^\omega.
\end{equation}
Here $\ell_{x\to y}^\omega$ denotes the trivial apparent through-path carried by the final tensor-decoupled bare wire.

The formalization of compatible inheritance also yields a backward-generation algorithm for Coecke flow lines. Intuitively, rather than first enumerating all apparent through-paths in the initial string diagram $D_0$ and then testing them one by one, we may start from the terminal bare wire and successively take inverse images while moving backwards through the rewriting movie. In this way, all paths certifiable by the rewriting movie can be generated directly. 
More explicitly, suppose that the final-frame diagram has been rewritten into the form $    D_n
    \simeq
    B_{x\to y}\otimes R,
$ where $B_{x\to y}$ is a bare-wire diagram representing a simple morphism and is tensor-decoupled from the residual subdiagram $R$. The bare-wire diagram itself trivially carries a unique apparent through-path $\ell_{x\to y}$, we have:
\begin{equation}
    \mathcal{L}(B_{x\to y})
    =
    \left\{
        \ell_{x\to y}
    \right\}.
\end{equation}
We now introduce the notation $    \mathcal{C}_k
    \subseteq
\mathcal{L}(D_k),$
where $\mathcal{C}_k$ denotes the set of certifiable apparent through-paths in the $k$th frame--more precisely, those apparent through-paths that can be certified by the remaining remaining suffix $    D_k
    \Rightarrow
    D_{k+1}
    \Rightarrow
    \cdots
    \Rightarrow
    D_n$ of the rewriting movie. At the final frame,
we set
\begin{equation}
    \mathcal{C}_n
    :=
    \left\{
        \ell_{x\to y}
    \right\}.
\end{equation}
Then, recursively for $    k=n-1,\ldots,0,$ define
\begin{align}
    \mathcal{C}_k
    &:=
    T_{\rho_k}^{-1}
    \bigl[
        \mathcal{C}_{k+1}
    \bigr]
    \nonumber\\
    &:=
    \left\{
        \Gamma_k\in\mathcal{L}(D_k)
        \,\middle|\,
        \exists\,
        \Gamma_{k+1}\in\mathcal{C}_{k+1}
        \text{ such that }
        \Gamma_k
        \;T_{\rho_k}\;
        \Gamma_{k+1}
    \right\}.
\end{align}
The resulting set
\begin{equation}
    \mathcal{C}_0
    \subseteq
    \mathcal{L}(D_0)
\end{equation}
is precisely the set of all genuine through-paths in the initial string diagram that can be certified by this rewriting movie.

For a Coecke flow line, however, the situation is slightly subtler. A Coecke flow line is a branch-independent apparent through-path $\Gamma$ in the initial protocol string diagram $D^{\mathrm{init}}$. In each fixed branch $\omega$, it is instantiated as $\Gamma_0^\omega$. We must therefore perform the same backward-generation procedure separately for every fixed branch $\omega\in\Omega$.
For each branch, set
\begin{equation}
    \mathcal{C}_{n_\omega}^\omega
    :=
    \left\{
        \ell_{x\to y}^\omega
    \right\},
\end{equation}
where $\ell_{x\to y}^\omega$ is the unique apparent through-path carried by the bare-wire diagram $B_{x\to y}^\omega$. Then define recursively 
\begin{equation}
    \mathcal{C}_k^\omega
    :=
    \left(
        T_{\rho_k^\omega}
    \right)^{-1}
    \bigl[
        \mathcal{C}_{k+1}^\omega
    \bigr],
    \qquad
    k=n_\omega-1,\ldots,0.
\end{equation}
The branch-independent path $\Gamma$ then defines a Coecke flow line from
$x$ to $y$ if and only if
\begin{equation}
    \Gamma_0^\omega
    \in
    \mathcal{C}_0^\omega
    \qquad
    \text{for every }
    \omega\in\Omega.
\end{equation}
In words, for every fixed classical branch, backward generation from the
tensor-decoupled bare wire in the final frame must reach the initial
instantiation of $\Gamma$ in that branch.

The same construction applies to the joint configurations of
Section~\ref{sec:joint-configurations}, with inverse images
taken under $\widehat{T}_{\rho}^{(m)}$.
Suppressing the branch label temporarily, let
$\boldsymbol{\ell}=(\ell_1,\ldots,\ell_m)$ be the configuration
carried by the distinct terminal bare-wire factors.
Set
\begin{equation}
    \mathcal{C}_n^{(m)}
    :=
    \{\boldsymbol{\ell}\},
    \qquad
    \mathcal{C}_i^{(m)}
    :=
    \left(\widehat{T}_{\rho_i}^{(m)}\right)^{-1}
    \left[\mathcal{C}_{i+1}^{(m)}\right],
    \qquad i=n-1,\ldots,0.
\end{equation}
Every $\mathcal{C}_i^{(m)}$ is a subset of
$\operatorname{Conf}_m(D_i)$.
Repeating this construction in each branch, a
branch-independent initial candidate is certified precisely
when its instantiation belongs to $\mathcal{C}_0^{(m),\omega}$
for every $\omega\in\Omega$.
This is the backward-generation counterpart of the forward
joint definition: the search retains complete configurations
whose representatives are compatible at every frame.
Independently generated single-line witnesses need not
satisfy this joint requirement.

\subsection{ Compatible-Inheritance Relations in ZX String Diagrams}\label{sec43}

The formalism developed above does not depend on any particular string-diagrammatic language. It requires only that, within a local diagram, we can well define apparent through-paths, boundary-leg pairs, boundary fibres, and the inheritance relation induced by a local rewrite. When one specializes Coecke-flow theory to a particular string-diagrammatic calculus, the essential task is then precisely to perform a boundary-fibre analysis of the elementary local movies of that calculus.
We shall illustrate this procedure using two concrete small-scale rewrite rules of the ZX calculus. The task is straightforward. For a given rewrite rule $    \rho:
    P_\rho
    \equiv
    Q_\rho,$ 
we first identify its external boundary $\partial\rho$, then classify the boundary-leg pairs $    e\in\binom{\partial\rho}{2},$ and finally compute the cardinalities of the boundary fibres on the two sides, $    \left|\mathcal{L}_{P_\rho}(e)\right|,$ $    \left|\mathcal{L}_{Q_\rho}(e)\right|.$ Once these data are known, the effect of the rewrite rule on Coecke flow lines is, in principle, completely determined.

\begin{figure}[htbp]
    \centering
    \includegraphics[width=1\textwidth]{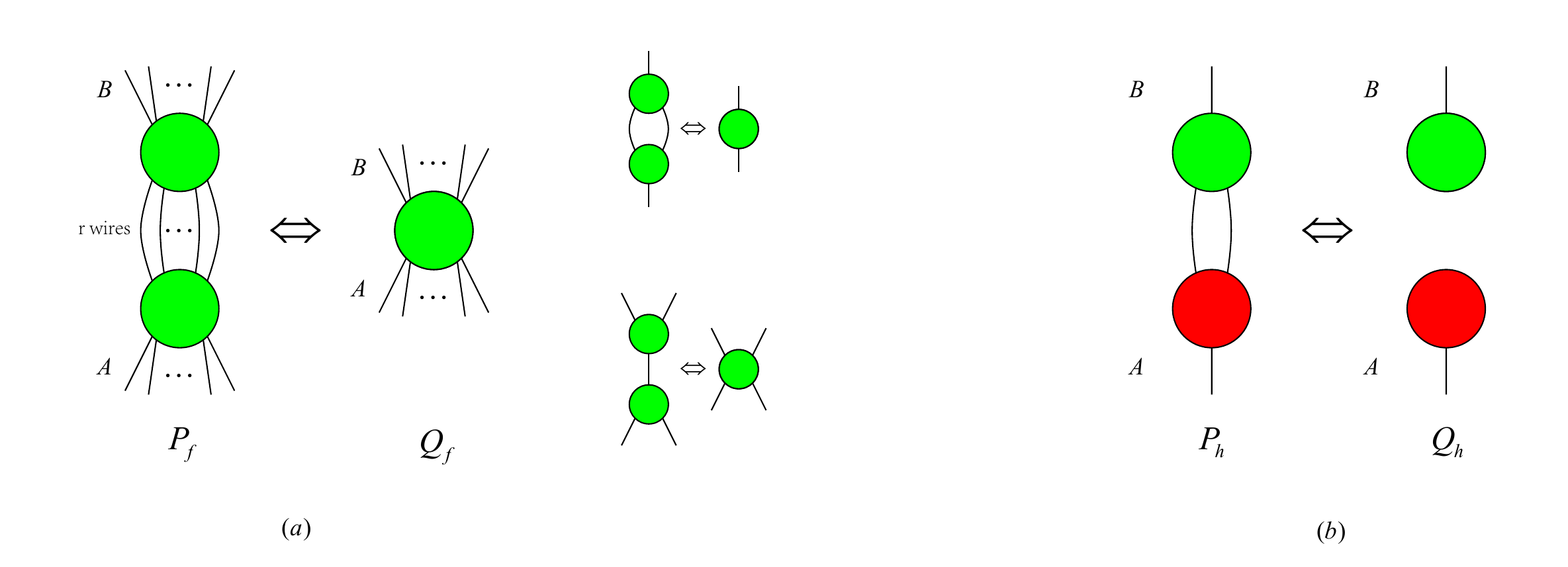}
    \caption{ Two local ZX rewrites illustrating compatible inheritance.
    (a) The generalized same-colour spider-fusion equation     $P_f\equiv Q_f$, whose two-spider side contains $r$ internal wires,     together with two more elementary fusion rules from which the     generalized equation can be assembled: a double-wire fusion with one     external boundary leg on each spider, and a single-wire fusion with two     external boundary legs on each spider.
    (b) The Hopf rewrite between the apparently connected side $P_h$ and     the disconnected side $Q_h$.}
    \label{fig-fib}
\end{figure}

Consider first the spider-fusion rule of the ZX calculus, whose reverse direction is the spider-splitting rule, as shown in Fig.~\ref{fig-fib}(a). We obtain the following table:
\begin{equation}\label{tabf}
\begin{array}{c|cc}
    \text{type of boundary-leg pair}
    &
    \left|\mathcal{L}_{P_f}(e)\right|
    &
    \left|\mathcal{L}_{Q_f}(e)\right|
    \\ \hline
    e\in A\times B
    & r & 1
    \\
    e\in\binom{A}{2}
    &1 ? & 1
    \\
    e\in\binom{B}{2}
    &1 ? & 1
\end{array}
\end{equation}
Here $P_f$ denotes the two-spider side and $Q_f$ the single-spider side. $A$ and $B$ are the sets of external legs attached to the two original spiders, respectively; and $r$ is the number of internal wires connecting the two spiders.
This table already displays the essential features of the rewrite rule.
First, spider fusion or splitting does not itself produce an empty fibre. It is therefore not a local mechanism that directly eliminates a through-path. 

Now let us consider a cross-side boundary pair $    e\in A\times B.$ The two-spider side contains $r$ local through-path representatives, whereas the single-spider side contains only one. 
Thus, if a certifying rewriting movie uses the fusion direction at some step, $    \text{two spiders}
    \longrightarrow
    \text{one spider},$ then, for this cross-side boundary pair $e$, the fibre cardinality changes as $    r\longrightarrow 1.$ This means that, whichever of the $r$ internal wires is traversed by the candidate apparent through-path on the $P_f$ side, that path is successfully inherited--and uniquely so--by the sole admissible apparent through-path representative on the $Q_f$ side. 
If, by contrast, the certifying movie uses the splitting direction at some step, $    \text{one spider}
    \longrightarrow
    \text{two spiders},$ then the same cross-side boundary pair gives $    1\longrightarrow r.$ 
This means the candidate apparent through-path on the $P_f$ side thereby branches into several admissible successors in this elementary movie. This is not a failure, but it results in a proliferation of possible witness lineages. Subsequent certification must continue to examine these admissible successors until at least one lineage is found that survives all the way to being carried by a terminal tensor-decoupled bare wire. Only then can the genuine through-goingness of the initial path be certified, so that it qualifies as a Coecke flow line. Evidently, such a one-to-many inheritance relation also has a nontrivial effect on the backward-generation algorithm.\footnote{We shall not elaborate on this point here, since the forward and backward descriptions are equivalent.}

Now let us consider a same-side boundary pair $    e\in\binom{A}{2}$ or $ e\in\binom{B}{2}.$ 
The question marks in the table (~\ref{tabf}) have been left deliberately. At first sight, one might expect that, on the $P_f$ side, a path need only pass directly through the spider on which its two boundary legs are incident, and hence that the corresponding fibre should have cardinality $1$. This reasoning, however, overlooks the possibility that the path may temporarily leave that spider, enter the other spider along one internal wire, return along a different internal wire, and finally exit through another external leg on its original side. Moreover, when sufficiently many internal wires are present, still longer alternating excursions may occur. 
Such complexity appears not to constitute substantive inheritance content carried by spider fusion itself. Rather, it resembles a redundancy produced by compressing a large derived equation into a single rewriting step. To remove this redundancy, we may decompose spider fusion into the two more elementary fusion rules also shown in
Fig.~\ref{fig-fib}(a).
In the first type, the two spiders are connected by only one internal wire. A route that leaves one spider cannot subsequently return to it, since no second internal wire is available. Every boundary-leg pair therefore gives unique inheritance. In the second type, the two spiders are joined by two internal wires, but each spider has only one external boundary leg. 
Since each spider has only one external boundary leg, when a route that crosses to the other spider along one internal wire and returns along the other comes back to the very spider at which it entered, the only external boundary leg available there is the one through which it entered. Under our standing convention that an apparent through-path may not traverse the same wire more than once, such recirculation is therefore excluded.
A generalized spider fusion involving spiders of arbitrary valence and an arbitrary number of internal wires can be expanded into a fine-grained rewriting movie composed of these two elementary rules. At every step of this movie, recirculation-type redundancy is prevented from appearing as an uncontrolled one-step boundary fibre. In this sense, the question marks in the coarse-grained table can be dispensed with.

As our next example, we investigate the so-called Hopf rule, as shown in
Fig.~\ref{fig-fib}(b). Its external boundary contains only two wire ports, and hence there is a unique boundary-leg pair, $    e=\{x,y\}.$ 
Let $P_h$ denote the apparently connected side and $Q_h$ the disconnected side. Their boundary-fibre data are
\begin{equation}
\begin{array}{c|cc}
    \text{unique boundary-leg pair}
    &
    \left|\mathcal{L}_{P_h}(e)\right|
    &
    \left|\mathcal{L}_{Q_h}(e)\right|
    \\ \hline
    e=\{x,y\}
    & 2 & 0
\end{array}
\end{equation}
Notice that an empty fibre occurs on the right-hand side: $    \mathcal{L}_{Q_h}(e)
    =
    \varnothing.$ This signifies the death of inheritance. If the direction of the rewrite in a certifying movie is $    P_h
    \longrightarrow
    Q_h,$ then a local apparent through-path already present on the apparently
connected side, $    \theta
    \in
    \mathcal{L}_{P_h}(e),$ has no admissible inheritor in the succeeding frame. The corresponding inheritance lineage therefore terminates at this step. Note that in the forward certification procedure, since the task is to determine whether a candidate apparent through-path chosen in the initial diagram realizes genuine through-goingness, the opposite rewrite direction, $
    Q_h
    \longrightarrow
P_h,$ need not be considered~\footnote{This direction nevertheless remains meaningful for the backward-generation algorithm. }, since the disconnected side has admitted no legal apparent through-path in the first place. 

The same boundary-fibre analysis can be carried out, rule by rule, for a set of elementary local ZX rewrite rules adopted as the axioms of a deductive system, thereby determining their respective mechanisms of compatible inheritance. We shall not mechanically repeat this analysis for every rule here. Reassuringly, the boundary fibres of the small ZX diagrams involved have cardinality $1$ or $0$ in most cases, so that the resulting structure is as simple as one might have expected.
Here are some additional comments. First, the axiomatic content of the ZX calculus depends on its intended expressive scope. When the processes under consideration are restricted to stabilizer processes, a comparatively small collection of ZX rewrite rules is sufficient. We list such a collection in Appendix~\ref{appd2}; it is commonly referred to as a Clifford ZX rule set~\cite{Backens:2013hto, Backens:2015nhm}. If the scope is enlarged to Clifford+$T$ processes~\cite{Jeandel:2018rqy}, or to general pure qubit processes~\cite{Jeandel:2019mez}, a correspondingly strengthened ZX axiomatization must be chosen.
Second, even after the intended semantics has been fixed, there remains some conventional freedom in the choice of which rewrite rules are taken as the axiomatic premises of the deductive system~\cite{Vilmart:2018hiq}.
Finding an irredundant, or even near-minimal, complete axiomatization of the ZX calculus is itself a nontrivial problem~\cite{Vilmart:2018hiq, Backens:2017jwp}.

For the joint configurations introduced in
Section~\ref{sec:joint-configurations}, the boundary-fibre
analysis must be supplemented by a simultaneous choice of
local representatives. Even when every relevant single-line
boundary fibre is nonempty, their representatives may fail
to coexist without sharing a wire. The local-to-global
construction therefore keeps the external remnants fixed
and checks the wire supports of all affected through-segments
together, as required by
Eq.~\eqref{eq:joint-inheritance}.

The permitted local rewrite steps need not form an
irredundant axiom set. In particular, we may admit the Hopf
identity directly as a local step, equipped with its
boundary-fibre relation, even though it is derivable from
other ZX identities. If a displayed step is instead used
as an abbreviation for a specified sequence of elementary
rewrites, joint inheritance must hold at every intermediate
frame of that sequence. Different derivations of the same
semantic equality need not certify the same joint
configurations. A derivation that fails this requirement
remains a valid ZX equality proof, but does not provide a
joint witness for the configuration under consideration.

\section{The Physical Meaning of Quantum Information Flow}\label{sec5}

\subsection{Coecke Flow Lines in Quantum Strategy Diagrams}\label{sec51}

The categorical quantum mechanics developed by Coecke and others, together with the subsequent study of ZX string diagrams, offers us a shift in quantum theory from state-vector centrism to morphism centrism. From this process-centered perspective, quantum structure naturally encompasses not only Hilbert-space vectors, which can be formalized as special morphisms from the tensor unit to a nontrivial system, $    \lvert\psi\rangle : I \longrightarrow A,$ but also concepts of growing importance in quantum information theory, such as quantum channels, which can be formalized as morphisms between nontrivial systems, $    \mathcal{E}:A\longrightarrow B.$

In a certain sense, the CQM elevates the structure of the interaction between the ``quantum world'' and the ``classical world''---long treated merely as background in quantum theory--into an explicit process-theoretic object.
The operations through which the classical world intervenes in quantum structures may be divided into two classes according to their functional roles in a process diagram. The first consists of \emph{inquiry-type operations}: these are what are ordinarily called measurements, and their purpose is to obtain a classical response. The second consists of what we shall call \emph{shaping-type operations}, whose purpose is actively to alter quantum structure. In a process diagram, the former are represented by morphisms from a nontrivial system to the tensor unit, namely effects, $    e:A\longrightarrow I,$ whereas the latter are represented by morphisms from one nontrivial system to
another, $    f:A\longrightarrow B.$
More explicitly, an effect corresponds to such an arrangement of experimental devices, which is designed to make a quantum structure phenomenologically present itself as one among a set of distinguishable, empirically accessible appearances. In a individual run, one of these appearances actually occurs and may subsequently be recorded as ordinary classical information. At the same time, however, this kind of operation changes the quantum structure, placing it in the corresponding conditioned process. 
In a sense, this structural change is the passive price paid for obtaining a classical response.
A shaping-type operation, by contrast, means that we actively arrange experimental components which, during the experimental run, do not output distinguishable classical appearances, but instead couple coherently into the quantum structure and play a role of ``shaping’’ it within the quantum process. The crucial point is that genuinely nontrivial shaping does not consist in a single local operation. Rather, it consists in an overall strategy that organizes multiple inquiry-type operations and shaping-type operations so as to direct the quantum structure toward a target structure. This is what is ordinarily called a \emph{protocol}.

The specifically dichotomized terminology just introduced will prove worthwhile. An ordinary protocol can now be described as follows: given an initial quantum structure, a strategy attempts to shape it into a target morphism. Let us consider two examples.
Quantum teleportation is a shaping strategy that takes a cup morphism--what is ordinarily called a Bell state--as its given initial quantum structure, and the identity morphism--namely the identity channel-- as its target morphism. Using the classical information obtained through inquiry, it coordinates the Bell-type inquiry~\footnote{Here we use the short expression ``Bell inquiry'' for an arrangement of experimental apparatus that reliably associates the four Bell-basis states with four distinguishable, empirically accessible appearances; in other words, what is ordinarily called a Bell-measurement device.}
 (i.e., a Bell measurement) at Alice's side with the shaping operation (i.e., a Pauli correction) at Bob's side, and thereby successfully shapes the given structure into the target morphism.
Similarly, entanglement swapping is a shaping strategy that takes two juxtaposed cup morphisms as its given initial quantum structure, and a new cup morphism connecting the two outer sites as its target morphism. Using the classical information obtained through inquiry, it coordinates the Bell-type inquiry  (i.e., a Bell measurement) at the intermediate sites with the shaping operation (i.e., a Pauli correction) at the outer ends (that is, Pauli corrections), and thereby successfully shapes the given structure into the target morphism.
We emphasize that the nontriviality of \emph{successful shaping} derives, to a considerable extent, precisely from the epistemic limitations inherent in the role of \emph{measurement} as an \emph{inquiry}. We cannot directly experience quantum structure. Nor is the ordinary classical information obtained through inquiry an isomorphic transcription of that structure.
More strongly, even when a set of inquiry apparatuses has established a stable correspondence between a quantum structure and several empirically accessible appearances, the mechanism by which one definite appearance rather than another occurs in a particular experimental run remains opaque.
A shaping strategy can nevertheless accept such limited inquiry responses, whose underlying mechanism remains unclear, and use the classical information obtained through inquiry to coordinate subsequent operations. It thereby reliably attains a preassigned target morphism, regardless of which response the inquiry actually presents. It can thereby ensure that, whichever response is actually presented, the prescribed target morphism is still reliably achieved. It is in this sense that \emph{shaping} expresses a form of active control, in contrast with the phenomenological and passive character of \emph{inquiry}.

We now turn to the physical meaning of Coecke flow.
In brief, within the context of quantum theory, a Coecke flow line is, relative to a shaping strategy that specifically produces a simple bare-wire morphism decoupled from the remainder, a constrained, quasi-local, line-like presentation of that target bare-wire factor within the original strategy diagram. By a \emph{bare-wire morphism}, we mean a morphism whose string-diagrammatic representation is a bare wire; in the present discussion, we typically have in mind the identity morphism and the cup morphism.

Strictly speaking, two distinct levels should be distinguished here.  
The first is a purely string-diagrammatic fact. Given a rewriting movie
\begin{equation}
    D_0
    \Rightarrow
    D_1
    \Rightarrow
    \cdots
    \Rightarrow
    D_n,
\end{equation}
we have
\begin{equation}
    \llbracket D_0\rrbracket
    =
    \llbracket D_1\rrbracket
    =
    \cdots
    =
    \llbracket D_n\rrbracket.
\end{equation}
Each of the apparently different string diagrams in the movie is merely a different representation of the same morphism. Suppose, therefore, that the final frame admits a tensor decomposition
\begin{equation}
    D_n
    \simeq
    T\otimes R,
\end{equation}
containing a simple bare-wire factor $T$. 
Since $D_0$ and $D_n$ represent, in essence, the same morphism, it is natural to expect that the simple bare wire factor $T$ should retain at least some constrained corresponding presentation in the more complicated initial presentation $D_0$. This is precisely what a Coecke flow line attempts to identify by tracing $T$ backwards through the compatible-inheritance relations induced by string-diagram rewriting.
\footnote{Categorical process theories expressed in the language of string diagrams have, in fact, developed into serious research programs across many disparate domains. In particular, the categorical research on Natural Language developed by Coecke and his collaborators explicitly introduced Frobenius structures in several of its important developments~\cite{Kartsaklis:2014meaning, Sadrzadeh:2013frobeniusI, Sadrzadeh:2016frobeniusII}. 
This suggests that whenever semantics-preserving string-diagram rewriting reveals a target structure in a simple presentation, the question of its constrained, quasi-local presentation within a more complicated presentation may, in principle, have significance beyond quantum theory. We shall not develop such possible generalizations here.}
Only at the second level do we assign to the initial string diagram $D_0$, within the context of quantum mechanics, the specialized physical semantics of a ``shaping’’ strategy diagram. Such a diagram presupposes some initial quantum structure and organizes inquiry devices, the ordinary classical information obtained through inquiry, and the corresponding shaping operations, so that the complete strategy reliably realizes the target morphism. At this level, the simple bare wire factor $T$ represents either a Bell pair or an identity channel.

\subsection{Interpretation of Several Examples}\label{sec52}

We next turn to several concrete examples, through which we shall gradually develop and comment on the physical meaning of Coecke flow.

Our first examples are taken from several of Coecke's early papers. They include the process network already encountered above (see Fig.~\ref{fig-flow}), consisting only of bipartite projector boxes together with ordinary local unitary-operation boxes, as well as quantum teleportation, entanglement swapping, and GHZ-assisted teleportation.

\subsubsection{Example 1: A Process Network of Bipartite Projector Boxes}\label{sec521}

\begin{figure}[htbp]
    \centering
    \includegraphics[width=1\textwidth]{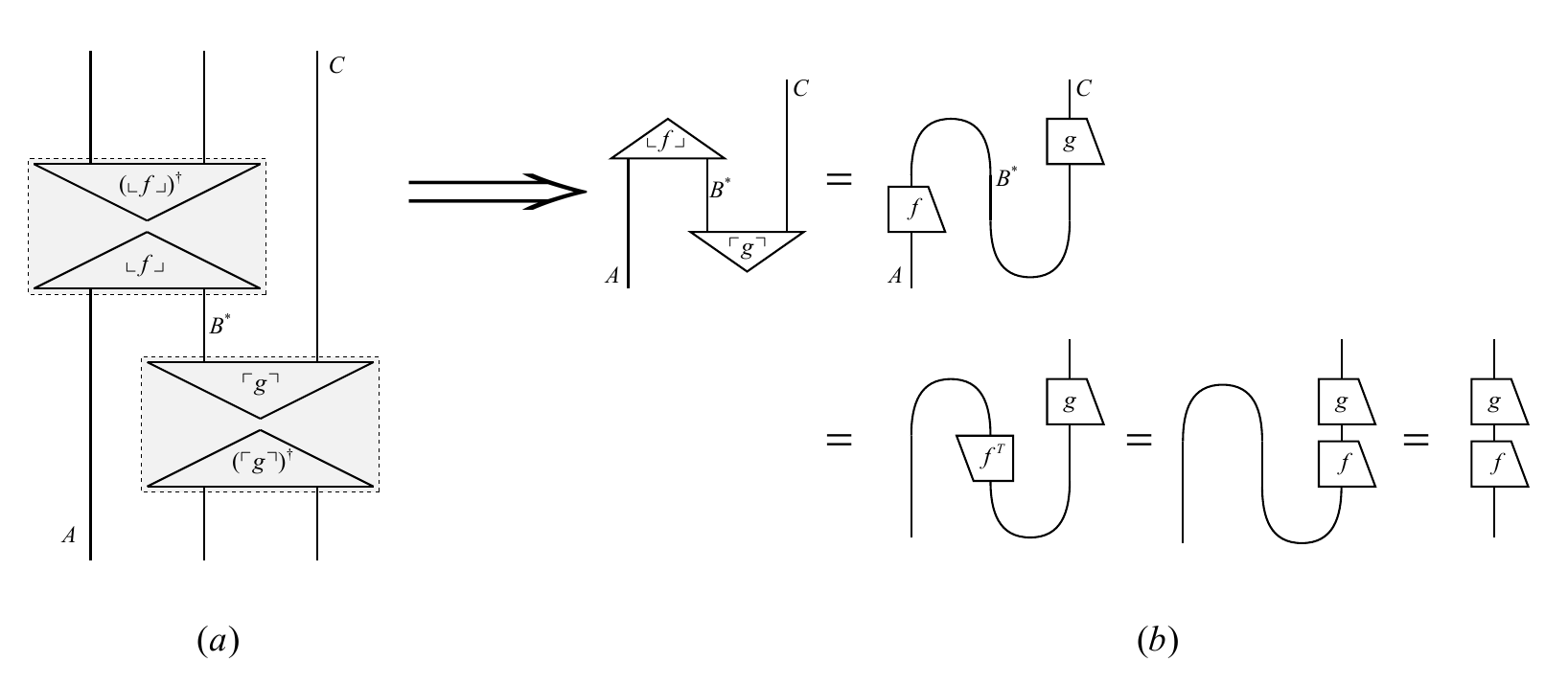}
    \caption{ (a) A fragment of a dagger compact closed string diagram composed of projector boxes.
(b) FIG.(a) is first brought, via C--J duality, into a standard form; the mate equations are then applied, and finally the snake equations are used. For more technical details, see Appendix \ref{appb2}. }
    \label{fig-qif}
\end{figure}
We reproduce in Fig.~\ref{fig-qif}(a) a typical local fragment of the diagram shown earlier in Fig.~\ref{fig-flow}. More precisely, it represents the standard dagger-compact string-diagrammatic form of a fixed branch of an entanglement specification network assembled from many projector boxes, together with some additional local unitary operations.
The essential point is that each projector box--a ket-bra in conventional Dirac notation--can always be constructed using either a name or a coname~\footnote{Here the corresponding name or coname is understood to be normalized. If an unnormalized convention is adopted, the resulting positive morphism differs from the corresponding rank-one projector only by a nonzero scalar factor.}, depending on the object-type convention adopted for the wires of the string diagram. More explicitly, using a name one may construct
\begin{equation}
    P_f
    :=
    \ulcorner f\urcorner
    \circ
    \bigl(\ulcorner f\urcorner\bigr)^\dagger
    :
    A^*\otimes B
    \longrightarrow
    A^*\otimes B ,
\end{equation}
whereas using a coname one may construct
\begin{equation}
    P_g
    :=
    \bigl(\llcorner g\lrcorner\bigr)^\dagger
    \circ
    \llcorner g\lrcorner
    :
    A\otimes B^*
    \longrightarrow
    A\otimes B^* .
\end{equation}
Accordingly, one always obtains a local pattern of the form shown in Fig.~\ref{fig-qif}(b), in which names and conames are locally glued together. Through the Choi-Jamiolkowski duality, together with repeated use of the mate identities and the tail-removal operation implemented by the snake equations, such a pattern can always be translated into a local one input-one output morphism. For more technical details, see Appendix \ref{appb2}.

Now consider a more complete Coecke flow line in Fig.~\ref{fig-flow}. We immediately encounter a nontrivial fact: the morphism represented by the entire diagram can be decomposed into the tensor product of the composition of the morphisms read successively along the Coecke flow line and the remainder of the diagram. 
More explicitly, let $    f_1,\ldots,f_n$ denote, in the order of the flow, the one input-one output morphisms induced whenever the Coecke flow line passes through a projector box. Then the morphism represented by the full diagram admits a clean decomposition of the form
\begin{equation}
    \llbracket D\rrbracket
    \simeq
    \bigl(f_n\circ\cdots\circ f_1\bigr)\otimes R ,
\end{equation}
where $    T
    =
    f_n\circ\cdots\circ f_1$
is the target morphism factor read off along the flow line, while $R$ denotes the remainder of the diagram.
Compared with our standard formulation in terms of a target bare-wire morphism factor, the target morphism factor $T$ here is, strictly speaking, a more general one-wire morphism than an identity or a cup. This does not, however, produce any essential structural difference. In many cases of practical interest, including the quantum teleportation protocol and the entanglement swapping protocol, the morphisms involved are invertible. One may then append a compensating operation $T^{-1}$ at the output boundary indicated by the Coecke flow line in the original strategy diagram, so that the target morphism factor ultimately shaped by the corrected strategy is exactly a bare-wire morphism.

Historically, it was precisely such clean examples, involving relatively simple initial quantum structures and relatively simple strategies, that Coecke first considered. As we have just seen, they exhibit a particularly strong form of genuine through-goingness in the strategy diagram: a target morphism factor is given exactly by composing, in the direction of the flow, the morphisms successively read off along a single path.
It was this unusually strong and elegant structural feature that motivated
Coecke to describe the phenomenon as a ``quantum information flow''.  Indeed, his original formulation amounts to the statement that, up to an overall nonzero scalar, the output state may be written as the result of applying to the input state the successive composition of the morphisms read off along the constrained path:
\begin{equation}
    \phi_{\mathrm{out}}
    =
    \bigl(f_n\circ\cdots\circ f_1\bigr)
    \bigl(\phi_{\mathrm{in}}\bigr).
\end{equation}
Quantum information therefore appears to flow ``following the line'', and this purely mathematical fact admits an interpretation in terms of a quantum information flow.

\subsubsection{Example 2: Teleportation through GHZ}\label{sec522}

\begin{figure}[htbp]
    \centering
    \includegraphics[width=0.9\textwidth]{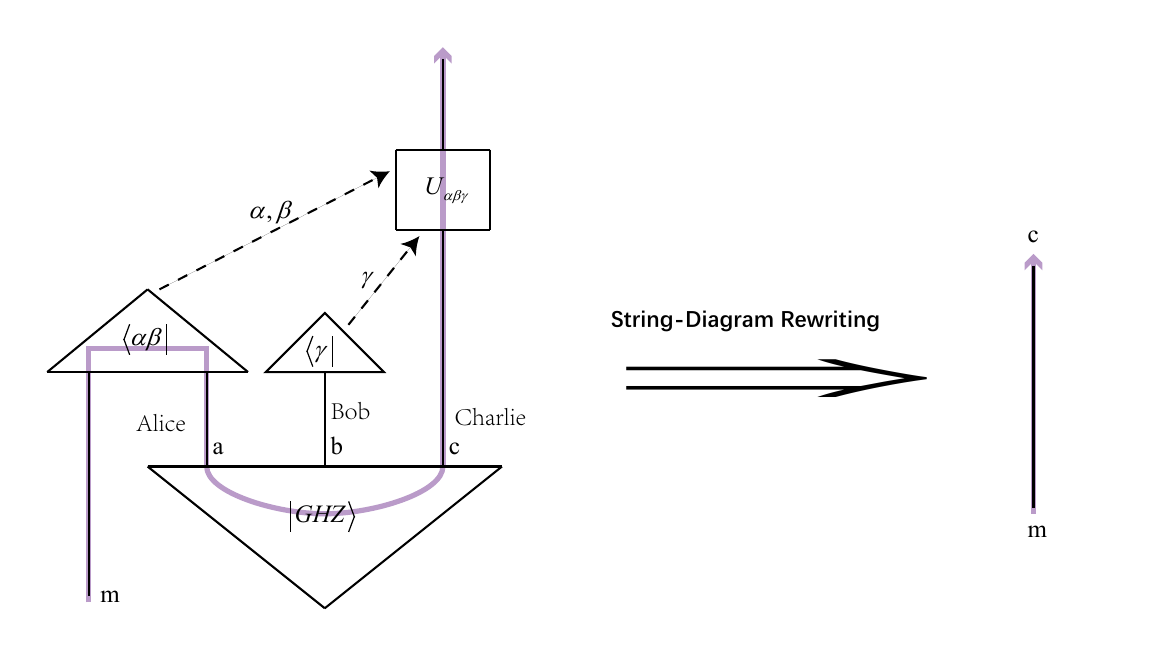}
   \caption{ GHZ-assisted teleportation and its Coecke flow line. }
\label{fig-ghz}
\end{figure}

Our next example is the GHZ-assisted teleportation protocol~\cite{Karlsson:1998opa,Hillery:1998yq}. We represent its entanglement specification network as the string diagram shown in Fig.~\ref{fig-ghz}, which we shall again regard as a strategy diagram.
Using the ZX calculus, one can prove that this initial string diagram can ultimately be reduced to a clean bare wire~\cite{Hillebrand:2011thesis}. A more detailed review of the protocol is given in Appendix~\ref{appd4}.
Here the initial quantum structure is a GHZ state, while the target morphism to be shaped is an identity morphism between $m$ and $c$, $    \mathrm{id}_{m\to c}.$ 
It means the successful teleportation of a quantum message $m$ from the sender Alice to the receiver Charlie. In particular, Ref.~\cite{Hillebrand:2011thesis} gives a more detailed ZX representation of the strategy diagram and proves, by ZX string-diagram rewriting for every classical branch, that the strategy is indeed equivalent to a bare wire whose morphism semantics is the identity. The technical details are reviewed in Appendix~\ref{appd4}.

This example provides particularly clear physical motivation for the systematic mathematical formalization of Coecke flow lines developed above. What we are doing, in effect, is extending Coecke's original notion of quantum information flow.
Coecke's initial idea of quantum information flow arose in special entanglement specification networks of the kind considered in Example~1, built from bipartite projector boxes and local one-wire operations. 
There seems, however, to be no reason to restrict the concept itself to this particular technical construction. 
Such networks should instead be regarded as the setting in which quantum information flow was first identified in a particularly transparent form.

More explicitly, whenever a more general strategy diagram physically realizes some form of quantum-information transfer and, at the level of string-diagrammatic calculus, reveals a corresponding minimal bare-wire factor, there is already a natural motivation for extending the notion of ``quantum information flow''. This is precisely what occurs in GHZ-assisted teleportation, where the initial quantum structure is no longer bipartite but genuinely multipartite.
We are therefore naturally led to ask whether, in such a more general setting, the terminal bare wire still possesses some through-going, line-like presentation in the initial strategy diagram. In the bipartite projector networks originally considered by Coecke, this through-goingness can be recognized directly by means of the elementary box-traversal rule.
For a general string diagram, however, that simple rule is no longer available. The systematic formalization developed above in terms of apparent through-paths, compatible inheritance, and Coecke flow lines was introduced precisely in order to formulate and answer this question rigorously for more general strategy diagrams.

For the present GHZ-assisted teleportation example, the purple line shown in Fig.~\ref{fig-ghz} is exactly such a genuine through-path, obtained by tracing back the terminal bare wire according to the mathematical procedure for Coecke flow lines developed in this paper.
We immediately encounter an important fact in this example beyond the bipartite-projector-network setting. Bob's auxiliary measurement node is not traversed by the purple flow line, even though that component is indispensable for reducing the complete strategy diagram to the final identity bare wire. Consequently, if one simply composes, in the direction of the flow, the morphisms successively encountered along this path, one plainly does \emph{not} recover the target bare wire morphism itself.

\subsubsection{Example 3: Teleportation and Entanglement Swapping}\label{sec523}

\begin{figure}[htbp]
    \centering
    \includegraphics[width=1\textwidth]{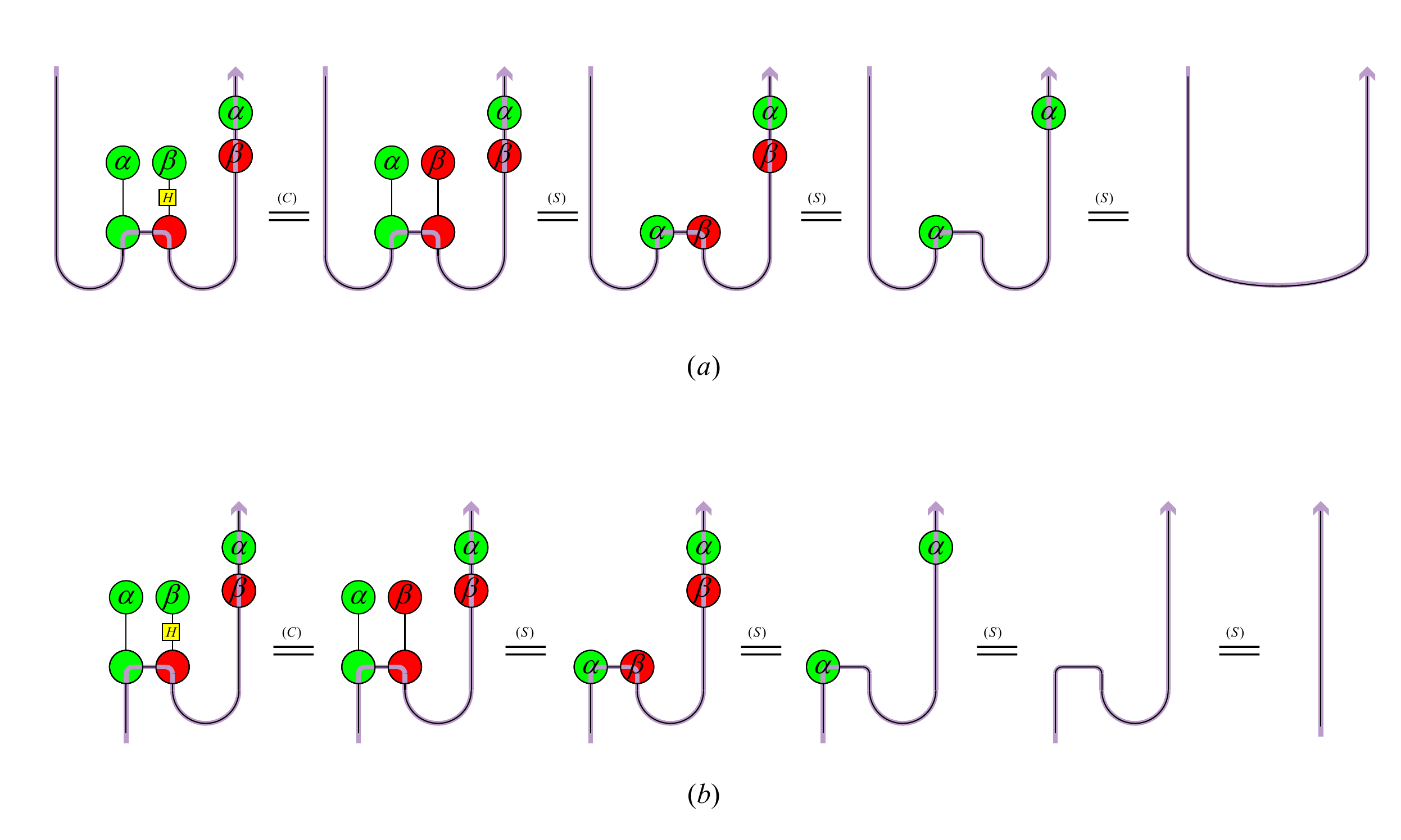}
\caption{(a) ZX string-diagram rewriting of entanglement swapping.
(b) ZX string-diagram rewriting of quantum teleportation. }
    \label{fig-swap}
\end{figure}

In fact, technical subtleties arise even in those protocols originally considered by Coecke, for which both the initial quantum structure and the strategy are exceptionally simple.
Figures~\ref{fig-swap}(a) and~\ref{fig-swap}(b) show, respectively, the ZX string-diagram rewritings of entanglement swapping and teleportation (redrawn FIG.~\ref {fig-zxte2} here for convenience), together with the corresponding Coecke flow lines, which manifestly obey the compatible-inheritance rules developed in this paper.
The first point to notice is that the string-diagram rewriting movie for entanglement swapping is, in essence, a ``bent'' version of that for teleportation.\footnote{This is precisely one of the virtues of categorical quantum mechanics, or more generally of string-diagrammatic calculus: it reveals common structure underlying processes that may initially appear quite different.} A frame-by-frame comparison shows that the two movies differ in only one respect at every stage: the former has an additional cup attached at the far left. Apart from this, their graphical structure and rewriting steps are identical. In this sense, the two protocols are structurally essentially the same. The difference is merely that the target bare wire morphism in entanglement swapping is a cup morphism (namely a Bell pair) between two remote qubits, whereas in teleportation the target bare wire morphism is an identity morphism, namely an identity channel.

Notice, however, that in the relatively coarse-grained string-diagrammatic representations used in examples such as Examples~1 and~2, these protocols initially appear to realize the most ideal situation: one may read off the morphism boxes successively encountered along the Coecke flow line, and their composition gives exactly the target factor. Once these coarse-grained morphism boxes--for example, the Bell effect--are further resolved into a more fine-grained string-diagrammatic language, such as the ZX diagrams used here, the situation changes subtly.
The flow line will now generally pass through only certain local constituents of the original coarse-grained morphism box. In Fig.~\ref{fig-swap}, for example, the flow line passes through only two of the small spiders constituting the Bell-effect morphism.
 Such spiders are generally multi-input--multi-output morphisms, while the flow line traverses only two of their ports. The categorical semantics does not, in general, automatically factor such a spider into a one input-one output morphism lying along the flow line, tensored with a lateral complementary factor.
Consequently, at this finer level of presentation, there need not exist a natural notion of simply ``composing the morphisms along the line.''
Nevertheless, whether one uses a coarse-grained string diagram or a more fine-grained representation such as a ZX diagram, the morphism semantics of the strategy diagram itself remains unchanged. Any legitimate string-diagrammatic rewriting must therefore ultimately reveal the same target bare wire factor, while the Coecke flow line in the initial diagram can always be rigorously traced back from the terminal bare wire through compatible inheritance.
The lesson is that, even in particularly clean examples with simple initial quantum structures and simple strategies, a better, presentation-independent statement is the following:

\begin{quote}
Around a Coecke flow line there exists a target-relevant effective region of limited thickness, with the line serving as its skeleton. When the process structure contained in this region is organized in the direction of the flow, it can be composed into the target bare-wire factor.
\end{quote}

In particular, at an appropriate scale of coarse-graining into boxes, this effective region may be compressed into a path of morphisms, whose effective morphisms, composed along the path, reproduce the target bare wire factor exactly.

We shall refer to this phenomenon specifically as \emph{strong quasi-locality in clean strategies}. A caution is nevertheless essential: whether such strong quasi-locality--in particular, the existence of a target-relevant effective region of bounded thickness--continues to hold for more complicated initial quantum structures and shaping strategies is not guaranteed by the definition of a Coecke flow line itself.

\section{Outlook and Conclusion}\label{sec6}

\subsection{Entanglement Distillation and Bit Threads}\label{sec61}

Our definition of Coecke flow allows a strategy to shape several mutually decoupled cup morphisms simultaneously. In the usual language of quantum information theory, this corresponds precisely to distilling several Bell states from a given initial quantum structure~\cite{Bennett:1995ra, Horodecki:2009zz, Bennett:1995tk}. Entanglement distillation therefore provides another very natural setting in which Coecke flow may be applied.

To keep the discussion within the formal framework established in this paper, we shall for the moment restrict attention to single-copy, exact distillation that succeeds deterministically in every classical branch. Suppose that an admissible shaping strategy ultimately reveals
\begin{equation}
    \eta_1\otimes\cdots\otimes\eta_k\otimes R,
\end{equation}
where each $\eta_i$ is a bare cup morphism and $R$ is the remainder decoupled from these target factors. The strategy has then shaped $k$ Bell-type target factors out of the initial quantum structure. To represent these target cups simultaneously, their
line-like representatives must admit a joint witness in
the sense of Section~\ref{sec:joint-configurations}.
When this condition holds, they constitute a joint
Coecke-flow configuration associated with the distillation
strategy.

For certain important and controllable cases, quantum information theory already provides a rather complete characterization of the reachability of such distillation processes. For example, deterministic single-copy LOCC transformations between bipartite pure states are completely characterized by the majorization relation between their Schmidt spectra~\cite{Nielsen:1999zza}. For bipartite pure stabilizer states, the stabilizer group admits a canonical decomposition in which the nonlocal sector consists of independent Bell-pair blocks; corresponding local Clifford operations can be constructed systematically to transform the state into a tensor product of Bell pairs and local residual factors~\cite{Fattal:2004frh, Audenaert:2005hch}.

For general multipartite entangled states and mixed states, however, the theory of entanglement distillation is far less complete, and many quantitative questions concerning optimal rates and finite-copy performance remain difficult in general~\cite{Horodecki:2009zz}. The present work does not attempt to solve these existing quantitative problems. Coecke flow suggests a question at a different level. Given a successful distillation strategy, can one further identify, for each target cup factor, a constrained, quasi-local, line-like presentation of that factor within the complete strategy string diagram, and regard such a presentation as an independent structural object of study? In this sense, quantum information flow may even provide a new, indirect way of probing the entanglement structure of an entangled state. The reason is that the very existence of such a flow ultimately presupposes a latent through-goingness supported by the entanglement structure intrinsic to the state itself, which is then activated by a particular shaping strategy.
From this point of view, the family of Coecke flows supported by an initial quantum structure under a given class of tasks and a given class of operations might tell us not only how many ``cup factors''--Bell states--can be distilled, but also something about how its intrinsic entanglement structure supports the shaping of those target factors.
In this sense, quantum information flow may even provide a new, indirect way of probing the entanglement structure of an entangled state. The reason is that the very existence of such a flow ultimately presupposes a latent through-goingness supported by the entanglement structure intrinsic to the state itself, which is then activated by a particular shaping strategy.

Strictly speaking, a morphism describing a quantum structure generally admits many mathematically equivalent string-diagrammatic presentations. When, however, a quantum structure is additionally equipped with some physically preferred graphical organization, the trajectory of a Coecke flow within that organization may acquire an intrinsic significance relative to the physical organization itself.
Holographic tensor-network states provide a particularly interesting example (see e.g. \cite{Swingle:2009bg, Swingle:2012wq, Pastawski:2015qua, Hayden:2016cfa, Bao:2018pvs}). They are holographic states with an explicit internal compositional structure: the tensor network is simultaneously a quantum process diagram representing the state and a discrete carrier of the bulk geometric organization supplied by holographic principles~\cite{Maldacena:1997re, Gubser:1998bc , Witten:1998qj }.
\footnote{By saying that a quantum structure is equipped with a physically preferred graphical organization, we do not mean to refer exclusively to holographic geometric states. More generally, several other examples immediately suggest themselves. Certain PEPS tensor networks, for instance, are quantum process diagrams defined on a given physical lattice~\cite{Verstraete:2004cf, Verstraete:2008cex}.
There are also quantum states that are interpreted as describing spatial geometry itself, such as spin-network states in loop quantum gravity ~\cite{Rovelli:1995ac, Rovelli:1994ge }.}
Particularly interestingly, in another work~\cite{companion} we found that, for
such ``holographic geometric states'', Coecke flow lines exhibit behavior
strikingly similar to that of so-called holographic bit threads~\cite{Freedman:2016zud, Cui:2018dyq, Headrick:2017ucz, Headrick:2022nbe}.

Holographic bit threads arise from a max-flow reformulation of the Ryu-Takayanagi prescription for holographic entanglement entropy~\cite{Ryu:2006bv, Ryu:2006ef, Hubeny:2007xt}, based on the max flow-min cut theorem.
Yet because they not only match the entanglement entropy directly, but also possess a strikingly geometric line-like appearance, they naturally invite the thought that they might capture additional geometric aspects of quantum entanglement. 
This interpretation, however, does not follow logically from the mathematical definition of bit threads. 
Coecke flow, on the other hand, starts from a completely different, process-theoretic point of departure and rigorously produces a class of through-going lines associated with the shaping of target entanglement factors. When the initial quantum structure also possesses a preferred physical geometry, these flow lines leave trajectories within that geometry.
In another parallel article~\cite{companion}, we investigate
the correspondence between Coecke-flow configurations in
holographic tensor-network states and the mathematical
constraints obeyed by bit threads, especially the
divergenceless condition and the density bound.
For this comparison, simultaneous trajectories are understood
as joint configurations at the elementary qubit-wire
resolution, with composite tensor-network bonds represented
by their specified bundles of elementary wires.
We believe that this similarity gives more substance to the possibility that the line-like geometry of bit threads may genuinely reflect certain aspects of the structure of quantum entanglement, rather than being merely an artifact of their graphical representation. In this sense, what might otherwise appear as a suggestive visual analogy becomes a question that can be seriously formulated and further tested.

\begin{figure}[htbp]
    \centering
    \includegraphics[width=0.9\textwidth]{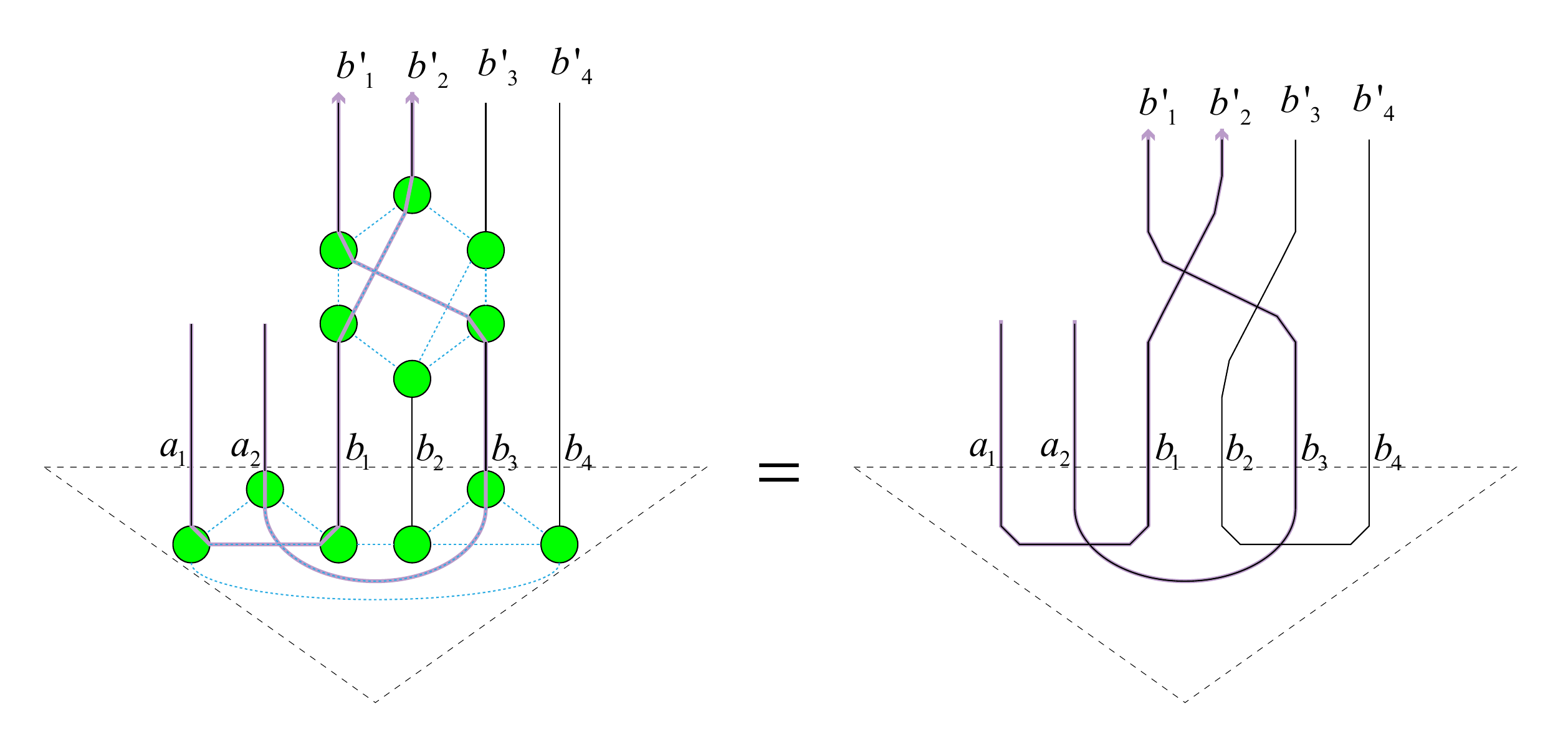}
    \caption{Coecke flows in a single-cell HaPPY example. The initial quantum structure is an $\mathrm{AME}(6,2)$ state represented by the lower triangular-prism ZX diagram, with external legs $a_1,a_2\in A$ and $b_1,b_2,b_3,b_4\in B$. The upper ZX subdiagram represents a $B$-local operation acting on $b_1,b_2,b_3$, while $A$ and $b_4$ are left unchanged; blue dashed edges denote wires carrying an $H$ gate. ZX rewriting shows that this shaping strategy produces two mutually decoupled target cup morphisms between $A$ and $B$, as exposed in the equivalent diagram on the right. The purple curves are the corresponding Coecke flow lines traced back from these target factors. In particular, their trajectories through the resource-state sector of the initial diagram exhibit the characteristic line-like behavior of holographic bit threads. }
    \label{fig-ame}
\end{figure}

To illustrate this idea, let us now present one elementary example. Figure~\ref{fig-ame} concerns a holographic tensor network that can be treated exactly within the ZX calculus: the $\mathrm{AME}(6,2)$ state used as a single cell in the HaPPY model~\cite{Pastawski:2015qua}. Its ZX representation is a triangular prism built from six spiders~\cite{Helwig:2013qoq}. Each blue dashed edge indicates an edge containing an $H$ gate.\footnote{All technical details concerning this example are given in Ref.~\cite{companion}. Here we use the example primarily to make the underlying idea transparent.}
The physical setting displayed in Fig. ~\ref{fig-ame}  is the following. We take the initial quantum structure to be a ``single-layer HaPPY disk,'' namely a single $\mathrm{AME}(6,2)$ state.\footnote{In contrast with the usual tensor-network convention, we draw all legs more carefully as upward-pointing. The topological deformation used in the drawing here makes no substantive difference.} Its external legs are divided into two sides, $    a_1,a_2\in A,$ $    b_1,b_2,b_3,b_4\in B.$ For ease of narration, we shall simply refer to these legs as ``experimenters'' belonging to two laboratories, $A$ and $B$.
We then construct the following strategy. The experimenters $b_1,b_2,b_3$ in laboratory $B$ jointly apply a $B$-local operation represented by the ``tilted tall triangular-prism ZX diagram'', while $b_4$ and all experimenters in laboratory $A$ act trivially. By string-diagram rewriting, one can prove that this strategy succeeds in shaping two target cup morphisms between $A$ and $B$. More importantly, together with the success of the shaping strategy, the rewriting also nontrivially exhibits the two corresponding Coecke flow lines, obtained by tracing the two target factors back into the original strategy diagram.
For readers familiar with holographic duality, it is easy to recognize that the trajectories of these Coecke flow lines within the resource-state sector of the initial strategy diagram in Fig.~\ref{fig-ame} have precisely the characteristic appearance of bit-thread trajectories. Reference~\cite{companion} further shows how this correspondence provides a physically motivated interpretation of the notorious nonuniqueness of holographic bit-thread configurations.

This comparison touches upon one of the most suggestive ideas in contemporary quantum-gravity research. The lesson of holographic duality is often summarized by the phrase ``gravity from quantum entanglement''. Yet, in a certain sense, quantum entanglement as a relational structure is itself still not fully understood. What holography seems rather to suggest is that certain highly organized patterns of quantum entanglement can, in special limits, reveal a structural form approaching the geometry of Einstein gravity. This may indicate that, even more generally, quantum entanglement as a relational structure possesses some still poorly understood, geometry-like level of organization.
Holographic states are, of course, not generic quantum states. On the contrary, they are highly special and are typically subject to strong conditions such as large $N$ and strong coupling. But precisely for this reason, they provide a refined and controllable sample through which some aspect of a deeper structure may become visible.

At present, our most thorough understanding of Coecke flow remains limited: it is a constrained, quasi-local, line-like presentation of a target bare wire factor within the complete strategy diagram. The increasingly influential entanglement-geometry perspective in contemporary physics nevertheless gives us reason to view this limited characterization with some optimism. Holographic theory has repeatedly shown that geometric objects can effectively characterize certain structural aspects of quantum entanglement even before those objects have been reduced to a complete microscopic interpretation.
From this perspective, the geometric character of the line-like presentation provided by Coecke flow may itself be capable of playing some role in characterizing the organization of entanglement. The similarity between bit threads and Coecke flows further strengthens this expectation. We emphasize, however, that what this similarity provides is a basis for further investigation, not an a priori logical proof.

\subsection{Quantum Information Flow beyond Dagger-Compact Structure and ZX String Diagrams}\label{sec62}

The historical development of quantum information flow is clearly rooted deeply in the dagger-compact process language developed by Coecke, and it subsequently acquired a richer expression in the ZX string diagrams that grew out of the same tradition. The present work naturally inherits this line of development. After the formalization developed above, however, one can see that the core notion of quantum information flow is not specific to the ZX calculus, or even to the dagger-compact framework itself. The more general structure on which it depends consists of a process diagram with a well-defined semantics, a string of string-diagram rewrites that preserve morphism semantics, and the compatible inheritance of some simple target factor through those rewrites. Quantum information flow can therefore, in principle, be extended beyond the original CQM framework and beyond ZX string diagrams.

From this more general standpoint, the CQM originally developed by Coecke etc., may, in a certain restricted sense, be regarded as CQM ``in the narrow sense''. Here, ``narrow'' does not mean that the framework is mathematically impoverished, nor that categorical quantum mechanics as a research programme is itself narrow. Rather, the term is intended to emphasize that the historically most mature and representative dagger compact-ZX realization arose from a rather definite and special physical context.
Indeed, the categorical quantum mechanics originally developed by Coecke did not place equal emphasis on describing the nontrivial internal organization of an initial quantum structure and on describing the laboratory strategy acting upon it. Much of the nontrivial content of the theory was concentrated on the latter: namely, on how nontrivial shaping strategies in the laboratory could be represented as process diagrams that can be composed, deformed, and reasoned about. By contrast, the initial quantum structure on which such a strategy acts was deliberately reduced to a particularly privileged and simple cup-type structure.
Of course, once the cup is singled out as the preferred form of the initial quantum structure, its corresponding cap is naturally privileged within the strategy as the matched form of inquiry (i.e., measurement).
This preference was thoroughly motivated by the class of problems that shaped the early development of the framework (teleportation and entanglement swapping being paradigmatic examples), yet it still bears a certain element of historical contingency. In this sense, the dagger-compact structure reflects a particular allocation of conceptual attention within the early CQM framework.
Clearly, this privileged status of the cup, although by no means arbitrary, should not automatically be elevated into an a priori paradigm for arbitrary quantum structures. In some cases, such as the stabilizer states encountered in Fig~\ref{fig-ame}, the ZX string-diagrammatic language subsequently developed on top of the dagger-compact skeleton is already capable not only of representing the laboratory strategy acting on the prepared quantum structure, but also of revealing in considerable detail the nontrivial internal organization of that prepared structure itself. The two can therefore be treated naturally within a single string diagram.

The essential point, however, is that neither dagger-compact categories nor the ZX categories obtained by further equipping them with the corresponding Frobenius structures constitute, a priori, a preferred intrinsic syntax for an arbitrary prepared quantum structure. Important nontrivial quantum states are already known whose internal organization is more naturally described by other kinds of categorical language. 
Spin-network states, for example, are naturally described in terms of representation categories of compact groups~\cite{ Baez:1994hx }, whereas string-net condensed states are naturally described in terms of fusion categories~\cite{ Levin:2004mi, Kirillov:2011mk}.
It follows that, in principle, within the framework of quantum information flow considered here, the string-diagrammatic language used to describe the prepared quantum structure need not be identical to the string-diagrammatic language used to describe the laboratory strategy. The two might instead be joined to form a hybrid categorical setting, without thereby altering the concept of quantum information flow itself. We leave this possibility open for future investigation.

\subsection{Conclusion}\label{sec63}

The main purpose of this work has been to reformulate a highly suggestive graphical intuition of quantum information flow from Coecke's early work as a string-diagrammatic concept that can be rigorously tested. 
In the original picture, one finds, on the one hand, information-flow paths identified in the initial entanglement specification network according to local rules, and, on the other hand, the bare wires ultimately revealed after semantics-preserving rewriting of the protocol string diagram. What we have attempted to provide is precisely the formal bridge that was previously missing between these two line-like descriptions.

To this end, we introduced the notions of apparent through-path, compatible inheritance, and genuine through-goingness, and defined a Coecke flow line as a branch-independent initial through-path that, in every fixed classical branch, can be traced through a sequence of compatible inheritances to a decoupled target bare wire factor. By means of local boundary fibres, this inheritance relation can further be localized to elementary string-diagrammatic rewrite rules, and the local-to-global gluing principle then determines the corresponding inheritance relation for complete apparent through-paths. The corresponding backward algorithm also shows that one may start from a terminal bare wire and generate, in reverse, all flow lines that can be certified. The ZX calculus provides a concrete and calculable example of this general structure.

In the physical setting of quantum theory, we have argued that the most robust interpretation of a Coecke flow is as a constrained, quasi-local, line-like presentation, within the complete protocol diagram, of a target bare wire morphism factor realized by this protocol strategy. 
For particularly clean protocols, such a presentation may even exhibit a stronger form of quasi-locality, but this stronger property is not guaranteed by the definition itself.

A further direction is to investigate the relation to flow
and generalized flow (gflow) in measurement-based quantum
computation~\cite{DanosKashefi:2006,BrowneEtAl:2007}.
These structures organize measurement order and correction
dependencies. In the standard $(X,Y)$-plane setting, an open
graph with gflow admits a vertex-disjoint path cover, whose
paths serve as wires in circuit
decompositions~\cite{MiyazakiEtAl:2015}.
This suggests a concrete question: under what choices of
ZX presentation and rewriting witness can such paths be
realized as the joint Coecke-flow configurations defined
in Section~\ref{sec:joint-configurations}?
The comparison should include any output compensation
needed to expose the target bare-wire factors, and should
distinguish the selected paths from the set-valued correction
data of gflow. Establishing such a relation could help
clarify how the geometry of target-factor representatives
is constrained by the correction structure of a protocol.

More generally, the formalization developed here does not depend on the ZX calculus and, in principle, does not depend on dagger-compact structure either. Its essential requirements are only a process diagram with a well-defined semantics, semantics-preserving local rewrites, and compatible inheritance of a target factor through those rewrites. We therefore hope that the present work provides a sufficiently clear conceptual foundation for ``quantum information flow'' to become once again an independently investigable structural object, while leaving room for further study of its role in entanglement structure and in broader categorical process theories.

\newpage
\begin{appendix}

\appendix

\section{Category Theory and String Diagrams }\label{appa}

\subsection{Categories}\label{appa1}

A category $\mathcal{C}$ consists of the following data:
\begin{itemize}
    \item objects $A,B,C,\ldots$;
    \item for each pair of objects $A,B$, a collection of morphisms from $A$ to $B$, denoted by $\mathcal{C}(A,B)$. We write a morphism $f\in\mathcal{C}(A,B)$ simply as
    \begin{equation}
        f:A\to B;
    \end{equation}
    \item associative composition: if $f:A\to B$, $g:B\to C$,     then there is a composite morphism
    \begin{equation}
        g\circ f:A\to C,
    \end{equation}
    satisfying
    \begin{equation}
        (h\circ g)\circ f = h\circ (g\circ f);
    \end{equation}
    \item identity morphisms: for each object $A$, there is an identity morphism
    \begin{equation}
        1_A:A\to A,
    \end{equation}
    such that, for every morphism $f:A\to B$,
    \begin{equation}
        f\circ 1_A = 1_B\circ f = f.
    \end{equation}
\end{itemize}

In what follows, we shall use these categorical terms to describe quantum mechanics. For physicists, this language is in fact more natural than it may appear at first sight. Objects may be understood as systems, while morphisms may be understood as operations or processes. Under this interpretation, composition in a category corresponds precisely to the successive execution of processes.

\subsection{ Monoidal Categories and String Diagrams }\label{appa2}

For quantum theory, in order to express the fact that processes can be combined ``in parallel”, we need to further equip a category with a tensor-product structure. A category equipped with such a structure is called a monoidal category. Monoidal categories play a foundational role in applied category theory. A monoidal category allows us to combine two systems $A$ and $B$ into a joint system $A\otimes B$, and to combine two processes $ f:A\to C$, $ g:B\to D$ in parallel as
\begin{equation}
    f\otimes g:A\otimes B\to C\otimes D.
\end{equation}

\begin{figure}[htbp]
    \centering
    \includegraphics[width=1\textwidth]{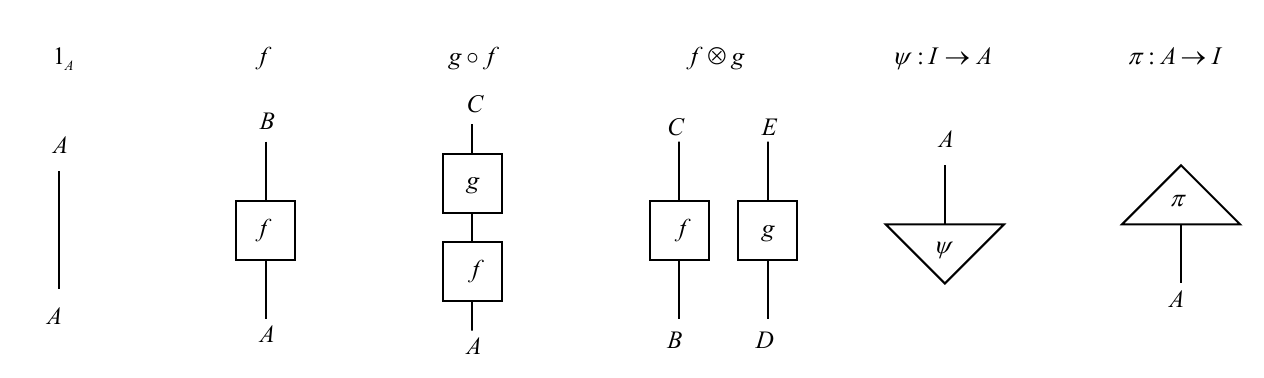}
    \caption{ Graphical representations of various morphisms in string diagrams.}
    \label{fig-stri}
\end{figure}
We now introduce the notion of string diagrams~\cite{Joyal:1991esw}. String diagrams are a class of embedded graphs with boundary, which can be used to represent morphisms in various monoidal categories. In a string diagram, wires represent objects, or more precisely identity morphisms on objects; boxes represent general morphisms; vertical stacking represents sequential composition; and horizontal juxtaposition represents tensor product. See Figure~\ref{fig-stri}. We follow the convention that diagrams are read from bottom to top. Notice in particular that the unit object $I$ is represented graphically by ``no wire'', but it is nevertheless important for the following objects: a state $ \psi:I\to A$ is an ``output wire with no input'', corresponding to the Dirac ket semantics, while an effect $ \pi:A\to I$ is an ``input wire with no output'', corresponding to the Dirac bra semantics.

The reader will immediately recognize that the ``string diagrams'' defined here are very close to what is usually called a ``graphical calculus''. This includes Penrose's brilliant tensor-diagrammatic notation proposed in 1971~\cite{Penrose:1971}, the tensor networks that are now ubiquitous in the quantum information and many-body physics communities, and also the ZX calculus to be introduced later. In fact, from the modern point of view, they are indeed all concrete instances of string diagrams under suitable semantics. One may say that ``graphical calculus'' is a broader and more phenomenological umbrella term, whereas by ``string diagrams'' in the present paper we more specifically mean those graphical languages whose skeleton is given by the compositional structure of sequential and parallel composition, and which can be understood within a monoidal-categorical semantic framework. In a broad graphical calculus, one often implicitly assumes that certain topological deformations of diagrams obviously do not change their meaning. The advantage of category theory is that it places these implicit rules back into an explicit structural background. In particular, we refer to the structures that will matter for CQM in this paper: monoidal, symmetric monoidal, and further dagger symmetric monoidal and dagger compact closed structures. Only after the relevant coherence theorems have been established~\cite{MacLane:1998, Joyal:1991esw, Kelly:1980, Selinger:2011,Selinger:2007eep} can we strictly say that, for string diagrams carrying these structures, topological deformations in the appropriate sense do not change the morphisms represented by the diagrams.

\section{Categorical Quantum Mechanics}\label{appb}

\subsection{Dagger Compact Closed Categories}\label{appb1}

\begin{figure}[htbp]
    \centering
    \includegraphics[width=1\textwidth]{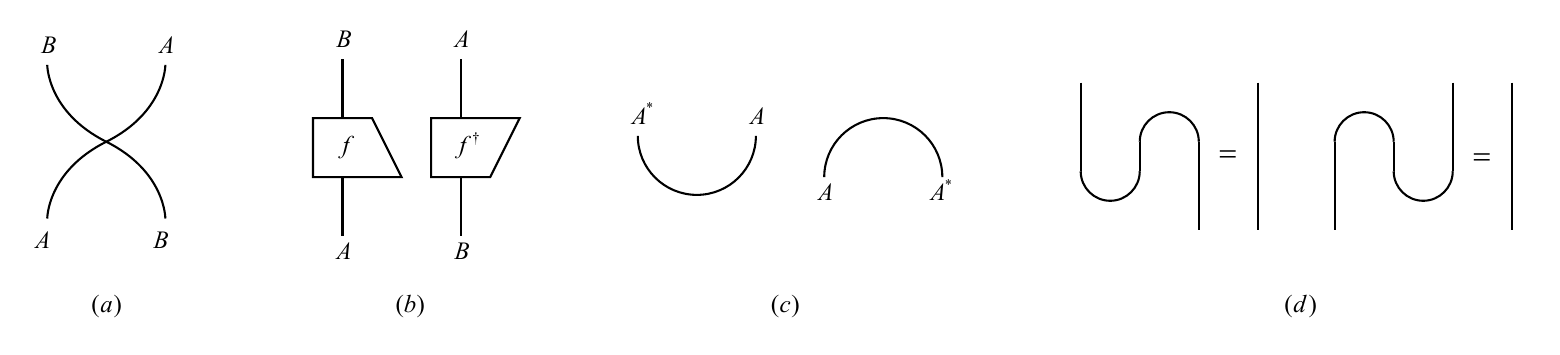}
    \caption{ (a) The crossing wire $\sigma$. 
(b) A morphism and its dagger. 
(c) The cup and the cap. 
(d) The snake equations. }
    \label{fig-comp}
\end{figure}

Starting from monoidal categories, we next successively add symmetry, dagger, and compact closed structures. In other words, we consider increasingly structured monoidal categories. Each strengthening adds, on the one hand, new axiomatic content at the algebraic level and, on the other hand, new graphical components or new legitimate deformation rules in the language of string diagrams. The reason for doing so is that the syntactic background needed for CQM is not exhausted by the serial and parallel composition of processes; it must also be able to express further structural features of finite-dimensional pure processes in ordinary quantum information theory. We are finally led to dagger compact closed categories, or simply dagger compact categories, introduced by Abramsky and Coecke under the name ``strongly compact closed categories''~\cite{Abramsky:2004doh, Selinger:2007eep}.  This structure will be seen to be indispensable for the graphical explanation of quantum phenomena such as teleportation, and it is also the underlying graphical mechanism on which Coecke historically relied in forming the intuition of ``quantum information flow''.

We now briefly introduce the above structures in order.

In the graphical language, passing from a monoidal category to a symmetric monoidal category (SMC) amounts to adding a new graphical component, namely the crossing wire $\sigma$:
\begin{equation}
    \sigma_{A,B}: A\otimes B \to B\otimes A ,
\end{equation}
as shown in Figure~\ref{fig-comp} (a). The symmetry $\sigma$ must satisfy the corresponding naturality and coherence conditions. The intuitive meaning of naturality here is the following: applying processes separately to two systems and then swapping them should give the same result as first swapping the two systems and then applying the corresponding processes to the swapped positions.

A dagger SMC is a symmetric monoidal category equipped with a dagger structure. In Hilbert-space semantics, $f^\dagger$ corresponds to the Hermitian adjoint. More explicitly, in a $\dagger$-SMC, every morphism $f:A\to B$ has a dagger $ f^\dagger:B\to A$ satisfying
\begin{equation}
    (g\circ f)^\dagger = f^\dagger\circ g^\dagger,\qquad
    (f\otimes g)^\dagger = f^\dagger\otimes g^\dagger,\qquad
    (f^\dagger)^\dagger = f.
\end{equation}
Graphically, the dagger operation on a morphism is drawn as a flipping of the box. For visual clarity, boxes representing morphisms are therefore drawn in shapes that are not symmetric under vertical reflection, as shown in Figure~\ref{fig-comp} (b). The corresponding graphical calculus and coherence results for dagger symmetric monoidal categories are discussed in \cite{Selinger:2007eep}.

Finally, to say that a category is a dagger compact closed category \cite{Selinger:2007eep, Abramsky:2004doh} means that it is a symmetric monoidal category equipped with a compatible dagger, and that for every object $A$ one is given a dual object $A^*$, together with
\begin{itemize}
    \item a cup, also called a unit or coevaluation,
    \begin{equation}
        \eta_A:I\to A^*\otimes A;
    \end{equation}
    \item a cap, also called a counit or evaluation,
    \begin{equation}\label{cap}
        \varepsilon_A:A\otimes A^*\to I.
    \end{equation}
\end{itemize}
These data satisfy the snake equations, which will be seen to be the root of the mechanism behind quantum information flow:
\begin{align}
    (\varepsilon_A\otimes 1_A)\circ (1_A\otimes \eta_A)
    &= 1_A, \\
    (1_{A^*}\otimes \varepsilon_A)\circ (\eta_A\otimes 1_{A^*})
    &= 1_{A^*}.
\end{align}
See Figure~\ref{fig-comp} (c) and (d).

A $\dagger$-compact structure also requires the dagger to be compatible with the cup-cap structure. Intuitively, the cap is the dagger of the cup; in other words, they are graphical vertical reflections of each other. However, in our convention, $\varepsilon_A$ is not literally equal to $\eta_A^\dagger$, because
\begin{equation}
    \eta_A^\dagger:A^*\otimes A\to I,
\end{equation}
and (\ref{cap}) have different input orders. Therefore, although they can still be understood graphically as saying that the cap is the dagger of the cup, at the level of typed formulae there is one symmetry map in between. There is likewise a coherence theorem for dagger compact closed categories: an equation expressed in the dagger compact categorical language holds if and only if it holds in the graphical language up to isotopy~\cite{Selinger:2007eep, Abramsky:2004doh }.

We should also foreshadow that the category $\mathbf{ZX}$ to be introduced later is in fact a concrete category with a dagger compact structure. Here one should keep in mind that ``dagger compact'' refers to a structural property. The category ZX, as a concrete category, not only has this dagger compact structure, but also carries $ZX$-spiders, phases, Hadamard gates, and the specific relations among these generators.

\subsection{C-J Duality and Mate}\label{appb2}

\begin{figure}[htbp]
    \centering
    \includegraphics[width=1\textwidth]{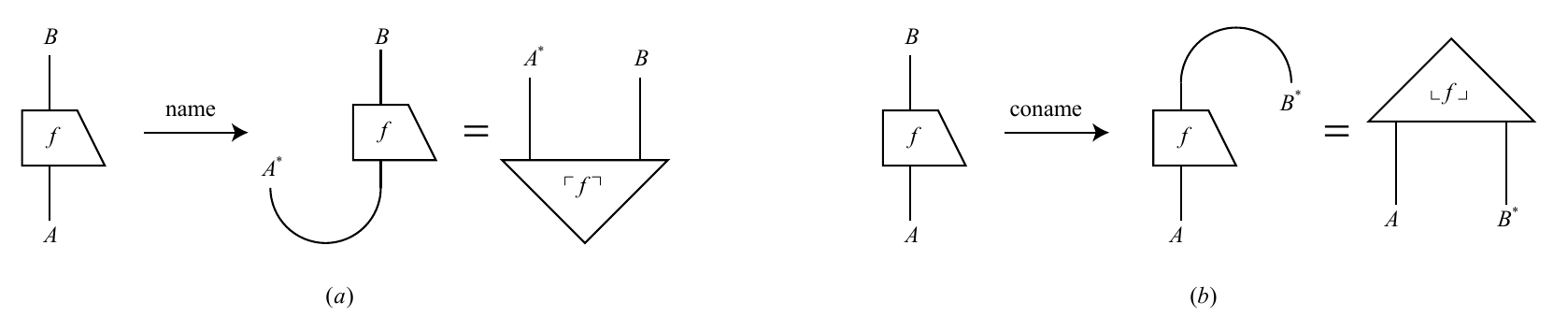}
    \caption{(a) Name. (b) Coname.  }
    \label{fig-name}
\end{figure}

As shown in Figure~\ref{fig-name} (a), for any morphism $f:A\to B$, its name is defined as
\begin{equation}
    \ulcorner f \urcorner
    :=
    (1_{A^*}\otimes f)\circ \eta_A
    :
    I\to A^*\otimes B .
\end{equation}
Intuitively, the name turns a map into a bipartite state. More explicitly, under the name operation, the input leg $A$ of the original morphism $f:A\to B$ is bent to the output side. It therefore appears not as $A$ itself, but as the dual object $A^*$.

The key point is that the dagger compact closed structure guarantees a crucial natural isomorphism, namely the map-state duality~\cite{Abramsky:2004doh, Selinger:2007eep},
\begin{equation}
    \mathrm{Hom}(A,B)\cong \mathrm{Hom}(I,A^*\otimes B).
\end{equation}
This means that, through the inverse ``unname'' operation, any bipartite state, after choosing the relevant identifications of the tensor factors, can be uniquely associated with a one-input-one-output morphism $A\to B$.

The reader will recognize that this is precisely the familiar Choi-Jamiolkowski isomorphism ~\cite{Choi:1975nug, Jamiolkowski:1972pzh}, especially in finite-dimensional Hilbert spaces. In that setting, it says nothing more mysterious than the following: an operator $f:A\to B$ is sent, by acting on one half of a Bell state, to
\begin{equation}
    |f\rangle
    :=
    (1_{A^*}\otimes f)\,|\mathrm{Bell}\rangle
    \in A^*\otimes B .
\end{equation}
Conversely, the unname operation reconstructs, from a given vector $|\psi\rangle\in A^*\otimes B$, an operator $f_\psi$. It is useful to emphasize here that an arbitrary bipartite vector $|\psi\rangle$ does correspond to some linear operator $f_\psi$, but unless $|\psi\rangle$ itself satisfies strong additional conditions, this operator $f_\psi$ is not required to be unitary, isometric, invertible, or norm-preserving. In other words, the morphism $f_\psi$ obtained by un-naming an arbitrary $|\psi\rangle\in A^*\otimes B$ is, in general, merely a general linear map.

Similarly, one can define the corresponding coname of $f$ by
\begin{equation}
    \llcorner f \lrcorner
    :=
    \varepsilon_B\circ (f\otimes 1_{B^*})
    :
    A\otimes B^* \to I .
\end{equation}
Its graphical representation is shown in Figure~\ref{fig-name} (b). Likewise, the coname expresses a natural bijection between a map and a bipartite effect (namely a bra in Dirac notation):
\begin{equation}
    \mathrm{Hom}(A,B)\cong \mathrm{Hom}(A\otimes B^*,I).
\end{equation}

\begin{figure}[htbp]
    \centering
    \includegraphics[width=1\textwidth]{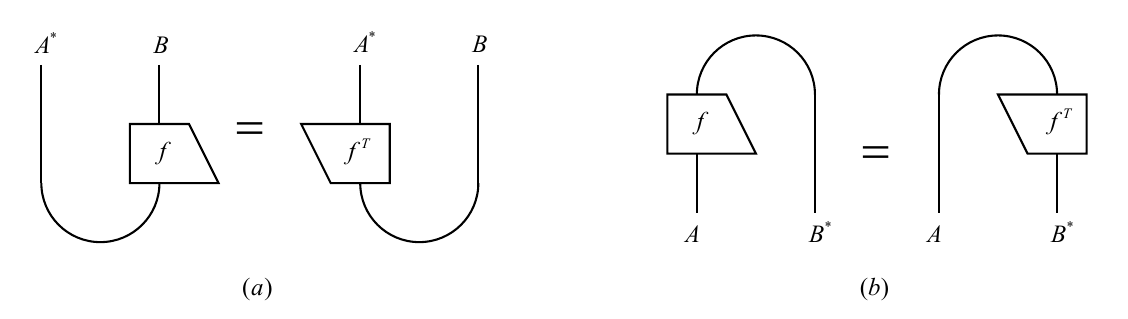}
    \caption{ Mate formulae.  }
    \label{fig-mate}
\end{figure}

As shown in Figure~\ref{fig-mate}, we next introduce an operation which, in the string-diagrammatic representation, amounts to ``sliding the box $f$ along a cup or a cap to the other leg''. This operation is usually called the mate, or dual transpose, of a morphism $f:A\to B$:
\begin{equation}
    f^T:B^*\to A^* .
\end{equation}
It is induced by the compact closed structure and is determined by either of the following equivalent formulae:
\begin{equation}\label{mate1}
    (f^T\otimes 1_B)\circ \eta_B
    =
    (1_{A^*}\otimes f)\circ \eta_A,
\end{equation}
\begin{equation}\label{mate2}
    \varepsilon_B\circ (f\otimes 1_{B^*})
    =
    \varepsilon_A\circ (1_A\otimes f^T).
\end{equation}
The string-diagrammatic representations of these two formulae are shown respectively in Figure~\ref{fig-mate} (a) and Figure~\ref{fig-mate} (b). They mean, respectively, that when $f$ is slid from the right leg of a cup to the left leg, it becomes $f^T$, and that when $f$ is slid from the left leg of a cap to the right leg, it also becomes $f^T$. Notice that the graphical representation of $f^T$ is not the same as the direct vertical flipping of the original trapezoid that represents the dagger operation; the sharp corner of the trapezoid points in a different direction. As our notation suggests, in finite-dimensional Hilbert spaces with the standard cup, $f^T$ is concretely the matrix transpose. If this is further combined with the dagger, one may define
\begin{equation}
    \overline{f}:=(f^\dagger)^T:A^*\to B^*,
\end{equation}
which, in the standard Hilbert-space realization, corresponds to entrywise complex conjugation.

\subsection{From Entanglement Specification Networks to String Diagrams}\label{appb3}

Our dedicated formalization of ``quantum information flow’’ begins with the naive picture of an entanglement specification network, in which a quantum information flow behaves like ``a tour route through a sightseeing map (a protocol network)’’, wherein the `` scenic spots’’ contain both resource-state nodes and operation nodes.  
 We shall, however, make this picture rigorous in the language of string diagrams. More explicitly, we first formalize a tour route running from an input end to an output end in an entanglement specification network as a continuous apparent through-path in the initial string diagram, entering through one port and eventually exiting through another. On the basis of this formalized notion of an apparent through-path, we can then define genuine quantum information flow.

\begin{table}[htbp]
    \centering
    \renewcommand{\arraystretch}{1.35}
    \setlength{\tabcolsep}{6pt}
    \small
    \begin{tabularx}{\textwidth}{
        |>{\raggedright\arraybackslash}X
        |>{\raggedright\arraybackslash}X
        |>{\raggedright\arraybackslash}X|
    }
        \hline
        \rowcolor{gray!15}
        \textbf{General string diagram}
        &
        \textbf{ZX string diagram}
        &
        \textbf{Entanglement specification network}
        \\
        \hline

        Wire (object)
        &
        ZX wire
        &
        System wire
        \\
        \hline

        Open string diagram composed of abstract boxes (morphism)
        &
        ZX string diagram composed of concrete generators such as red and green spiders
        &
        Protocol network containing both resource-state nodes and operation nodes
        \\
        \hline

        Diagram with empty input (state)
        &
        ZX diagram with empty input
        &
        Resource-state sector
        \\
        \hline

        Diagram with empty output (effect)
        &
        ZX diagram with empty output
        &
        Projector box corresponding to a fixed measurement branch
        \\
        \hline

        Sequential composition (morphism composition)
        &
        Sequential composition of ZX subdiagrams
        &
        One operation followed by another operation
        \\
        \hline

        Parallel juxtaposition (tensor product)
        &
        Parallel juxtaposition of ZX subdiagrams
        &
        Two operations occurring in parallel
        \\
        \hline

        Apparent through-path
        &
        Apparent through-path
        &
        ``Tour route''
        \\
        \hline
    \end{tabularx}
    \caption{Correspondence between general string diagrams, ZX string diagrams, and entanglement specification networks.}
    \label{tab:string-zx-network}
\end{table}

Table~\ref{tab:string-zx-network} gives a detailed correspondence between entanglement specification networks and general string diagrams, especially ZX string diagrams. The central idea is that each fixed-branch entanglement specification network can be viewed as an initial string diagram to be rewritten. Accordingly, a ``tour route'' in it, which will serve as a candidate route for quantum information flow, can be formalized as an apparent through-path in this initial string diagram. 

Let us add a subtle point concerning the correspondence between entanglement specification networks and string diagrams. The experimental temporal order displayed in an entanglement specification network is indeed often highly aligned with the syntactic order of composition displayed in the corresponding string diagram. Nevertheless, the two should still be carefully distinguished. The reason why they are highly aligned is not mysterious: a physical protocol is built by composing local processes, and categorical syntax is precisely designed to characterize how processes are composed. Therefore, under our bottom-to-top reading convention, operations that occur earlier in the experiment often correspond to local processes placed lower in the string diagram, while operations that occur later often correspond to local processes placed higher in the string diagram. Similarly, sequential composition looks like ``happening next'', while tensor juxtaposition looks like ``happening at the same time''.\footnote{Of course, tensor juxtaposition here should not be understood as strict physical simultaneity. More accurately, it expresses the absence of sequential dependence. Interestingly, from this perspective, categorical graphical language is naturally inclined to organize processes in terms of local gluing and causal dependence, which to some extent alleviates the difficulty of ``distant simultaneity'' in special relativity.} However, the before--after order in an experimental procedure is constrained by very concrete physical implementation conditions, whereas the syntactic order of composition only concerns how the entire morphism expression is built.

\section{String-Diagrammatic Calculus of Quantum Teleportation}\label{appc}

\subsection{Review of Quantum Teleportation}\label{appc1}

Quantum teleportation tells the following story. Suppose Alice holds an unknown single-qubit state
\begin{equation}
    |\chi\rangle_1
    =
    \alpha |0\rangle_1+\beta |1\rangle_1,
    \qquad
    |\alpha|^2+|\beta|^2=1 .
\end{equation}
Alice wants to ``transmit'' this unknown state to Bob. The method is as follows: first, Alice and Bob share a Bell pair in advance; second, Alice performs a Bell measurement on the message qubit and her half of the Bell pair; third, Alice sends two classical bits to Bob through classical communication to inform him of the measurement outcome; fourth, Bob applies the corresponding conditional correction according to a pre-agreed correction table. We now prove that this indeed successfully implements quantum teleportation.

As shown in Figure~\ref{fig-tele}, let the Bell state shared in advance by Alice and Bob be the standard one
\begin{equation}
    |\Phi^+\rangle_{23}
    =
    \frac{1}{\sqrt{2}}
    \bigl(
        |00\rangle_{23}+|11\rangle_{23}
    \bigr).
\end{equation}
Here particle $2$ is held by Alice, while particle $3$ is held by Bob. Therefore the initial state of the total system is
\begin{align}
    |\Psi\rangle_{123}
    &=
    |\chi\rangle_1\otimes |\Phi^+\rangle_{23} \notag\\
    &=
    \bigl(\alpha |0\rangle_1+\beta |1\rangle_1\bigr)
    \otimes
    \frac{1}{\sqrt{2}}
    \bigl(
        |00\rangle_{23}+|11\rangle_{23}
    \bigr) \notag\\
    &=
    \frac{1}{\sqrt{2}}
    \bigl(
        \alpha |000\rangle
        +
        \alpha |011\rangle
        +
        \beta |100\rangle
        +
        \beta |111\rangle
    \bigr).
\end{align}
The key trick is to expand the first two qubits, $1,2$, in the Bell basis. We use
\begin{align}
    |\Phi^+\rangle
    &=
    \frac{1}{\sqrt{2}}
    \bigl(|00\rangle+|11\rangle\bigr),
    &
    |\Phi^-\rangle
    &=
    \frac{1}{\sqrt{2}}
    \bigl(|00\rangle-|11\rangle\bigr),
    \\
    |\Psi^+\rangle
    &=
    \frac{1}{\sqrt{2}}
    \bigl(|01\rangle+|10\rangle\bigr),
    &
    |\Psi^-\rangle
    &=
    \frac{1}{\sqrt{2}}
    \bigl(|01\rangle-|10\rangle\bigr).
\end{align}
After rearranging, one obtains the standard form of the teleportation formula:
\begin{align}
    |\chi\rangle_1|\Phi^+\rangle_{23}
    =
    \frac{1}{2}
    \bigl[
    &|\Phi^+\rangle_{12}
        \bigl(\alpha |0\rangle+\beta |1\rangle\bigr)_3
    +
    |\Phi^-\rangle_{12}
        \bigl(\alpha |0\rangle-\beta |1\rangle\bigr)_3
    \notag\\
    &+
    |\Psi^+\rangle_{12}
        \bigl(\beta |0\rangle+\alpha |1\rangle\bigr)_3
    +
    |\Psi^-\rangle_{12}
        \bigl(-\beta |0\rangle+\alpha |1\rangle\bigr)_3
    \bigr].
\end{align}
Using the following four Pauli operators,
\begin{equation}
    I=
    \begin{pmatrix}
        1&0\\
        0&1
    \end{pmatrix},
    \qquad
    Z=
    \begin{pmatrix}
        1&0\\
        0&-1
    \end{pmatrix},
    \qquad
    X=
    \begin{pmatrix}
        0&1\\
        1&0
    \end{pmatrix},
    \qquad
    XZ=
    \begin{pmatrix}
        0&1\\
        -1&0
    \end{pmatrix},
\end{equation}
one can further write the above formula compactly as
\begin{equation}
    |\chi\rangle_1|\Phi^+\rangle_{23}
    =
    \frac{1}{2}
    \bigl(
    |\Phi^+\rangle_{12} I|\chi\rangle_3
    +
    |\Phi^-\rangle_{12} Z|\chi\rangle_3
    +
    |\Psi^+\rangle_{12} X|\chi\rangle_3
    +
    |\Psi^-\rangle_{12} XZ|\chi\rangle_3
    \bigr).
\end{equation}
This formula manifestly shows that, together with the correction table shown below, the quantum teleportation protocol is successfully implemented.
\begin{center}
\begin{tabular}{c|c|c}
\hline
Alice's Bell-measurement outcome & Bob's quantum state & Correction applied by Bob \\
\hline
$|\Phi^+\rangle$ & $|\chi\rangle$ & $I$ \\
$|\Phi^-\rangle$ & $Z|\chi\rangle$ & $Z$ \\
$|\Psi^+\rangle$ & $X|\chi\rangle$ & $X$ \\
$|\Psi^-\rangle$ & $XZ|\chi\rangle$ & $ZX$ \\
\hline
\end{tabular}
\end{center}

More formally, one can analyze quantum teleportation using the language of quantum channels. First, write the input as an arbitrary density matrix $\rho_1$. Denote the Bell projectors corresponding to the four possible outcomes of Alice's Bell measurement by
\begin{equation}
    P_k = |\beta_k\rangle\langle \beta_k|,
    \qquad
    k=00,01,10,11.
\end{equation}
Denote the corresponding Bob-side corrections by the four Pauli operators
\begin{equation}
    U_k\in\{I,Z,X,ZX\}.
\end{equation}
Then the whole protocol, regarded as an input--output map, can be written as
\begin{equation}
    \mathcal{T}(\rho)
    =
    \sum_k
    U_k\,
    \operatorname{Tr}_{12}
    \Bigl[
        (P_k\otimes I_3)
        \bigl(
            \rho_1\otimes
            |\Phi^+\rangle\langle \Phi^+|_{23}
        \bigr)
    \Bigr]
    U_k^\dagger .
\end{equation}
One can prove that, for any input state $\rho$,
\begin{equation}
    \mathcal{T}(\rho)=\rho .
\end{equation}
That is,
\begin{equation}
    \mathcal{T}=\mathrm{id}.
\end{equation}
In fact, one can prove an even stronger version of the statement that ``the overall channel of quantum teleportation is an identity channel'': for any input entangled with a reference system, the protocol also preserves that entanglement. Interested readers may consult, for example, Ref.~\cite{Nielsen:2012yss}.

\subsection{Quantum Teleportation from the CQM Perspective}\label{appc2}

\begin{figure}[htbp]
    \centering
    \includegraphics[width=1\textwidth]{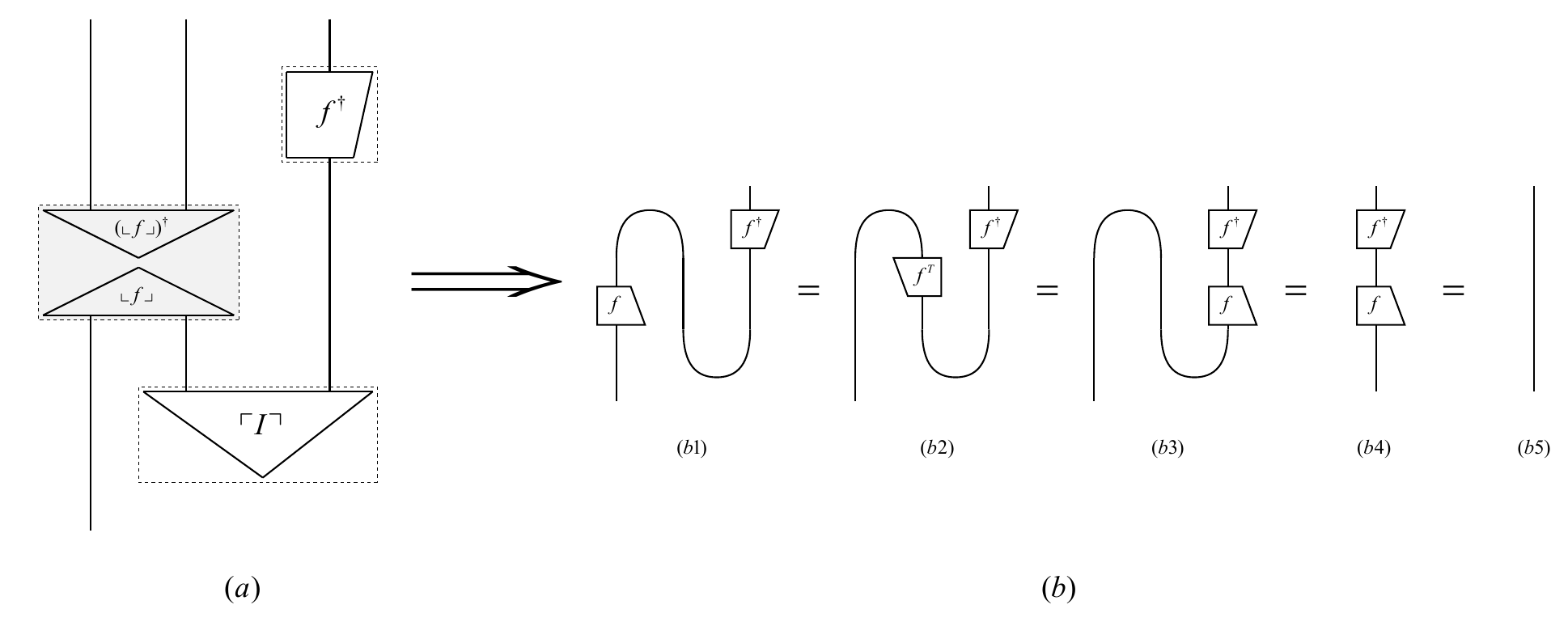}
    \caption{String-diagrammatic derivation of quantum teleportation in a fixed Bell-measurement branch.}
    \label{fig-tele2}
\end{figure}

We now provide a more detailed explanation of the string-diagrammatic calculus of quantum teleportation discussed in Section~\ref{sec22}. For convenience, the relevant diagram is reproduced in Figure~\ref{fig-tele2}. The derivation is completely parallel to that in Appendix~\ref{appb2}, except that it now involves more concrete morphisms. Our purpose, however, is also to use teleportation as an example to display the one-to-one correspondence between algebraic manipulations and string-diagrammatic transformations.

First, the shared Bell resource state is simply the cup itself, or equivalently $\ulcorner I \urcorner$. Second, the Bell kets appearing in the projectors are precisely the names of the four Pauli maps:
\begin{equation}
    \ulcorner I \urcorner=|\Phi^+\rangle,\qquad
    \ulcorner Z \urcorner=|\Phi^-\rangle,\qquad
    \ulcorner X \urcorner=|\Psi^+\rangle,\qquad
    \ulcorner XZ \urcorner=|\Psi^-\rangle .
\end{equation}
In the actual argument, we only need the corresponding four Bell bras, which are the conames:
\begin{equation}
    \llcorner I \lrcorner=\langle\Phi^+|,\qquad
    \llcorner Z \lrcorner=\langle\Phi^-|,\qquad
    \llcorner X \lrcorner=\langle\Psi^+|,\qquad
    \llcorner XZ \lrcorner=\langle\Psi^-| .
\end{equation}

Now fix one branch $f\in\{I,Z,X,XZ\}$, and suppose that Bob places the corresponding correction $c_f$ on his side. Reading the standard string diagram from bottom to top, the algebraic expression for the overall morphism corresponding to the whole branch depicted in Figure~\ref{fig-tele2} (a) is
\begin{equation}
    T_f
    :=
    c_{f,B}\circ
    (\llcorner f \lrcorner_{MA}\otimes 1_B)
    \circ
    (1_M\otimes \eta_{AB}) .
    \label{eq:Tf-def}
\end{equation}
Here we use the letters $M,A,B$ to denote the previous qubits $1,2,3$, respectively, and $\llcorner f \lrcorner_{MA}$ denotes the Bell effect corresponding to this branch on Alice's side.

We now simplify Eq.~\eqref{eq:Tf-def} step by step. First, substitute the explicit expansion of the coname:
\begin{equation}
    \llcorner f \lrcorner_{MA}
    =
    \varepsilon_{MA}\circ(f_M\otimes 1_A),
\end{equation}
which gives
\begin{align}
    T_f
    &=
    c_{f,B}
    \circ
    (\varepsilon_{MA}\otimes 1_B)
    \circ
    (f_M\otimes 1_A\otimes 1_B)
    \circ
    (1_M\otimes \eta_{AB}) .
    \label{eq:Tf-expand-coname}
\end{align}
The corresponding string-diagrammatic move is the following: the coname is decomposed, via C--J duality, into a box $f$, attached to the left leg of the cap, as shown in Figure~\ref{fig-tele2} (b1).

Next, applying the mate equation~(\ref{mate2}) to the cap gives
\begin{align}
    T_f
    &=
    c_{f,B}
    \circ
    (\varepsilon_{MA}\otimes 1_B)
    \circ
    (1_M\otimes f_A^T\otimes 1_B)
    \circ
    (1_M\otimes \eta_{AB}) .
    \label{eq:Tf-cap-mate}
\end{align}
The corresponding string-diagrammatic move is that Alice's box  $f$ crosses the cap bend, moving from the message leg $M$ to its dual leg $A$, and correspondingly becomes $f^T$, as shown in Figure~\ref{fig-tele2} (b2).

Then, applying the mate equation (\ref{mate1}) to the cup of the shared Bell resource, we obtain
\begin{align}
    T_f
    &=
    c_{f,B}
    \circ
    (\varepsilon_{MA}\otimes 1_B)
    \circ
    (1_M\otimes 1_A\otimes f_B)
    \circ
    (1_M\otimes \eta_{AB})
    \notag\\
    &=
    c_{f,B}
    \circ
    f_B
    \circ
    (\varepsilon_{MA}\otimes 1_B)
    \circ
    (1_M\otimes \eta_{AB}) .
    \label{eq:Tf-after-cup}
\end{align}
The corresponding string-diagrammatic move is that the box crosses the cup bend and arrives at Bob's leg $B$, where it changes back from $f^T$ to $f$ (Figure~\ref{fig-tele2} (b3)).

Notice that we can now apply the snake equation:
\begin{equation}
    (\varepsilon_{MA}\otimes 1_B)
    \circ
    (1_M\otimes \eta_{AB})
    =
    1_{M\to B},
\end{equation}
to the snake tail in Eq.~\eqref{eq:Tf-after-cup}.
Thus,
\begin{equation}
    T_f=c_{f,B}\circ f_B .
\end{equation}
This step corresponds to Figure~\ref{fig-tele2} (b4), namely straightening the snake.

Therefore, as long as Bob chooses the correction satisfying~\footnote{Since in the teleportation example the branch labels $f\in\{I,Z,X,XZ\}$ are unitary Pauli maps, this is equivalently
\begin{equation}
    c_f=f^\dagger.
\end{equation}}
\begin{equation}
    c_f=f^{-1},
\end{equation}
we obtain
\begin{equation}
    T_f=1_{M\to B}.
\end{equation}
In this way, we arrive at Figure~\ref{fig-tele2} (b5), namely a clean straight line running from bottom to top.

Note that the CQM treatment of teleportation certainly does not give a final answer different from what ordinary quantum information theory has long told us: teleportation, as an overall input--output process, is equivalent to an identity channel. However, the classical state-centred narrative does tend to direct our attention to the question of ``how the state gets there''. By contrast, process-centred CQM turns the outcome of each measurement branch into a concrete morphism label $f$. Teleportation thereby becomes a story about the propagation of this morphism label $f$. This kind of morphism ``ID card'' is transported to the output side along an internal channel supported by cups, caps, and snakes in the dagger compact closed structure. There it meets the corresponding correction, cancels with it, and finally leaves us with an identity ``bare wire''. The physical meaning of quantum information flow is precisely this propagation of morphism-type data.

\section{ZX Calculus}\label{appd}

\subsection{The Relation between CQM, Dagger Compact Closed Categories, and the ZX Calculus}\label{appd1}

Directly speaking, the category $\mathbf{ZX}$ is an example of a concrete dagger compact category~\cite{Coecke:2008lcg, Duncan:2009ocf, vandeWetering:2020giq, Coecke:2017dti}. Its objects are represented by parallel quantum wires, while its morphisms are ZX diagrams with specified input and output boundaries. However, before introducing it, it is necessary for us to first clarify the subtle relation among CQM, Dagger Compact Closed Categories, and the ZX Calculus.

We first emphasize that the dagger compact closed categories introduced in Appendix~\ref{appa} are not ``the definition of quantum mechanics''. Rather, they specifically capture the finite-dimensional pure-process fragment of quantum theory, where finite-dimensional Hilbert spaces are taken as systems and linear maps between them as processes. This is precisely the level at which the tensor-network manipulations in the present paper are carried out. In other words, CQM provides a structural syntax for characterizing quantum processes. This syntax, however, depends on the level of theory under discussion. The finite-dimensional and infinite-dimensional cases are not the same, and the structures needed for the pure-process fragment and for general quantum channels are also different. 
Let us briefly mention that, for CPTP maps, namely completely positive trace-preserving maps, which characterize physically allowed maps between general density matrices, one should further perform the so-called CPM construction~\cite{Selinger:2007eep} or CP* construction~\cite{Coecke:2014exa} based on a dagger compact closed category. In the infinite-dimensional case, the situation is more subtle, because infinite-dimensional Hilbert spaces usually do not have duals with cap-cup structures of the kind appearing in dagger compact closed categories. Nevertheless, progress on CQM for infinite-dimensional Hilbert spaces can be found in Refs. ~\cite{Gogioso:2017xth, Gogioso:2017xli }.

Now we emphasize the relation between the ZX calculus and dagger compact closed categories. 
Metaphorically speaking, the relation between dagger compact closed structure and the ZX language is like the relation between a syntactic skeleton and one concrete dialect built on that skeleton. In the dagger compact skeleton, we already have serial and parallel composition of processes, dagger, cups and caps, and the snake identities. Hence we can perform operations such as ``flipping boxes'', ``bending wires'', and map--state duality. ZX first inherits this whole graphical skeleton, rather than constructing a completely different world. On top of this skeleton, however, ZX chooses the concrete ``vocabulary'' of qubit theory. More precisely, it adds several classes of basic generators: $Z$ (green)-spiders, $X$ (red)-spiders, Hadamard gates and phase parameters. The red and green spiders correspond to two complementary observable structures. Each observable structure can be characterized by a commutative special dagger Frobenius algebra, which, in an abstract categorical setting, expresses which states can be copied and deleted like classical data. The Hadamard gate plays the role of a bridge that connects and transforms these two structures into each other. In addition, each observable structure is equipped with a phase group.
Besides these concrete pieces of ``vocabulary'', ZX also imposes additional ``syntactic habits''. In other words, it introduces equations specific to these generators. These equations go beyond the purely topological rules of the dagger compact skeleton, such as the snake equations. In calculations, they appear as rewriting rules, such as spider fusion and colour change; see Appendix~\ref{appd2} for details. 
Finally, ZX also specifies a semantic interpretation into qubit linear algebra. More explicitly, the reason why the ZX calculus is not merely an abstract graphical language is that it interprets diagrams as linear maps between qubit Hilbert spaces through a functor
\begin{equation}
    [\![\cdot]\!]:\mathbf{ZX}\to \mathbf{Qubit}.
\end{equation}
Thus, serial composition of diagrams corresponds to composition of linear maps, juxtaposition of diagrams corresponds to tensor product, and the basic graphical generators correspond to concrete matrices. The point is that ZX is not quantum mechanics by itself; rather, it acquires the meaning of qubit quantum mechanics through a standard interpretation functor.

For the purpose of the present paper, we focus on ZX diagrams in which all spider phases are integer multiples of $\pi/2$. These are usually called Clifford ZX diagrams, and the ZX rewriting rules listed in the appendix are mainly for this Clifford fragment. Any Clifford circuit gives rise to a Clifford ZX diagram. Moreover, Backens proved that the ZX calculus is complete for the Clifford fragment of pure-state quantum mechanics ~\cite{Backens:2013hto, Backens:2015nhm }. This means that, in pure-state qubit quantum mechanics involving stabilizer states, Clifford gates, and measurements of Pauli observables, every equality that holds at the level of matrices can be derived purely diagrammatically.

\subsection{A Review of the Rewriting Rules of the ZX Calculus}\label{appd2}

\begin{figure}[htbp]
    \centering
    \includegraphics[width=1\textwidth]{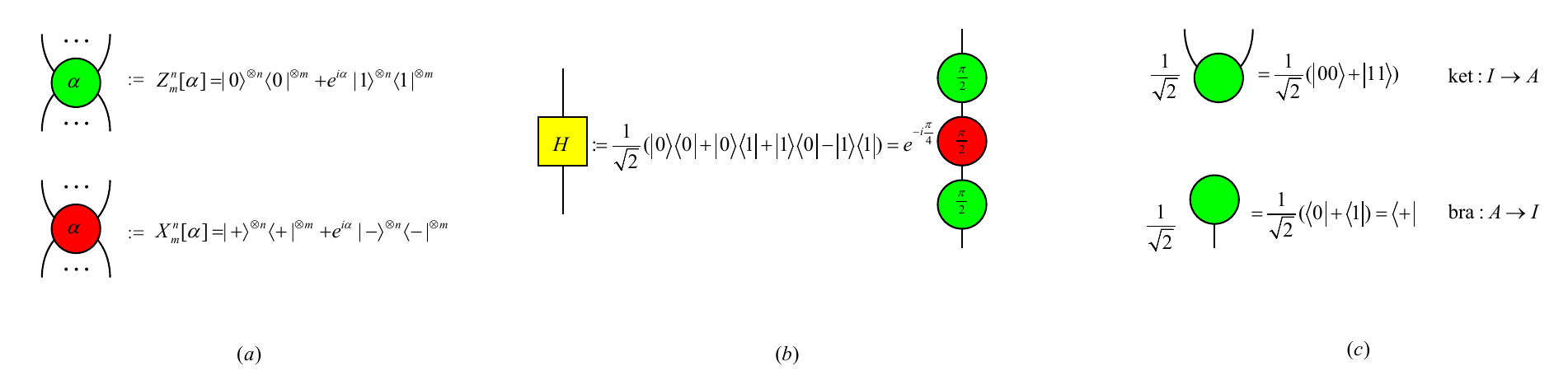}
    \caption{ (a) Definitions of green and red spiders. 
(b) Euler decomposition of the Hadamard gate. 
(c) Examples of states and effects.}
    \label{fig-spid}
\end{figure}
From a practical point of view, the reader may regard the ZX calculus reviewed below simply as a way of decomposing the ``blind boxes'' introduced earlier in the CQM context into finer-grained elementary components. The most important of these are green $Z$-spiders and red $X$-spiders.
The simplest way to introduce $Z$-spiders and $X$-spiders is shown in Figure~\ref{fig-spid} (a). They are represented in ket-bra form using, respectively, the eigenbasis of the Pauli operator $Z$, usually called the computational basis in quantum computation, $ \{|0\rangle,|1\rangle\}$, and the eigenbasis of $X$, $ \left\{     |+\rangle=\frac{1}{\sqrt{2}}(|0\rangle+|1\rangle),     \quad
    |-\rangle=\frac{1}{\sqrt{2}}(|0\rangle-|1\rangle)     \right\}$: 

\begin{equation}
    Z_m^n[\alpha]
    =
    |0\rangle^{\otimes n}\langle 0|^{\otimes m}
    +
    e^{i\alpha}
    |1\rangle^{\otimes n}\langle 1|^{\otimes m},
\end{equation}
\begin{equation}
    X_m^n[\alpha]
    =
    |+\rangle^{\otimes n}\langle +|^{\otimes m}
    +
    e^{i\alpha}
    |-\rangle^{\otimes n}\langle -|^{\otimes m}.
\end{equation}
Throughout this paper, our drawing convention remains consistent with the CQM convention: diagrams are read from bottom to top. Thus, a spider with $m$ input wires and $n$ output wires represents a morphism, namely a linear map from $m$ qubits to $n$ qubits,
\begin{equation}
    (\mathbb{C}^2)^{\otimes m}\to (\mathbb{C}^2)^{\otimes n}.
\end{equation}
In particular, we can also describe two special kinds of morphisms: states, with empty input, namely kets in the quantum-mechanical context; and effects, with empty output, namely bras. Examples are shown in Figure~\ref{fig-spid} (c). Let us add that when the phase is zero, it is usually omitted from the spider. Scalar factors are more subtle. In many situations, they do not affect practical calculation and reasoning, just as is often the case in quantum-circuit or tensor-network contexts. Nevertheless, one way to keep track of them is to use a black diamond to represent $\sqrt{2}$.~\footnote{The number $2$ appears here because it is the Hilbert-space dimension of a single qubit.} We shall see this convention in later diagrams.

\begin{figure}[htbp]
    \centering
    \includegraphics[width=1\textwidth]{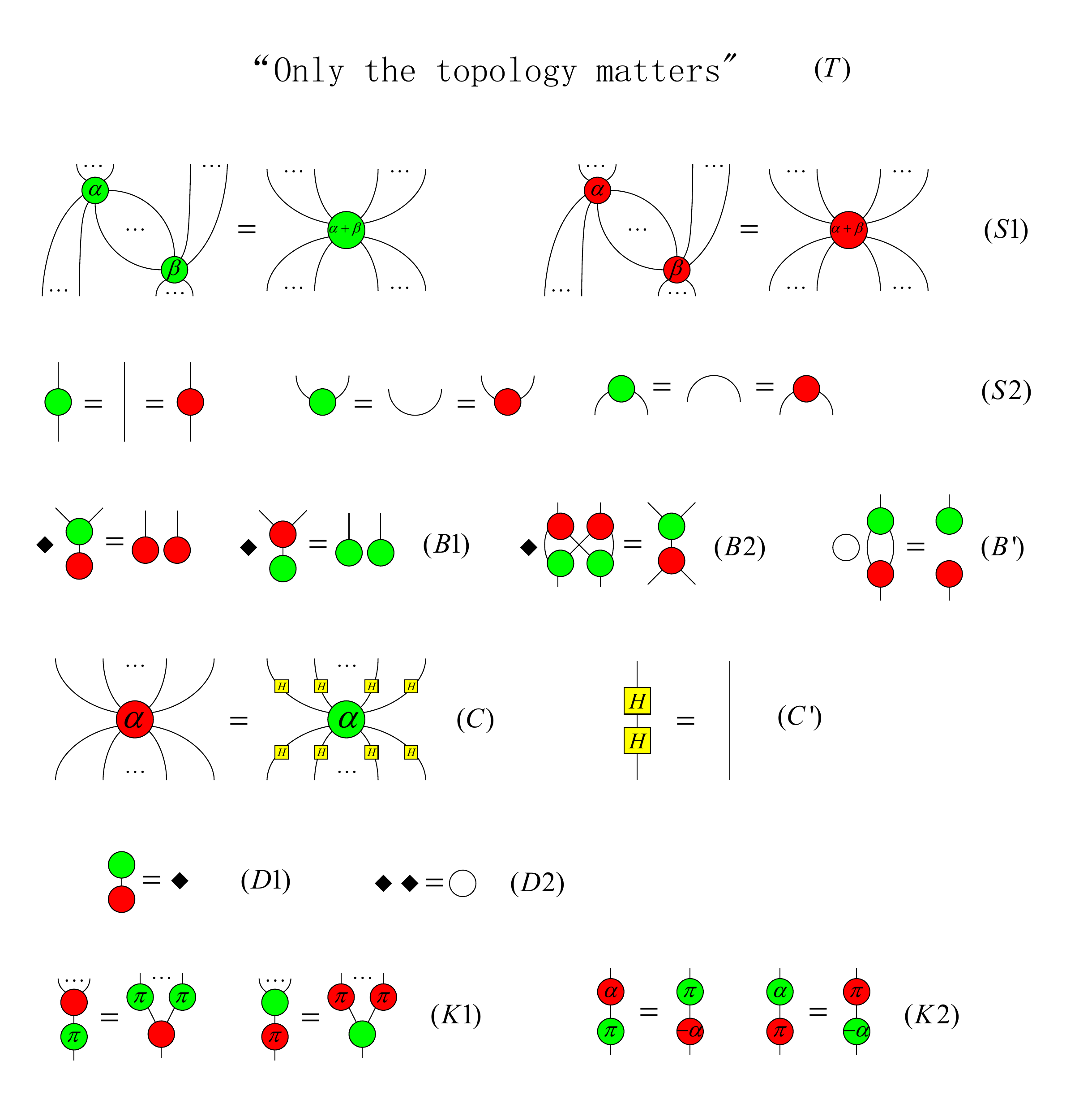}
    \caption{  Rules for the ZX-calculus}
    \label{fig-rule}
\end{figure}
For the purpose of the present paper, we shall focus only on the ZX rules that are complete for the stabilizer fragment of pure-state quantum mechanics, also called the basic ZX rules. We first give, as shown in Figure~\ref{fig-spid} (b), the ZX-spider decomposition, or Euler decomposition, of the Hadamard gate. The Hadamard gate switches between the $Z$-basis and the $X$-basis. It satisfies
\begin{equation}
    H|0\rangle=|+\rangle,
    \qquad
    H|1\rangle=|-\rangle,
\end{equation}
\begin{equation}
    H|+\rangle=|0\rangle,
    \qquad
    H|-\rangle=|1\rangle.
\end{equation}

The full set of ZX rewriting rules applicable to the stabilizer fragment of pure-state quantum mechanics is summarized in Figure~\ref{fig-rule}.\footnote{The classification of the rules here mainly follows the pedagogical style of Ref.~ \cite{Coecke:2013tutorial}, which may be more suitable for introducing the ZX calculus into the holographic context for the first time. These rules can also be organized in a more axiomatic style; see e.g. Ref.~\cite{Kissinger:2024pqs }.}
We now briefly explain several of the rules in the figure. The $T$-rule means that, for ZX diagrams, only the connectivity of the diagram is essential, not its particular planar drawing. In essence, this is the meta-rule guaranteed by the coherence of dagger compact categories: after fixing the boundary, isotopic string diagrams represent the same morphism.
The $S$-rule is the fusion rule for spiders of the same colour. In particular, $(S2)$, as a special case, gives the ZX-spider representations of cups, caps, and identity morphisms.
Rule ($B1$) can be roughly read as: ``green copies red points'', and ``red copies green points''. The black diamond represents the scalar $\sqrt{2}$. Rule ($B2$) is a strong exchange principle for loops with alternating red and green points. Using ($B1$) and ($B2$), one can prove the Hopf law ($B'$). Intuitively, this law says that pairs of parallel edges between spiders of different colours can be cancelled. Its importance lies in the fact that it reflects the strong complementarity between the two Frobenius structures behind the ZX calculus.
The $K$-rule concerns the special property of spiders with phase $\alpha=\pi$. The $C$-rule states that the Hadamard gate can serve as an explicit colour-changing operation. In particular, its special case is the ($C'$)-rule, usually called the Hadamard cancellation rule.


\subsection{Teleportation with a Bell Resource State: ZX Simplification in a Fixed Branch}\label{appd3}

The ZX-diagrammatic representation of quantum teleportation is shown in Figure~\ref{fig-zxte}. We first explain the components of the initial string diagram, and then show how, in a fixed branch, this initial ZX string diagram is simplified step by step, by means of ZX rewriting rules, into the final trivial ZX diagram consisting only of a bare wire.

First, the ZX string-diagrammatic representations of the bras of the four Bell basis states are shown in Figure~\ref{fig-fine}(a). The corresponding Bell effects can be uniformly written as
\begin{equation}
    \langle B_{\alpha,\beta}|
    =
    \frac{1}{\sqrt{2}}
    \bigl(
        \langle 00|
        +
        e^{-i\alpha}\langle 11|
    \bigr)
    \bigl(1\otimes X^{\beta/\pi}\bigr),
    \qquad
    \alpha,\beta\in\{0,\pi\}.
\end{equation}
The four choices of $(\alpha,\beta)$ give precisely
\begin{equation}
    \langle B_{0,0}|=\langle\Phi^+|,
    \qquad
    \langle B_{\pi,0}|=\langle\Phi^-|,
    \qquad
    \langle B_{0,\pi}|=\langle\Psi^+|,
    \qquad
    \langle B_{\pi,\pi}|=\langle\Psi^-|.
\end{equation}

Next, we verify that Figure~\ref{fig-fine}(c) gives the correct ZX representations of Bob's four Pauli correction maps. The relevant one-input-one-output spiders are
\begin{equation}
    Z_1^1[\alpha]
    =
    |0\rangle\langle 0|
    +
    e^{i\alpha}|1\rangle\langle 1|,
    \qquad
    X_1^1[\beta]
    =
    |+\rangle\langle +|
    +
    e^{i\beta}|-\rangle\langle -|.
\end{equation}
For $\alpha,\beta\in\{0,\pi\}$, these specialize to
\begin{equation}
    Z_1^1[\alpha]
    =
    \begin{cases}
        I, & \alpha=0,\\
        Z, & \alpha=\pi,
    \end{cases}
    \qquad
    X_1^1[\beta]
    =
    \begin{cases}
        I, & \beta=0,\\
        X, & \beta=\pi.
    \end{cases}
\end{equation}
Therefore, the composite diagram in Figure~\ref{fig-fine}(c) represents
\begin{equation}
    U_{\alpha,\beta}
    :=
    X_1^1[\beta]\circ Z_1^1[\alpha]
    =
    X^{\beta/\pi}Z^{\alpha/\pi}.
\end{equation}
The four choices of $(\alpha,\beta)$ give precisely
\begin{equation}
    U_{0,0}=I,
    \qquad
    U_{\pi,0}=Z,
    \qquad
    U_{0,\pi}=X,
    \qquad
    U_{\pi,\pi}=XZ,
\end{equation}
which are the four standard Pauli maps appearing in teleportation.

\begin{figure}[htbp]
    \centering
    \includegraphics[width=1\textwidth]{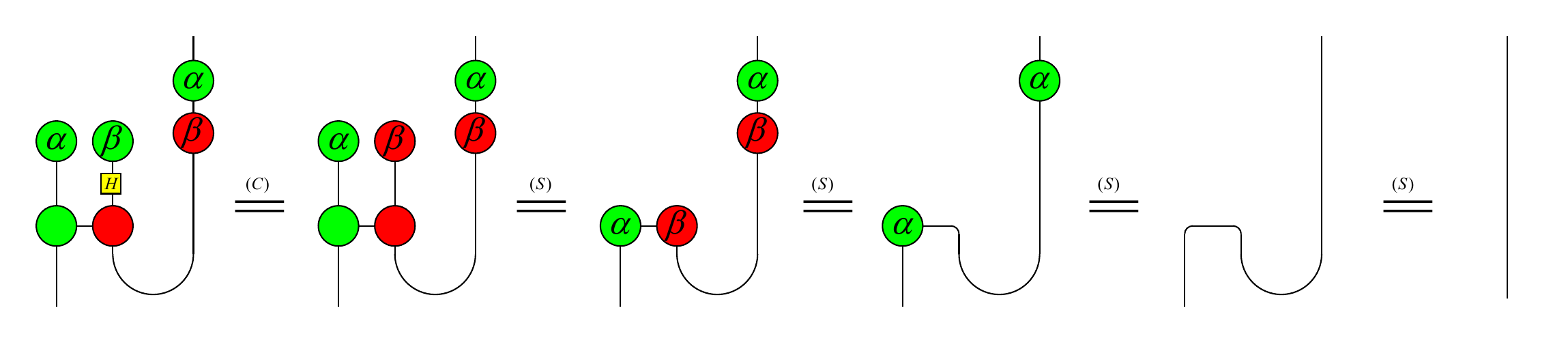}
    \caption{ZX-calculus derivation of the teleportation protocol in a fixed branch }
    \label{fig-zxte2}
\end{figure}

Together with the fact that the Bell resource state itself is represented by a cup, this confirms the correctness of the initial ZX string diagram. Figure~\ref{fig-zxte2} then shows how this initial ZX string diagram is simplified step by step to the final result by means of the ZX rewriting rules reviewed in Appendix~\ref{appd2}. 
We particularly emphasize that, in Figure~\ref{fig-zxte2}, the story of ``sliding the box $f$ along cup-cap'' in the CQM string diagram of Appendix~\ref{appc2} still exists. The only difference is that we now obtain a finer-grained representation of the morphism $f$, which appeared as a black-box morphism in the CQM skeleton: it is refined into a composite of a green phase spider and a red phase spider. Thus, the sequence of local ZX rewrites technically realizes the sliding of the box $f$ in CQM.

\subsection{Teleportation with a GHZ Resource State: ZX Simplification in a Fixed Branch}\label{appd4}

The details of the GHZ-assisted teleportation protocol shown in Fig.~\ref{fig-ghz} are as follows~\cite{Karlsson:1998opa,Hillery:1998yq}:

\begin{enumerate}
    \item The experimenter Alice holds the message qubit $m$ and leg $a$ of     the GHZ resource state. She performs a Bell-basis measurement on these     two qubits.

    \item The experimenter Bob is not the receiver of the message, but acts     instead as an assistant. He performs an $x$-basis measurement on the GHZ     leg $b$.

    \item Alice communicates the outcome of her Bell measurement to the     experimenter Charlie through a classical channel; Bob likewise     communicates the outcome of his $x$-basis measurement to Charlie.

    \item According to these two pieces of classical information, Charlie     applies the corresponding local correction to the GHZ leg $c$. After this correction, Alice's message qubit is successfully     recovered by Charlie.
\end{enumerate}
Denote Alice's Bell-measurement outcome by $\langle\Phi^\pm|$ and $\langle\Psi^\pm|$, and Bob's $x$-basis measurement outcome by $\langle x^\pm|$. Charlie's local correction is then given by
\begin{equation}
\begin{array}{c|cccc}
&
\langle\Phi^+|
&
\langle\Phi^-|
&
\langle\Psi^+|
&
\langle\Psi^-|
\\ \hline
\langle x^+|
&
I
&
\sigma_z
&
\sigma_x
&
\sigma_x\sigma_z
\\
\langle x^-|
&
\sigma_z
&
I
&
\sigma_x\sigma_z
&
\sigma_x
\end{array}
\end{equation}

Ref.~\cite{Hillebrand:2011thesis} gives the technical details of a ZX-calculus proof of GHZ-assisted teleportation. The essential point is that, in the language of the ZX calculus, for every fixed branch the initial ZX string diagram corresponding to the entanglement specification network can be simplified step by step to a bare wire, thereby exhibiting the quantum information flow. We review some of the key ingredients below.

\begin{figure}[htbp]
    \centering
    \includegraphics[width=0.4\textwidth]{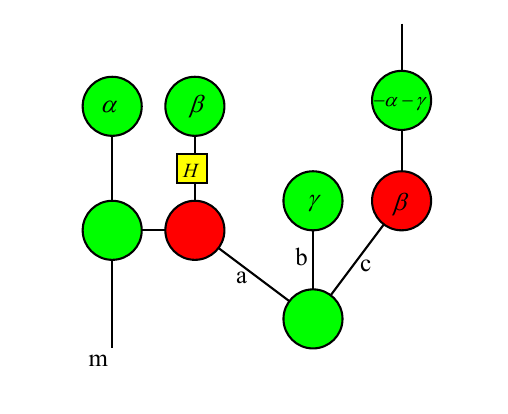}
    \caption{    ZX representation of the GHZ-assisted teleportation protocol in a fixed     branch. The phase labels $\alpha$, $\beta$, and $\gamma$ encode the     corresponding measurement outcomes.     }
    \label{fig-gin}
\end{figure}

We first rewrite the protocol diagram of Fig.~\ref{fig-ghz} more explicitly as a ZX string diagram, as shown in Fig.~\ref{fig-gin}. The ZX representation of a Bell measurement has already been introduced in Appendix~\ref{appd3}. The ZX representation of the GHZ state is equally simple: up to a normalization factor, it is just a zero-phase green spider with three upward-pointing legs. More explicitly,
\begin{equation}
    Z_{0}^{3}[0]
    =
    |0\rangle^{\otimes 3}
    +
    |1\rangle^{\otimes 3}
    =
    |000\rangle
    +
    |111\rangle .
\end{equation}

We now translate the operations performed by the experimenters systematically into the language of phase-labelled spiders. More precisely, the four possible outcomes of Alice's Bell measurement can be encoded by two phase parameters $    \alpha,\beta\in\{0,\pi\},$
while Bob's $x$-basis measurement outcome is encoded by $    \gamma\in\{0,\pi\}.$ 
Thus, every fixed branch corresponds to a definite set of phase data $(\alpha,\beta,\gamma)$, and the local correction that Charlie must ultimately apply can likewise be written as a combination of phase spiders in the ZX calculus. The following table summarizes this translation, expressing both the measurement outcomes and the local corrections of the laboratory description in terms of phase labels suitable for ZX rewriting:
\begin{equation}
\begin{array}{ccc|c|c|c}
\alpha & \beta & \gamma
&
\text{Bell state}
&
x\text{-basis state}
&
\text{local operation}
\\ \hline
0   & 0   & 0
& \langle\Phi^+|
& \langle x^+|
& I
\\
0   & 0   & \pi
& \langle\Phi^+|
& \langle x^-|
& \sigma_z
\\
0   & \pi & 0
& \langle\Psi^+|
& \langle x^+|
& \sigma_x
\\
0   & \pi & \pi
& \langle\Psi^+|
& \langle x^-|
& \sigma_x\sigma_y
\\
\pi & 0   & 0
& \langle\Phi^-|
& \langle x^+|
& \sigma_z
\\
\pi & 0   & \pi
& \langle\Phi^-|
& \langle x^-|
& I
\\
\pi & \pi & 0
& \langle\Psi^-|
& \langle x^+|
& \sigma_x\sigma_z
\\
\pi & \pi & \pi
& \langle\Psi^-|
& \langle x^-|
& \sigma_x
\end{array}
\end{equation}

\begin{figure}[htbp]
    \centering
    \includegraphics[width=1\textwidth]{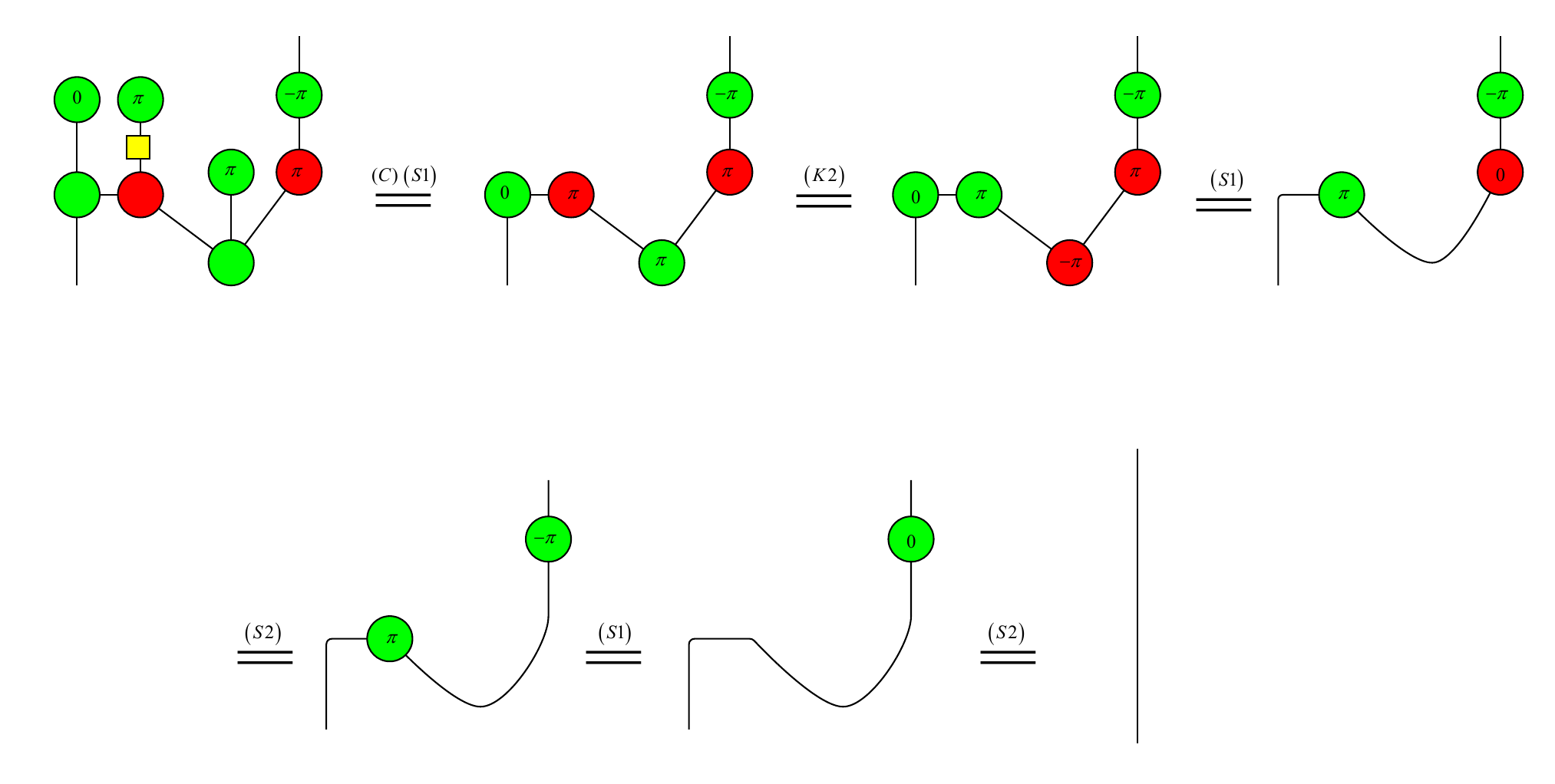}
    \caption{     ZX simplification of the fixed branch     $(\alpha,\beta,\gamma)=(0,\pi,\pi)$ of the GHZ-assisted teleportation     protocol. Successive semantics-preserving rewrites reduce the initial     diagram to a bare wire connecting the message input to the receiving     output.    }
    \label{fig-show}
\end{figure}

Everything is now in place. We next show how a chain of ZX-calculus rewrites reveals the existence of quantum information flow in the entanglement specification network. More precisely, for every fixed branch $    (\alpha,\beta,\gamma)\in\{0,\pi\}^{3},$ 
the corresponding initial ZX string diagram can ultimately be simplified, step by step, to a bare wire connecting the message input to the message receiver. Figure~\ref{fig-show} displays this simplification for the branch $    (\alpha,\beta,\gamma)=(0,\pi,\pi).$ Detailed proofs for all branches may be found in Ref.~\cite{Hillebrand:2011thesis}.

\end{appendix}

\newpage{\pagestyle{empty}\cleardoublepage}

\end{document}